\documentclass[11pt]{article}
\pdfoutput=1
\usepackage{jheppub}

\usepackage{amsmath} 
\usepackage{multirow} 
\usepackage{amsfonts}
\usepackage{amssymb,graphics,psfrag}
\usepackage{array,epsfig,multirow,graphicx,mathtools}
\usepackage{hyperref}
\newcommand{\be}{\begin{equation}}
\newcommand{\ee}{\end{equation}}
\newcommand{\bea}{\begin{eqnarray}}
\newcommand{\eea}{\end{eqnarray}}
  
\def\lam{{\lambda}}

\title{Topological vertex for Higgsed 5d $T_N$ theories}

\author[a]{Hirotaka Hayashi,}
\author[a, b]{Gianluca Zoccarato}

\affiliation[a]{Instituto de F\'{\i}sica Te\'orica UAM/CSIS, Cantoblanco, 28049 Madrid, Spain}
\affiliation[b]{Departamento de F\'{\i}sica Te\'orica, Universidad Aut\'onoma de Madrid, 28049 Madrid, Spain}

\emailAdd{h.hayashi@csic.es}
\emailAdd{gianluca.zoccarato@csic.es}

\abstract{We analyse the computation of the partition function of 5d $T_N$ theories in Higgs branches using the topological vertex. The theories are realised by a web of $(p,q)$ 5-branes whose dual description may be given by an M-theory compactification on a certain local non-toric Calabi-Yau
threefold. We explicitly show how it is possible to directly apply the topological vertex to the non-toric geometry.
Using this novel technique, which considerably simplifies the computation by the existing method, we are able to compute the partition function of the higher rank $E_6$, $E_7$ and $E_8$ theories. Moreover we show how in some specific cases similar results can be extended to the computation of the partition function of 5d $T_N$ theories in the Higgs branch using the refined topological vertex. These cases require a modification of the refined topological vertex.}

\begin{document}

\makeatletter
\let\old@fpheader\@fpheader
\renewcommand{\@fpheader}{\old@fpheader\hfill
IFT-UAM/CSIC-15-037}
\makeatother

\maketitle


%
%

\section{Introduction}

In recent years there has been a lot of progress in understanding superconformal field theories in various dimensions. Some of them may not admit any Lagrangian description in the ultraviolet (UV). Nevertheless they have played an important role in dualities and dynamics of supersymmetric gauge theories. In particular, the four-dimensional  $T_N$ theory\footnote{See \cite{Tachikawa:2015bga} for a review.}, which is ``non-Lagrangian'' and  has $SU(N)^3$ flavour symmetry, plays the central role in so-called class $\mathcal{S}$ theories \cite{Gaiotto:2009we}. Moreover Higgsing of the $T_N$ theory yields a large class of superconformal field theories as well. The five-dimensional version of $T_N$ theories has been constructed in \cite{Benini:2009gi} by a web of 5-branes. The flavour symmetry of the theory can be visualised by the presence of 7-branes attached to semi--infinite 5-branes \cite{DeWolfe:1999hj, Yamada:1999xr}. The Higgs branch of the theory is also explicitly realised as a space of the motion of fractionated 5-branes between the 7-branes. Moving the pieces of the 5-branes off to infinity corresponds to giving a large vacuum expectation value (vev) to hypermultiplets, and we obtain a different theory in the infrared (IR). We shall call such a theory as a Higgsed 5d $T_N$ theory and the corresponding diagram as a Higgsed diagram or non-toric diagram. The latter name indicates that its dual description corresponds to an M-theory compactification on a certain non--compact non--toric Calabi-Yau threefold.  This class of theories includes for instance $Sp(N)$ gauge theories with $N_f \leq 7$ fundamental hypermultiplets and one anti--symmetric hypermultiplet. It has been expected for a long time that these theories have an UV fixed point where the global symmetry is enhanced to $E_{N_{f+1}}$ \cite{Seiberg:1996bd, Morrison:1996xf, Douglas:1996xp, Intriligator:1997pq}. The enhancement of the global symmetries is supported from the computation of the five-dimensional superconformal index \cite{Kim:2012gu, Iqbal:2012xm, Hwang:2014uwa}. It is clear therefore that using webs of 5-branes it is possible to realise a large class of five-dimensional superconformal field theories.

One of the crucial physical questions is to compute the Nekrasov partition functions of the five-dimensional theories. Since the $T_N$ theory does not admit a Lagrangian description,\footnote{Recently, it has been noticed that a certain mass deformation of the $T_N$ theory induces an RG flow to a linear quiver theory \cite{Aganagic:2014oia, Bergman:2014kza, Hayashi:2014hfa}.} it is difficult to perform the localisation computation, initiated in \cite{Nekrasov:2002qd, Nekrasov:2003rj, Pestun:2007rz}, to obtain its partition function. However, the refined topological vertex \cite{Awata:2005fa, Iqbal:2007ii} provides a powerful tool to compute the partition function of the $T_N$ theory because its web diagram is dual to a toric Calabi--Yau threefold. After the removal of some extra factors independent of the Coulomb branch moduli it was possible to compute the partition function of the $T_N$ theory \cite{Hayashi:2013qwa, Bao:2013pwa}. The enhancement of the flavour symmetries in its superconformal index\footnote{As a matter of fact it is possible to see that the enhancement of the flavour symmetry already happens at the level of the Nekrasov partition function after a suitable shift of the Coulomb branch moduli \cite{Mitev:2014jza}.} was also checked in \cite{Hayashi:2013qwa, Bao:2013pwa, Bergman:2014kza}, and a crucial point in obtaining enhancement of the global symmetry of five-dimensional superconformal field theories was the correct removal of the extra factors independent of the Coulomb branch moduli \cite{Bergman:2013ala, Hayashi:2013qwa, Bao:2013pwa, Bergman:2013aca, Taki:2013vka, Hwang:2014uwa, Zafrir:2014ywa, Hayashi:2014wfa, Bergman:2014kza}\footnote{References \cite{Tachikawa:2015mha, Zafrir:2015uaa} discuss the enhancement of the global symmetry from the viewpoint of fermionic zero modes around one-instanton operators.}.

On the other hand the web diagrams corresponding to Higgsed 5d $T_N$ theories are not dual to toric geometries and therefore simple application of the refined topological vertex does not work. The formalism to compute their partition function was developed in \cite{Hayashi:2013qwa, Hayashi:2014wfa}\footnote{A different approach was taken in \cite{Diaconescu:2005ik, Diaconescu:2005mv} to compute unrefined topological string amplitudes for certain non--toric geometries.} by making use of the technique in \cite{Dimofte:2010tz, Taki:2010bj, Aganagic:2011sg, Gaiotto:2012uq, Gaiotto:2012xa, Aganagic:2012hs}. The prescription to compute the partition function of Higgsed 5d $T_N$ theories is a three step process. The starting point is to compute the partition function of the UV theory which embeds the IR theory by use of the refined topological vertex. That is possible if the UV theory is realised by a web diagram dual to a toric geometry. After this some of the parameters and moduli of the UV theory are suitably tuned in the partition function in order to open the Higgs branch. The result of this process is the partition function of the IR theory plus some contributions due to singlet hypermultiplets that are present in the Higgs vacuum and need to be removed to obtain the final result.  %

Although the prescription to compute the partition functions of Higgsed 5d $T_N$ theories is general it requires several steps to carry out the calculation. In particular if the diagram involves some complicated Higgsing the computation becomes more and more difficult. Since we have a web diagram which corresponds to a Higgsed 5d $T_N$ theory, it is desirable to develop a method by which we can directly apply the topological vertex to the Higgsed web diagram. The first observation was made in section 6.3 of \cite{Hayashi:2013qwa} where it was experimentally found that the refined topological vertex may be applied to some very special non-toric diagrams with some modification. In this paper, we will explore the validity of the application of the topological vertex to Higgsed diagrams in the full generality. Quite interestingly we will find that the standard topological vertex is fully applicable to Higgsed diagrams. Furthermore it will not be necessary to remove singlet hypermultiplets after Higgsing when the topological vertex is applied directly to the Higgsed diagrams, and the extra factors present in the partition function can be readily read off from the Higgsed diagram and easily removed. This new method therefore greatly simplifies the computation compared to the previous prescription. We will also exemplify the new prescription by computing the partition functions of higher rank $E_6, E_7, E_8$ theories\footnote{In fact, we will see that the constraint of the corresponding web diagrams implies the mass deformation of the theories yields $Sp(N)$ gauge theories with $N_f = 5, 6, 7$ fundamental hypermultiplets and one massless anti-symmetric hypermultiplet.}. These examples clearly show the simplification of our new method. 

Our analysis can be also extended to the refined topological vertex. Remarkably for the case with coincident external horizontal legs of a Higgsed diagram it is possible to show that one can use the Higgsed diagram to compute its partition function. However, in this case, it is necessary to use a new form of the refined topological vertex. A special limit of the case reduces to the computation in \cite{Hayashi:2013qwa}, and our new refined topological vertex 
proves the observation made in \cite{Hayashi:2013qwa}. We will also comment on some special cases of coincident external vertical and diagonal legs of a Higgsed diagram.

The organisation of this paper is as follows. In section \ref{sec:vertex}, we will present the general prescription to apply the topological vertex to Higgsed diagrams. In section \ref{sec:examples} we illustrate our method by computing the partition functions of higher rank $E_6, E_7,E_8$ theories checking also agreement with the results obtained by field theory computation. We then extend our analysis to the refined topological vertex in section \ref{sec:refined}. Appendices collect the technical details of the computational method of the refined topological vertex, the $Sp(N)$ Nekrasov partition functions as well as a short review of identities of Schur functions.

\section{Topological vertex for Higgsed 5d $T_N$ theories}
\label{sec:vertex}

In this section we discuss how it is possible to define the topological vertex to compute the topological string partition function of the 5d $T_N$ theory in the Higgs branch.
We start by showing how the topological vertex can be defined in the simple example of Higgsing parallel external legs of the diagram and then discuss the more general
case. Finally we conclude this section by discussing how the decoupled factors that appear in the
partition function can be identified in the diagram and appropriately subtracted. The computation using the refined topological vertex and the decoupled factor, including their definitions, for a toric diagram is summarised in appendix \ref{sec:app.top}.

\subsection{Higgs branch of $T_N$ theories}

\begin{figure}[t]
\begin{center}
\includegraphics[width=60mm]{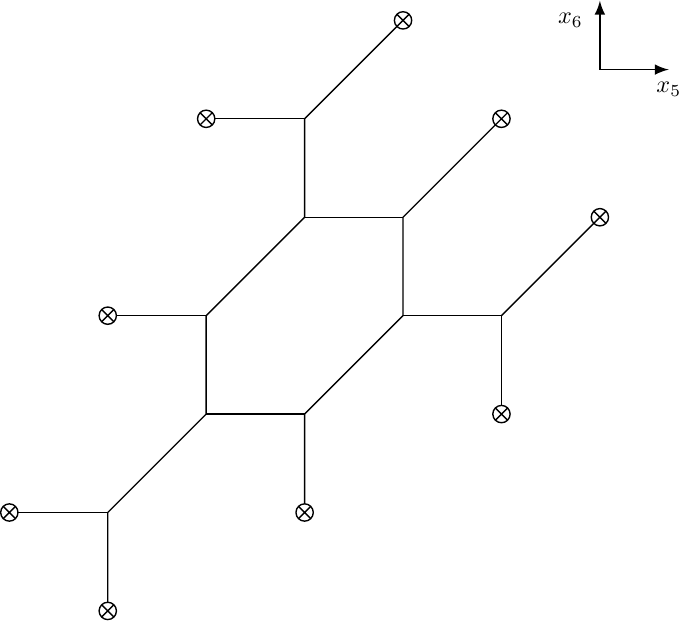}
\end{center}
\caption{The web diagram for the $T_3$ theory. Each $\otimes$ represents a 7-brane.}
\label{fig:T3}
\end{figure}

While there are different ways to engineer 5d $T_N$ theories in string and M-theory the perspective that allows to understand better the Higgs branch of these
theories involves webs of $(p,q)$ 5-branes in type IIB string theory \cite{Benini:2009gi}. 
One of the advantages of the use of webs of $(p,q)$ 5-branes, introduced in \cite{Aharony:1997ju, Aharony:1997bh}, is that, by terminating the semi-infinite external 5-branes in the diagram on 7-branes, the global symmetry becomes manifest and realised on the 7-branes \cite{DeWolfe:1999hj, Yamada:1999xr}.
The brane configuration of the system is shown in Table \ref{tb:branes}. 
\begin{table}[t]
\begin{center}
\begin{tabular}{c|c c c c | c | c c | c c c}
 & 0 & 1 & 2 & 3 & 4 & 5 & 6 & 7 & 8 & 9\\
 \hline
 D5-brane & $\times$ & $\times$ & $\times$ & $\times$ & $\times$ & $\times$ & &&& \\
 NS5-brane & $\times$ & $\times$ & $\times$ & $\times$ & $\times$ &  & $\times$ &&& \\
 (1,1) 5-brane & $\times$ & $\times$ & $\times$ & $\times$ & $\times$ & \multicolumn{2}{c|}{angle}&&& \\
 7-brane & $\times$ & $\times$ & $\times$ & $\times$ & $\times$ &  &  & $\times$ & $\times$  & $\times$
\end{tabular}
\caption{The configuration of 5-branes and 7-branes in webs. The angle in the $(x_5, x_6)$-plane is related to the charge of a $(p, q)$ 5-brane. For example, the $(1, 1)$ 5-brane corresponds to a diagonal line.}
\label{tb:branes}
\end{center}
\end{table}
The $T_N$ theory can be realised by a web of 5-branes with $N$ external D5-branes, $N$ external NS5-branes and $N$ external $(1, 1)$ 5-branes.
As an example we show
 the web diagram of the $T_3$ theory in Figure \ref{fig:T3} where we can explicitly see $SU(3) \times SU(3) \times SU(3)$ flavour symmetries realised on the 7-branes. In writing a web diagram, our convention is that the horizontal direction represents $x_5$ and the vertical direction represents $x_6$, Accordingly, horizontal lines, vertical lines and diagonal lines correspond to D5-branes, NS5-branes and $(1, 1)$ 5-branes respectively. 

\begin{figure}[t]
\begin{center}
\includegraphics[width=90mm]{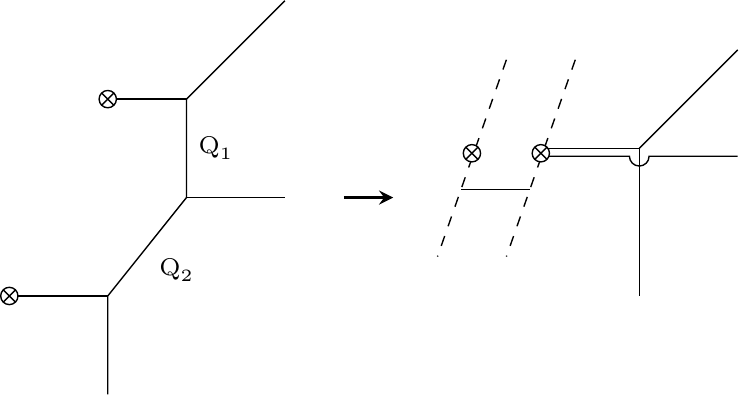}
\end{center}
\caption{An example of Higgsing by tuning the length of two 5-branes labelled as $Q_1, Q_2$. The broken lines represent the directions along $x^7, x^8, x^9$.}
\label{fig:Higgs}
\end{figure}

Moreover it is easy to realise the Higgs branch of 
these theories once 7-branes are introduced: after an appropriate tuning of the 
lengths of 5-branes in the diagrams it is possible to put some of external 5-branes on top of 
each other, and in this situation the piece of 5-brane hanging between the two 7-branes can move off the diagram as in Figure \ref{fig:Higgs}. 
Stripping off the piece of 5-branes corresponds to giving a large vev to hypermultiplets. The vev induces an RG flow and we obtain a different theory at low energy. We will call such an IR theory and an IR diagram as a Higgsed 5d $T_N$ theory and a Higgsed web diagram respectively. This particular class of Higgsing will be the main focus of this paper.

It is important to check that the Higgsed diagram preserves supersymmetry. For example only a single D5-brane can be connected to an NS5-brane without breaking supersymmetry \cite{Hanany:1996ie,Benini:2009gi}. Taking into 
account this and its $SL(2,\mathbb{Z})$ duals in a supersymmetric Higgsed diagram some of the 5-branes will be forced to jump over some other 5-branes like in the example in Figure  \ref{fig:Higgs}.

All the possible ways to put external 5-branes on 7-branes in a $T_N$ diagram can be nicely represented by a partition of $N$. A partition of $N$, namely a set of positive integers $[n_1, n_2, \cdots, n_k]$ such that $\sum_{i=1}^k n_i=N$, corresponds to a configuration where $n_i$ 5-branes are put on the same 7-brane. If $w_j$ of the $n_i$'s coincide in the partition, the flavour symmetry is $S[\prod_j U(w_j)]$. For instance the configuration of the external D5-branes of the $T_N$ theory is represented by the partition $[1, \cdots, 1]$ with $N$ $1$'s and this gives $SU(N)$ flavour symmetry as expected. The Higgsing shown in Figure \ref{fig:Higgs} represents changing the partition $[\dots,1, 1,\dots]$ to $[\dots,2,\dots]$. Quite interestingly these partitions
coincide with the ones which appear at a puncture of a Riemann surface characterising class $\mathcal{S}$ theory. Four-dimensional class $\mathcal{S}$ theories are constructed by compactifying a six-dimensional $(2, 0)$ theory on a Riemann surface with punctures \cite{Gaiotto:2009we}.
5-brane webs give  five-dimensional versions of class $\mathcal{S}$ theories and the partition which classifies the external 5-branes configuration corresponds to the Young diagram at the puncture. Due to this correspondence, we will often call the configuration of external 5-brane ending on 7-branes a puncture.

Another advantage of the 5-brane web is that we can compute the partition function of the 5d theory realised by the web diagram. The web diagram of the $T_N$ theory in type IIB theory is dual to a toric Calabi--Yau threefold in M-theory and in this dual picture the five-dimensional theory is realised as a low energy effective theory from the compactification of M-theory on the toric Calabi--Yau threefold. The partition function of the theory can be computed by the powerful technique of the refined topological vertex \cite{Awata:2005fa, Iqbal:2007ii} after eliminating what we call decoupled factor which is a contribution associated to strings between parallel external legs \cite{Hayashi:2013qwa, Bao:2013pwa, Bergman:2013aca}. Technical details of the refined topological vertex as well as the decoupled factor are summarised in appendix \ref{sec:app.top}.

The computation of the partition functions of the Higgsed 5d $T_N$ theories is more involved. Due to some jumps of 5-branes over other 5-branes Higgsed diagrams are not dual to toric Calabi--Yau threefolds and therefore we cannot directly apply the refined topological vertex formalism. However it is still possible to compute the partition function of the IR theory realised by a Higgsed diagram
by first computing the partition function of a UV theory which embeds the IR theory and then applying a suitable tuning condition to the parameters of the UV partition function. After eliminating some singlet hypermultiplets in the Higgs vacuum we obtain the partition function of the IR theory \cite{Hayashi:2013qwa, Hayashi:2014wfa}. For example the Higgsing in Figure \ref{fig:Higgs} requires a tuning
\begin{equation}
Q_1 = \left(\frac{q}{t}\right)^{\frac{1}{2}}\,, \qquad Q_2 = \left(\frac{q}{t}\right)^{\frac{1}{2}}\,, \label{refined.higgs1}
\end{equation} 
or 
\begin{equation}
Q_1 = \left(\frac{t}{q}\right)^{\frac{1}{2}}\,, \qquad Q_2 = \left(\frac{t}{q}\right)^{\frac{1}{2}}\,, \label{refined.higgs2}
\end{equation} 
where the two conditions give equivalent results of the Nekrasov partition function \cite{Hayashi:2013qwa, Hayashi:2014wfa}. Here $Q_{1,2} := e^{-L_{1,2}}$ where $L$ is the length of the 5-branes. In the dual description the length is the size of a corresponding two-cycle. Hence, we will often call K\"ahler parameter to refer to $Q$. $q, t$ are given by $q = e^{-\epsilon_2}, t = e^{\epsilon_1}$ where $\epsilon_1, \epsilon_2$ are the chemical potentials associated to the symmetries $SO(2) \times SO(2) \subset SO(4)$, and $SO(4)$ is the little group of the five-dimensional spacetime.

\subsection{Topological vertex, external legs}
\label{sec:top.external}
\begin{figure}[t]
\begin{center}
\includegraphics[width=50mm]{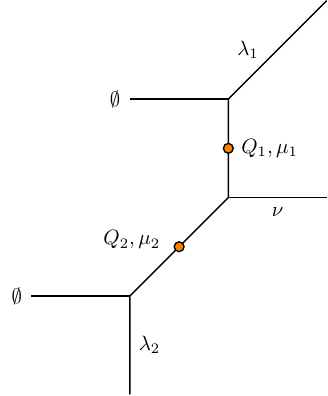}
\end{center}
\caption{Higgsing of parallel horizontal external legs in a $T_N$ diagram. The orange dots indicate the lines that are shrunk to zero length, and hence we use $Q =1$ in the computation of the topological string partition function.}
\label{fig:horz}
\end{figure}
While the procedure described in the previous subsection for the computation of the partition function of 5d $T_N$ theories in the Higgs branch is totally general it is 
quite clear that it is not an extremely efficient way to carry on this kind of computations. An improvement would be to directly have a formulation of the topological 
vertex that can be applied to web diagrams that are not dual to toric geometries but that can be obtained by toric web diagrams after a suitable tuning of its K\"ahler
parameters. Here we will show how this is indeed possible for the case of the topological vertex, discussing the simple example of placing external legs placed on top of 
each other and deferring the general case to the next subsection. 

The setup we have in mind is the one shown in Figure \ref{fig:horz}. The topological string partition function
for this kind of local diagram can be easily computed for the diagram is toric
\be\label{eq:part1}
Z= \sum_{\mu_1,\mu_2} (-1)^{|\mu_1|+|\mu_2|} C_{ \lambda_1\mu_1\emptyset}(q) C_{ \mu_2\mu_1^t \nu^t}(q) C_{ \mu_2^t\lambda_2 \emptyset}(q)\,.
\ee
The topological vertex is defined as \eqref{topological.vertex} with $t = q$. Note that we have set two K\"ahler parameters $Q_{1, 2}$ to $1$ since the corresponding two lines are shrunk to zero size. This corresponds to the unrefined version, $q=t$, of the tuning condition \eqref{refined.higgs1} or \eqref{refined.higgs2}. It is quite straightforward to perform the Young diagram summations on $\mu_1$ and $\mu_2$ using well known identities on Schur functions \eqref{eq:schurid1} and \eqref{eq:schurid2}, and after 
these summations the partition function \eqref{eq:part1} becomes
\be\label{eq:part2}\begin{split}
Z&= \sum_{\eta_1,\eta_2,\kappa_1,\kappa_2,\kappa_3} (-1)^{|\kappa_1|+|\kappa_2|+|\eta_1|+|\eta_2|} q^{\frac{||\lambda_2||^2-||\lambda_2^t||^2
+||\nu^t||^2}{2}}\tilde Z_{\nu}(q)\times\\
&\times s_{\lambda_1^t/\eta_1}(q^{-\rho})s_{\eta_1^t/\kappa_1^t}(q^{-\nu-\rho}) s_{\kappa_1/\kappa_3}(q^{-\rho})s_{\kappa_2/\kappa_3}
(q^{-\rho}) s_{\eta_2^t/\kappa_2^t}(q^{-\nu^t-\rho}) s_{\lambda_2 / \eta_2}(q^{-\rho})\times\\
&\times \prod_{i,j=1}^{\infty} \frac{(1-q^{i+j-1-\nu_{j}})(1-q^{i+j-1-\nu_{i}^t})}{(1-q^{i+j-1})}\,.
\end{split}\ee
The last terms in \eqref{eq:part2} are very important at this level, note in fact that if $\nu \neq\emptyset $ the product will always be zero. This greatly simplifies the 
expression and using the identity \eqref{eq:schur2} of Schur functions we can arrive to the simpler expression
\be\begin{split}
Z&= \sum_{\kappa_3}  q^{\frac{||\lambda_2||^2-||\lambda^t_2||^2}{2}}
 s_{\lambda_1^t/\kappa_3}(q^{-\rho}) s_{\lambda_2 / \kappa_3}(q^{-\rho}) \prod_{i,j=1}^{\infty} (1-q^{i+j-1 }) =  C_{\lambda_1 \lambda_2 \emptyset}(q)\prod_{i,j=1}^{\infty} (1-q^{i+j-1 }) \,.
\end{split}\ee
We see therefore that, up to an infinite product factor, the partition function reduces to a simple topological vertex. Since the infinite product factor will eventually 
contribute only to decoupled factors in the partition function of a diagram we will cancel it already at this level and propose that the partition function can be computed simply
applying the usual topological vertex rule in the Higgsed diagram.

Note that we have carried out the computation with a specific choice of the ordering of the Young diagrams in the topological vertices. A similar computation can be 
performed for other different choices of orderings in the topological vertices obtaining a result which is always consistent with the cyclic symmetry of the topological
vertex.

\subsection{Topological vertex, general case}
\begin{figure}[t]
\begin{center}
\includegraphics[width=50mm]{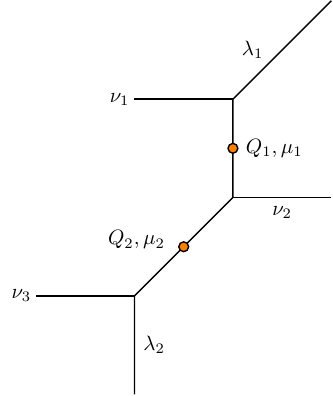}
\end{center}
\caption{Higgsing of parallel horizontal legs in a $T_N$ diagram. The orange dots indicate the lines that are shrunk to zero length.}
\label{fig:horz2}
\end{figure}

The procedure described in the previous subsection was limited to a simple case, namely the case in which the branes that are placed on top of each other are external
in the diagram. However, while in order to enter the Higgs branch of the $T_N$ theory placing external branes on top of each other is the starting point, the propagation
of the generalised s-rule \cite{Hanany:1996ie,Benini:2009gi} inside the diagram leads in some cases to a situation in which some internal branes are placed on top of each other, so that it is necessary to
have a rule for the topological vertex for this more complicated case. We will follow the same strategy of the previous computation, we show in Figure \ref{fig:horz2} the diagram
we are considering in this case.

The topological string partition function for the local diagram can be computed as usual using the topological vertex
\be
Z(\nu_1, \nu_3; \nu_2, \lambda_1, \lambda_2)= \sum_{\mu_1,\mu_2} (-Q_1)^{|\mu_1|}(-Q_2)^{|\mu_2|} C_{\lambda_1^t\mu_1  \nu_1^t}(q) C_{\mu_2 \mu_1^t \nu_2^t}(q) C_{\mu_2^t \lambda_2 \nu_3^t}(q)\,,
\ee
where for the moment we did not impose the tuning condition $Q_1 = Q_2 =1$. Again it is quite straightforward to perform the summations on the Young diagrams $\mu_1$ and $\mu_2$ giving the following result
\be\label{eq:part3}\begin{split}
Z(\nu_1, \nu_3; \nu_2, \lambda_1, \lambda_2)&= \sum_{\eta_1,\eta_2,\kappa_1,\kappa_2,\kappa_3} (-Q_1)^{|\eta_1|+|\kappa_2|-|\kappa_3|}  (-Q_2)^{|\eta_2|+|\kappa_1|-|\kappa_3|}q^{\frac{||\lambda_2||^2-||\lambda^t_2||^2+||\nu^t_1||^2+||\nu^t_2||^2+||\nu^t_3||^2}{2}}\\
&\times s_{\lambda_1/\eta_1}(q^{-\nu^t_1-\rho})s_{\eta_1^t/\kappa_1^t}(q^{-\nu_2-\rho}) s_{\kappa_1/\kappa_3}(q^{-\nu_3^t-\rho})\\
&\times s_{\kappa_2/\kappa_3}(q^{-\nu_1-\rho}) s_{\eta_2^t/\kappa_2^t}(q^{-\nu_2^t-\rho}) s_{\lambda_2/ \eta_2}(q^{-\nu_3-\rho})\\
&\times \tilde Z_{\nu_1}(q)\tilde Z_{\nu_2}(q)\tilde Z_{\nu_3}(q)\prod_{i,j=1}^{\infty} \frac{(1-Q_1q^{i+j-1 -\nu_{1,i}-\nu_{2,j}})(1-Q_2 q^{i+j-1-\nu_{2,i}^t-\nu_{3,j}^t})}{(1-Q_1 Q_2\,q^{i+j-1-\nu_{1,i}-\nu_{3,j}^t})}\,.
\end{split}\ee
As in the previous case it is extremely important to carefully look at the last factors in \eqref{eq:part3}, and in this case we find that it is important to define properly how to impose the tuning condition
$Q_1=Q_2=1$. We start by setting $Q_1=Q_2 = Q$ and then finally take the limit for $Q$ going to 1. We observe the following behaviour of the infinite product term
\bea
\lim_{Q\rightarrow 1}\prod_{i,j=1}^\infty \frac{(1-Q q^{i+j-1 -\nu_{1,i}-\nu_{2,j}})(1-Q q^{i+j-1-\nu_{2,i}^t-\nu_{3,j}^t})}{(1-Q^2\,q^{i+j-1-\nu_{1,i}-\nu_{3,j}^t})} &=& \left[\lim_{Q\rightarrow 1}\left(\frac{1-Q}{1-Q^2}\right)^n\right] F^{(\nu_1, \nu_3)}_{\nu_2}(q)\nonumber\\
&=& \frac{1}{2^n}\; F^{(\nu_1, \nu_3)}_{\nu_2}(q), \label{limit1}
\eea
where $n$ is the number of zero terms in the product $\prod_{i,j=1}^\infty (1-Q q^{i+j-1-\nu_{1,i}-\nu_{3,j}^t})$ when we set $Q=1$. When $\nu_1 = \nu_2^t$ or $\nu_3 = \nu_2^t$, the explicit expression of $F^{(\nu_1, \nu_3)}_{\nu_2}(q)$ is 
\be
\prod_{i,j=1}^\infty (1-Q q^{i+j-1-\nu_{1,i}-\nu_{1,j}^t})\quad \text{ or }\quad\prod_{i,j=1}^\infty (1-Q q^{i+j-1-\nu_{3,i}-\nu_{3,j}^t}),
\ee
respectively. It is important to note that \eqref{limit1} can be non--zero even if both $\nu_1 = \nu_2^t$ and $\nu_3 = \nu_2^t$ are not satisfied.

As opposed to the intuition from the  Higgsed web diagram in Figure \ref{fig:horz2}, neither $\nu_1 = \nu_2^t$ nor $\nu_3 = \nu_2^t$ is required for a non-zero result of \eqref{eq:part3}. Even in the case with  $\nu_1 = \nu_2^t$ or $\nu_3 = \nu_2^t$, the infinite product factor gives the odd ``weight'' $\frac{1}{2^n}$ as in \eqref{limit1}. When we simply focus on the case of $\nu_1 = \nu_2^t$ of \eqref{eq:part3},  it is possible to use the identity \eqref{eq:schur1} and arrive at the following result
\be\begin{split}
Z(\nu_1, \nu_3; \nu_1^t, \lambda_1, \lambda_2) &=q^{\frac{||\lambda_2||^2-||\lambda^t_2||^2+||\nu_1||^2+||\nu_1^t||^2+||\nu_3||^2}{2}}\tilde Z^2_{\nu_1}(q)\tilde Z_{\nu_3}(q)\\&\times \frac{1}{2^n}\prod_{i,j=1}^\infty (1-q^{i+j-1 -\nu_{1,i}-\nu^t_{1,j}})\sum_{\kappa_3} 
s_{\lambda_1/\kappa_3}(q^{-\nu_3^t-\rho})s_{\lambda_2/\kappa_3}(q^{-\nu_3-\rho}) \\
&=\frac{1}{2^n}
q^{\frac{||\nu_1||^2+||\nu_1^t||^2}{2}}\tilde Z^2_{\nu_1}(q) \, C_{\lambda_1^t \lambda_2 \nu_3^t}(q)\prod_{i,j=1}^\infty (1-q^{i+j-1 -\nu_{1,i}-\nu^t_{1,j}})\,,
\end{split}\ee
where we also used that $\tilde Z_\nu(q) = \tilde Z_{\nu^t} (q)$. At this point note that
\be\begin{split}
\prod_{i,j=1}^{\infty} (1-q^{i+j-1 -\nu_{i}-\nu^t_{j}})&= \prod_{i,j=1}^\infty (1-q^{i+j-1}) \prod_{s \in \nu} (1-q^{l_\nu(s)+a_\nu(s)+1}) (1-q^{-l_\nu(s)-a_\nu(s)-1})\\
&=(-1) ^{|\nu|}\prod_{i,j=1}^\infty (1-q^{i+j-1}) \prod_{s \in \nu} (1-q^{l_\nu(s)+a_\nu(s)+1}) ^2q^{-l_\nu(s)-a_\nu(s)-1}\\&= (-1) ^{|\nu|}\prod_{i,j=1}^\infty (1-q^{i+j-1}) \tilde Z_\nu^{-2}(q) q^{-\frac{||\nu||^2+||\nu^t||^2}{2}}\,,
\end{split}\ee
so that in the end we arrive at the quite simple result
\be\label{weight.vertex}
Z(\nu_1, \nu_3; \nu_1^t, \lambda_1, \lambda_2) =  \frac{1}{2^n} (-1)^{|\nu_1|}\, C_{\lambda_1^t \lambda_2 \nu_3^t}(q)\,\prod_{i,j=1}^\infty (1-q^{i+j-1 })\,,
\ee
Therefore, we obtain the single topological vertex as in section \ref{sec:top.external} but the weight $ \frac{1}{2^n}$ appears in the final expression. Also \eqref{weight.vertex} is not exactly equal to \eqref{eq:part3} with $Q_1 = Q_2 = 1$ since we ignore the cases where $\nu_1 \neq \nu_2^t$. The case with $\nu_3 = \nu_2^t$ also yields the same expression with $\nu_1$ exchanged with $\nu_3$ and we again have the weight $\frac{1}{2^n}$.

While these facts may lead to think that it is not possible to define a variation of the vertex rule for the diagram in Figure \ref{fig:horz2} we will now argue that, after the properly gluing the remaining
contributions involving the Young diagrams $\nu_1$ and $\nu_3$, the rule for the computation is rather simple. Since we assume that both $\nu_1$ and $\nu_3$ are non-trivial, those legs should be glued to some other legs in a complete diagram. Hence, in order to compute the partition function of the Higgsed diagram, it is enough if we obtain the same result after performing the Young diagram summations of $\nu_1, \nu_3$. In fact, it will turn out that the computation by replacing \eqref{eq:part3} with the single topological vertex as in \eqref{weight.vertex} but without $\frac{1}{2^n}$ exactly yields the same result as the original one of \eqref{eq:part3} after summing up the Young diagrams $\nu_1, \nu_3$. 

\begin{figure}[t]
\begin{center}
\includegraphics[width=50mm]{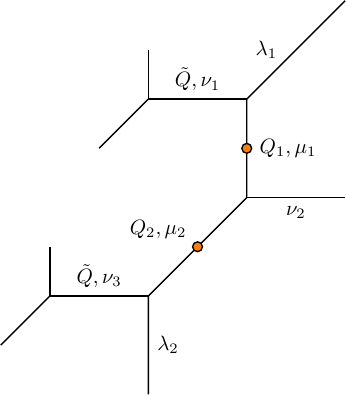}
\end{center}
\caption{Higgsing of parallel horizontal legs in a $T_N$ diagram with some additional parts of the diagram glued. The orange dots indicate the lines that are shrunk to zero length.}
\label{fig:horz3}
\end{figure}

\begin{figure}[t]
\begin{center}
\includegraphics[width=110mm]{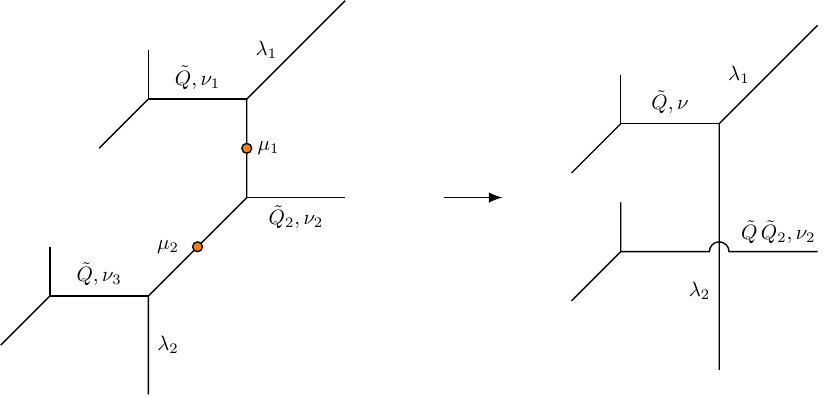}
\end{center}
\caption{The replacement of the vertical Higgsed strip part with a trivalent vertex after gluing two horizontal legs.}
\label{fig:newv}
\end{figure}

To this end we consider the diagram in Figure \ref{fig:horz3} and compute its local 
contribution to the topological string partition function. The result is the following one
\bea
Z(\nu_2, \lambda_1, \lambda_2)&=&\sum_{\nu_1, \nu_3}(-\tilde Q)^{|\nu_1|+|\nu_3|}C_{\emptyset \emptyset \nu_1}(q)C_{\emptyset \emptyset \nu_3}(q)Z(\nu_1, \nu_3; \nu_2, \lambda_1, \lambda_2) \nonumber\\
&=&q^{\frac{||\nu_2^t||^2+||\lambda_2||^2-||\lambda^t_2||}{2}}\tilde{Z}_{\nu_2}(q)\nonumber \\&&\sum_{\nu_1,\nu_3}
\frac{1}{2^n}\;(-\tilde Q)^{|\nu_1|+|\nu_3|}Z_1^{\{\nu_1,\nu_3\}}(\nu_2,\lambda_1,\lambda_2)\,,
\label{comp1}
\eea
where we defined
\be\begin{split}
Z_1^{\{\nu_1,\nu_3\}}(\nu_2,\lambda_1,\lambda_2)&= \sum_{\eta_1,\eta_2,\kappa_1,\kappa_2,\kappa_3} (-1)^{|\eta_1|+|\eta_2|+|\kappa_1|+|\kappa_2|-2|\kappa_3|} \\
&\times s_{\lambda_1/\eta_1}(q^{-\nu^t_1-\rho})s_{\eta_1^t/\kappa_1^t}(q^{-\nu_2-\rho}) s_{\kappa_1/\kappa_3}(q^{-\nu_3^t-\rho})\\&\times s_{\kappa_2/\kappa_3}
(q^{-\nu_1-\rho}) s_{\eta_2^t/\kappa_2^t}(q^{-\nu_2^t-\rho}) s_{\lambda_2/ \eta_2}(q^{-\nu_3-\rho})\\
&\times q^{\frac{||\nu_1||^2+||\nu^t_1||^2+||\nu_3||^2+||\nu^t_3||^2}{2}}\tilde Z^2_{\nu_1}(q)\tilde Z^2_{\nu_3}(q)\; F^{(\nu_1, \nu_3)}_{\nu_2}(q)\,,
\label{Z.double}
\end{split}\ee
We then compare the result \eqref{comp1} with another computation by replacing the vertical strip diagram in Figure \ref{fig:horz3} with the single topological vertex as in Figure \ref{fig:newv}. 
The topological string partition function computed from the local diagram of the right figure in Figure \ref{fig:newv} is 
\bea
Z'(\nu_2, \lambda_1, \lambda_2) &=&\tilde{Q}^{|\nu_2|} C_{\emptyset\emptyset\nu_2^t}(q)\sum_{\nu}(-\tilde{Q})^{|\nu|}C_{\emptyset\emptyset\nu}(q)C_{\lambda_1^t\lambda_2\nu^t}(q)\nonumber\\
&=&\tilde{Q}^{|\nu_2|}\left[q^{\frac{||\nu_2^t||^2}{2}}\tilde{Z}_{\nu_2}(q)\right]q ^{\frac{||\lambda_2||^2-||\lambda^t_2||}{2}}\sum_{\nu}(-\tilde{Q})^{|\nu|}Z_2^{\{\nu\}} (\lambda_1,\lambda_2), \label{Z.single.glue}
\eea
where we omitted a factor $(-\tilde{Q}_2)^{|\nu_2|}$ since this is not taken into account in \eqref{comp1} either. We also defined
\be
Z_2^{\{\nu\}} (\lambda_1,\lambda_2) =q^{\frac{||\nu||^2+||\nu^t||^2}{2}}\tilde Z^2_{\nu}(q)\sum_{\eta} s_{\lambda_1/\eta}(q^{-\nu^t-\rho}) s_{\lambda_2/\eta}(q^{-\nu-\rho})\,, \label{Z.single}
\ee

We argue that the summation over two Young diagrams $\nu_1$ and $\nu_3$ by \eqref{Z.double} can be precisely reproduced by the single Young diagram summation of $\nu$ by \eqref{Z.single} up to a irrelevant infinite product. 
More specifically, we propose that\footnote{Actually we observe that something stronger than \eqref{eq:vertnont} holds. In fact per each Young diagram $\nu$ there exist a set
of pairs of Young diagrams $\{\nu_1,\nu_3\}$ such that the sum on the left hand side of \eqref{eq:vertnont} restricted to these pairs correctly reproduces $Z_2^{\{\nu\}}(\lambda_1,\lambda_2)$. In the following explicit examples we will see the occurrence  of this phenomenon. }
\be\label{eq:vertnont}
\sum_{\substack{\nu_1,\nu_3\\|\nu_1|+|\nu_3|=\kappa}}\frac{1}{2^n}\;(-\tilde Q)^{|\nu_1|+|\nu_3|} Z_1^{\{\nu_1,\nu_3\}}(\nu_2,\lambda_1,\lambda_2) = 
\tilde Q^{|\nu_2|}\sum_{\substack{\nu\\|\nu|=\kappa-|\nu_2|}} (-\tilde Q)^{|\nu|}Z_2^{\{\nu\}}(\lambda_1,\lambda_2)\prod_{i,j=1}^\infty(1-q^{i+j-1})\,.
\ee
Note that the summation in \eqref{eq:vertnont} is taken over for fixed $\kappa$. We have checked this relation for different choices of $\lambda_1$, $\lambda_2$ and $\nu_2$ and for different values of $\kappa$ finding always perfect agreement. The appearance of the $\tilde Q^{|\nu_2|}$ in \eqref{eq:vertnont} is consistent with the fact that in Figure \ref{fig:newv} in the Higgsed diagram the $\nu_2$ summation
is associated with a leg whose K\"ahler parameter is $\tilde Q \tilde Q_2$. In other words, $\tilde Q^{|\nu_2|}$ factor of \eqref{eq:vertnont} reproduces the $\tilde Q^{|\nu_2|}$ factor appearing in \eqref{Z.single.glue}.

In order to obtain clearer picture of how the relation \eqref{eq:vertnont} holds, let us see it more explicitly\footnote{In the examples we are going to discuss we shall forget about the infinite product appearing in \eqref{eq:vertnont}.}. For example, \eqref{Z.single} with $\nu=[2]$ for $\lambda_1=[1], \lambda_2=[1,2]$  is reproduced by the relation
\be
\frac{1}{2}Z_1^{\{[2],\emptyset\}}(\emptyset,[1],[1,1])+\frac{1}{2}Z_1^{\{\emptyset,[2]\}}(\emptyset,[1],[1,1])=Z_2^{\{[2]\}}([1],[1,1])\,.\label{exp.young}
\ee
with 
\be
 Z_1^{\{[2],\emptyset\}}(\emptyset,[1],[1,1]) = Z_1^{\{\emptyset,[2]\}}(\emptyset,[1],[1,1]) = Z_2^{\{[2]\}}([1],[1,1]),
\ee
which is essentially due to the relation \eqref{weight.vertex}. The factors of $\frac{1}{2}$ in \eqref{exp.young} represent the weight $\frac{1}{2^n}$ with $n=1$. Note also that the other $\kappa = 2$ contribution from $Z_1^{\{[1],[1]\}}(\emptyset, [1], [1,1])$ is zero. Therefore, the effect of the weight $\frac{1}{2}$ in this case is precisely to reproduce the single Young diagram dependence from the double Young diagram dependence. Namely, the odd looking weight $\frac{1}{2^n}$ with $n=1$ is really necessary to reproduce the single Young diagram summation from the double Young diagram summations. 

One less trivial case with $n > 1$ is, for example, $(\nu_1, \nu_3) = ([2, 2], \emptyset)$ with $\nu_2, \lambda_1, \lambda_2$ being trivial. In this case, we have a weight $\frac{1}{2^2}$. However, there are other non-zero results with $\kappa=4$
 coming from $(\nu_1, \nu_3) = ([2,1],[1])$ and $([2],[1,1])$. Their weights are also $\frac{1}{2^2}$, and a non--trivial summation relation holds,
\be
Z_1^{\{[2,2],\emptyset\}}(\emptyset,\emptyset,\emptyset) = Z_1^{\{[2,1],[1]\}}(\emptyset,\emptyset,\emptyset) + Z_1^{\{[2],[1,1]\}}(\emptyset,\emptyset,\emptyset).  \label{relation}
\ee
Hence, the sum of the three terms becomes
\be
\frac{1}{4} \left[Z_1^{\{[2,2],\emptyset\}}(\emptyset,\emptyset,\emptyset)+ Z_1^{\{[2,1],[1]\}}(\emptyset,\emptyset,\emptyset)+Z_1^{\{[2],[1,1]\}}(\emptyset,\emptyset,\emptyset) \right]=\frac{1}{2}Z_1^{\{[2,2],\emptyset\}}(\emptyset,\emptyset,\emptyset)
\ee
The same phenomenon happens when $\nu_1$ is exchanged with $\nu_3$, and we finally obtain 
\be
\frac{1}{2}Z_1^{\{[2,2],\emptyset\}}(\emptyset,\emptyset,\emptyset) + \frac{1}{2}Z_1^{\{\emptyset, [2,2]\}}(\emptyset,\emptyset,\emptyset) =  Z_2^{\{[2,2]\}}(\emptyset,\emptyset)\,. \label{exp.young2}
\ee
We again see that the double Young diagram dependence becomes the single Young diagram dependence.

By using \eqref{eq:vertnont}, we have found that\footnote{In the following we shall remove the infinite product term that appears in \eqref{eq:vertnont} because this is only a contribution due to singlet hypermultiplets.}
\be
Z(\nu_2, \lambda_1, \lambda_2) = Z'(\nu_2, \lambda_1, \lambda_2)
\,. \label{double.vs.single}
\ee
Namely, after the summation, we obtain the same result up to the infinite product factor when we replace the vertical strip part of the diagram in Figure \ref{fig:horz3} with a trivalent vertex as in Figure \ref{fig:newv}. 
\begin{figure}[t]
\begin{center}
\includegraphics[width=110mm]{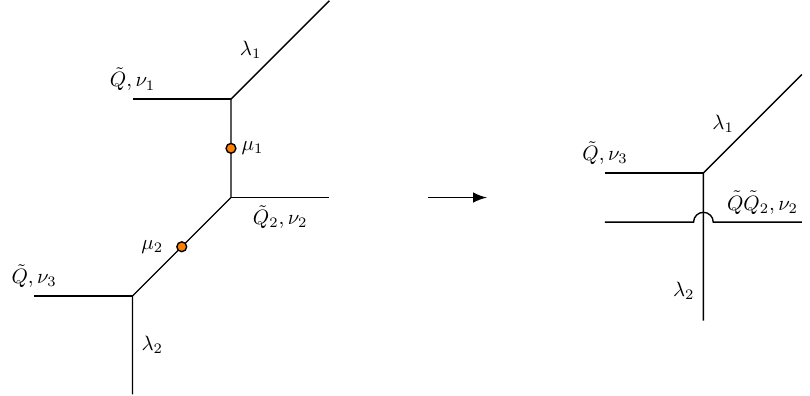}
\end{center}
\caption{The general procedure for the computation of the topological string partition function in a Higgsed diagram.}
\label{fig:newv2}
\end{figure}
This leads to a proposal that we can always perform the replacement of Figure \ref{fig:newv2} for the computation of the partition function of theories from Higgsed diagrams since both methods give the same result. Therefore, for the Higgsing in Figure \ref{fig:horz2}, we can again use the topological vertex\footnote{Or we can also use 
\be
Z(\nu_1, \nu_3; \nu_2, \lambda_1, \lambda_2)\rightarrow  (-1)^{|\nu_2|}\, C_{\lambda_1^t \lambda_2 \nu_1^t}(q)\,
\ee
with $\nu_2^t = \nu_3$. The two ways of the replacement give the same result in the end for the computation of a complete diagram.}
\be
Z(\nu_1, \nu_3; \nu_2, \lambda_1, \lambda_2)\rightarrow  (-1)^{|\nu_2|}\, C_{\lambda_1^t \lambda_2 \nu_3^t}(q)\, \label{replace.general}
\ee
with $\nu_2^t = \nu_1$. Namely, using this rule which pictorially we illustrate in Figure \ref{fig:newv2} it is therefore possible to directly compute the topological string partition function applying the topological vertex to the 
Higgsed diagram. 

Note moreover that the vertex \eqref{replace.general} that replaces the original diagram contains a peculiar factor of the form $(-1)^{|\nu|}$. The presence of this factor is actually perfectly
consistent and allows us to compute the topological string partition function in the Higgsed diagram with the usual rules of the topological vertex. We show in Figure \ref{fig:newv2} how the K\"ahler parameters are assigned in the case of gluing some legs, and the presence of the additional $(-1)^{|\nu|}$ factor simply allows to compute the topological string partition function introducing the usual factor $(-Q)^{|\nu|}$ where we called $Q$ the total K\"ahler parameter of the two glued legs. In the example of Figure \ref{fig:newv2}, the $Q$ is $\tilde{Q}\tilde{Q}_2$.

\begin{figure}[t]
\begin{center}
\includegraphics[width=120mm]{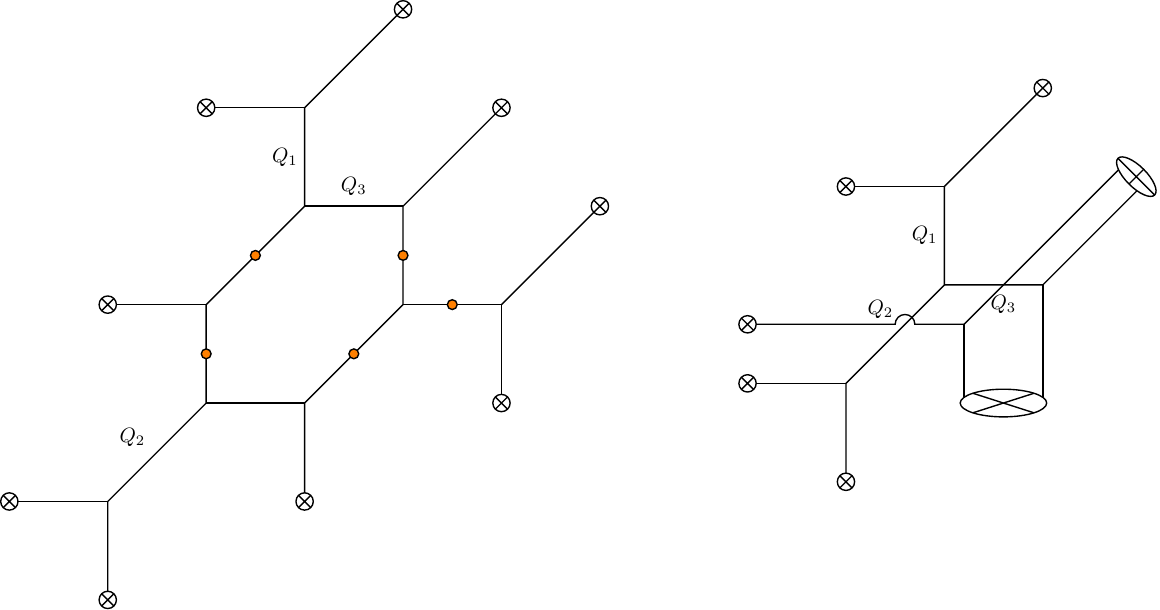}
\end{center}
\caption{On the left: a particular Higgs branch of a $T_3$ diagram. On the right: the diagram that allows to compute the topological string partition function with a single gluing.}
\label{fig:higT3}
\end{figure}

We would also like to discuss an explicit example to further corroborate our conjecture. The diagram we consider is the one in Figure \ref{fig:higT3}. We can first try to compute the result by simply 
applying the topological vertex to the Higgsed web diagram of the right figure of Figure \ref{fig:higT3}. Since the Higgsed web diagram consists of a single $T_2$ diagram\footnote{The trivalent vertex with trivial representations on all the legs gives a trivial contribution. } it is possible to compute the topological string partition function performing all Young 
diagram summations and the result is simply \cite{Kozcaz:2010af, Hayashi:2013qwa, Bao:2013pwa}
\be
Z^{\text{Higgs }T_3\text{-1}} = \tilde{Z}_{T_2} = \prod_{i,j=1}^\infty \frac{(1-Q_1 Q_2 Q_3 q^{i+j-1})\prod_{k=1}^3 (1-Q_k q^{i+j-1})}{(1-Q_1 Q_2 q^{i+j-1})(1-Q_2 Q_3 q^{i+j-1})(1-Q_1 Q_3 q^{i+j-1})}\,.
\ee
Let us compare the result with the one computed by the original method using the tuned UV diagram of the left figure of Figure \ref{fig:higT3}. In this example it is also possible to directly use the result of the topological string partition function for a Higgsed $T_3$ computed in \cite{Hayashi:2014wfa} and perform an additional Higgsing in the 
diagram. The result is simple and can be written in terms of infinite products
\be
Z^{\text{Higgs }T_3\text{-2}} = \prod_{i,j=1}^\infty \frac{(1-q^{i+j-1})^2(1-Q_1 Q_2 Q_3 q^{i+j-1})\prod_{k=1}^3 (1-Q_k q^{i+j-1})}{(1-Q_1 Q_2 q^{i+j-1})(1-Q_2 Q_3 q^{i+j-1})(1-Q_1 Q_3 q^{i+j-1})}\,.
\ee
We see therefore that up to some irrelevant singlet hypermultiplets the two results perfectly agree providing further evidence for the procedure we described.


Finally one last comment regarding the cyclic symmetry of the topological vertex. While we have done our computation with a very specific choice of ordering of the Young
diagrams in the topological vertex it is possible to do the computation with all other possible orderings and the result is always consistent with the cyclic symmetry of the
topological vertex (although it is not necessary to appeal to it to do the computation).
\\

\subsection{Decoupled factors for Higgsed 5d $T_N$ theories}
\label{sec:decoupled}
In this section we discuss how to correctly identify the decoupled factors in the Higgsed diagram. It is nice if a simple variant of the usual rule for the identification
of these factors can be applied to Higgsed diagrams leading to a very simple procedure for the subtraction of these contributions in the partition function. We will see in this subsection that this is indeed the case. Moreover
we will see that a great advantage of using the vertex rule for the Higgsed diagram is that it is no longer necessary to identify and cancel the contributions of singlet hypermultiplets
because these contributions are not present at all in the Higgsed diagram.
\begin{figure}[t]
\begin{center}
\includegraphics[width=40mm]{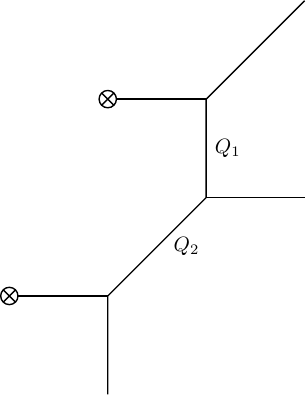}
\end{center}
\caption{Diagram contributing to decoupled factors in the topological string partition function.}
\label{fig:dec}
\end{figure}

We start by recalling the usual rule for the identification of decoupled factors in a toric diagram which is more easily described in the M-theory dual \cite{Hayashi:2013qwa,Bao:2013pwa}: decoupled factors can be identified as the contributions
coming from M2-branes wrapping curves with zero intersection number with any compact divisor of the geometry. In particular these curves can be continuously moved
off to infinity and the states carry charge only under global symmetries of the 5d theory. We show in Figure \ref{fig:dec} the basic example of a diagram containing decoupled factors whose 
contribution to the topological string partition function has the following form
\be
Z_{dec} = \prod_{i,j=1}^\infty (1-Q_1 Q_2 q^{i+j-1})^{-1}\,.
\ee
Since decoupled factors always involve pairs of parallel external legs we will always refer to them as decoupled factors coming from parallel external legs.
The claim of \cite{Bergman:2013ala, Hayashi:2013qwa,Bao:2013pwa, Bergman:2013aca} is that the correct partition function of the 5d field theory can be computed from the topological string partition function by appropriately cancelling
these contributions, namely the 5d Nekrasov partition function has the following form
\be
Z_{Nek} = \frac{Z_{top}}{Z_{dec}}\,.
\ee
\begin{figure}[t]
\begin{center}
\includegraphics[width=100mm]{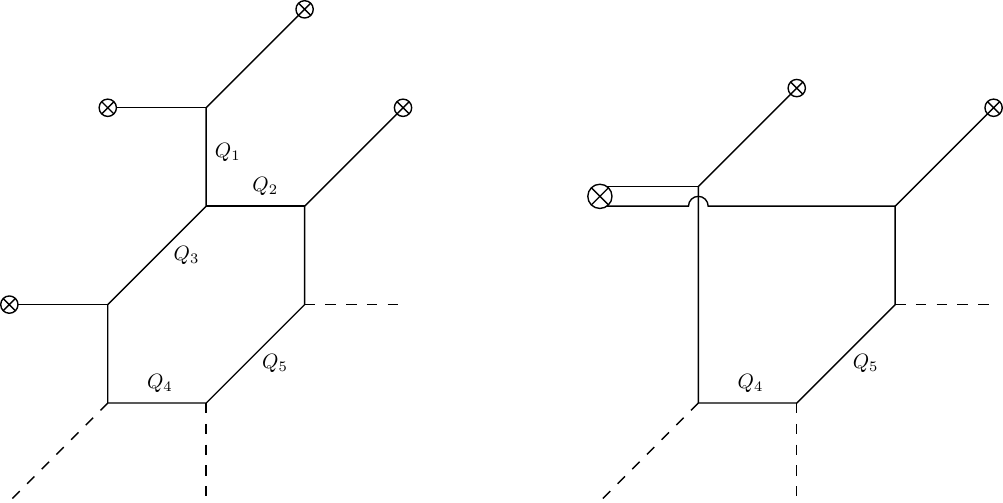}
\end{center}
\caption{On the left: Diagram with decoupled factors coming from parallel diagonal and horizontal legs. On the right: The diagram after Higgsing corresponding to putting external horizontal legs together.}
\label{fig:dec2}
\end{figure}

We would like to discuss how the contributions of decoupled factors can be computed in the case of a Higgsed diagram. We will look directly at the example of putting
a pair of parallel external 5-branes on top of each other for this already contains all the relevant information. The diagram we consider is in Figure \ref{fig:dec2}, and in particular
we will be interested in the contribution of decoupled factors coming from parallel diagonal legs. We see that in the toric diagram (therefore before putting the horizontal
5-branes on top of each other) the only contribution to the decoupled factor in the local part of the diagram we are considering is simply
\be\label{eq:dec1}
Z_{dec,\,/\!/} = \prod_{i,j=1}^\infty (1-Q_1 Q_2 q^{i+j-1})^{-1}\,.
\ee
We would like now to look at the particular case in which we put the two parallel horizontal external legs on top of other, and in order to do this we need to tune the K\"ahler
parameters of the diagram so that $Q_1 = Q_3 =1$. After entering the Higgs branch the contribution \eqref{eq:dec1} will still be present, however the curve with K\"ahler 
parameter $Q_1 Q_2$ is no longer present in the diagram and it seems a bit subtle to correctly identify the contribution of such decoupled factor directly in the Higgsed 
diagram. In fact in the case in which $Q_1 = Q_3 =1$ because of the geometric relations in the diagram it is true that $Q_2 = Q_4 Q_5$ and the contribution to the topological
string partition function of the decoupled factor is simply
\be
Z_{dec,\,/\!/} = \prod_{i,j=1}^\infty (1-Q_2 q^{i+j-1})^{-1}=\prod_{i,j=1}^\infty (1-Q_4 Q_5 q^{i+j-1})^{-1}\,.
\ee
We see therefore that, while the original curve that contributed to the decoupled factor is no longer present in the Higgsed diagram, it is true that a different curve
is still present hosting the same contribution and therefore the contributions due the decoupled factors can be easily identified in the Higgsed diagram as well.
The rule we have described is actually sufficient to correctly compute the decoupled factors in any Higgsed diagram for more complicated cases can always be
studied by simple iteration of this rule.

\begin{figure}[t]
\begin{center}
\includegraphics[width=50mm]{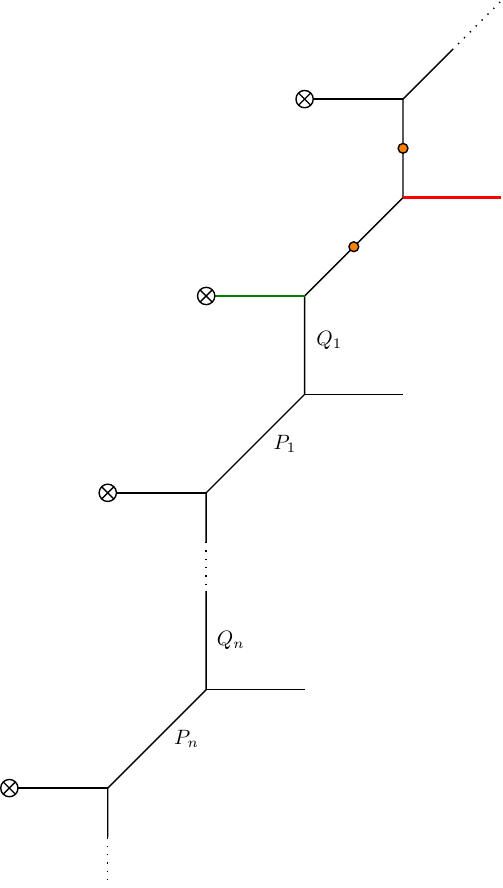}
\end{center}
\caption{The diagram which illustrates the cancellation between decoupled factors and singlet hypermultiplets. The decoupled factors coming from strings between the green leg and the other external legs cancels the contribution of singlet hypermultiplets coming from strings between the red leg and the external leg after the Higgsing. The orange dots indicates the lines becoming zero size. }
\label{fig:dec4}
\end{figure}

Another fundamental piece in the correct computation of the partition function in the Higgs branch of 5d theories is the correct identification of the contributions of 
singlet hypermultiplets to the partition function for these contributions need to be subtracted as well from the partition function in order to get the correct result. The
rule that was used in \cite{Hayashi:2013qwa,Hayashi:2014wfa} to correctly identify these contributions is that these hypermultiplets will come from M2-branes wrapping curves connecting pairs of 
legs in the diagram that become external after entering the Higgs branch. Remarkably these contributions are totally absent if the partition function is computed using
the usual topological vertex rule for Higgsed diagrams\footnote{We stress that it is crucial to drop the infinite product term in \eqref{double.vs.single} in order not to have
any contribution of singlet hypermultiplets in the case of Higgsed diagrams.}. The absence of these contributions is due to a cancellation of contributions between a part of the decoupled factors and singlet hypermultiplets and we will show now the basics of this cancellation. 

We show in Figure \ref{fig:dec4} the situation we are considering. Note that after
putting the pair of external 5-branes on top of each other the 5-brane coloured in red will become an external brane and therefore there will be contributions due to 
singlet hypermultiplets in the partition function that originate from curves connecting the new external 5-brane and the 5-branes that were external before entering the Higgs
branch. The contributions due to these hypermultiplets in the partition function is easily computed
\be
Z_{hyper} = \prod_{i,j=1}^\infty (1-q^{i+j-1})^2 \,\prod_{k=1}^n \left(1- q^{i+j-1}\prod_{l=1}^k Q_l P_l \right)\,.
\ee
However the topological string partition function also contains contributions due to the decoupled factors that come from curves connecting the external 5-branes
of the original diagram before entering the Higgs branch, and in particular if we consider the contributions that involve only curves connecting the 5-brane coloured in 
green in Figure \ref{fig:dec4}  we see that this contribution is
\be
Z_{dec} = \prod_{i,j=1}^\infty (1-q^{i+j-1})^{-1} \,\prod_{k=1}^n \left(1- q^{i+j-1}\prod_{l=1}^k Q_l P_l \right)^{-1}\,.
\ee
Therefore we see that the total contribution in the topological string partition function is
\be
Z_{hyper}Z_{dec} = \prod_{i,j=1}^\infty (1-q^{i+j-1})\,.
\ee
Note that the infinite product term actually coincides with the one appearing in \eqref{double.vs.single} and this factor is dropped when computing the topological string
partition function for a Higgsed diagram. Therefore we see that contributions of singlet hypermultiplets are not present at all in this kind of computation. Since decoupled hypermultiplets appear only
when two legs are placed on top of each other this argument is sufficient to show that the contributions due to these hypermultiplets will be totally absent.
 In the Higgsed diagram therefore there will be no contributions due to decoupled hypermultiplets whereas some contributions due to decoupled factors will still be present. However these are nothing but the standard decoupled factors that can be easily identified from the Higgsed diagram instead of the original toric diagram, using the prescription described before in this section. Therefore, we need to only remove the 
contributions of the decoupled factors associated to the Higgsed diagram.

To summarise we see that it is possible to understand which are the contributions to the partition function due to decoupled factors by simple inspection of the Higgsed diagram. This is natural because
in general these contributions will appear in the topological string partition function which, as we have seen in the previous section, can be computed by mere application of the topological vertex
to the Higgsed diagram. Therefore the analysis of the Higgsed diagram is sufficient to correctly identify these contributions which, as it happens in the case of web diagrams dual to toric Calabi--Yau
threefolds, are given by strings between parallel external legs that are connected by a curve in the Higgsed diagram.
This for instance implies that if the Higgsed diagram is given by two disconnected diagrams the decoupled factor will be given by
the product of the decoupled factors of each diagram, and in the next section we will see some explicit examples in which this occurs.

\section{Examples}
\label{sec:examples}
In this section, we apply the technique of the topological vertex obtained in section \ref{sec:vertex} to certain Higgsed 5d $T_N$ theories. The analysis in section \ref{sec:vertex} shows that we can apply the rule of the topological vertex as well as the decoupled factors directly to Higgsed web diagrams although the web diagram is not dual to a toric Calabi--Yau threefold. Therefore, the computation is greatly simplified because we do not need to compute the topological string partition function for a larger web diagram which yields a UV theory. We will see it by explicitly computing the topological string partition function from web diagrams which realise the rank $2$ $E_6, E_7$ and $E_8$ theories.

\subsection{Rank $2$ $E_6$ theory} 

\begin{figure}[t]
\begin{center}
\includegraphics[width=60mm]{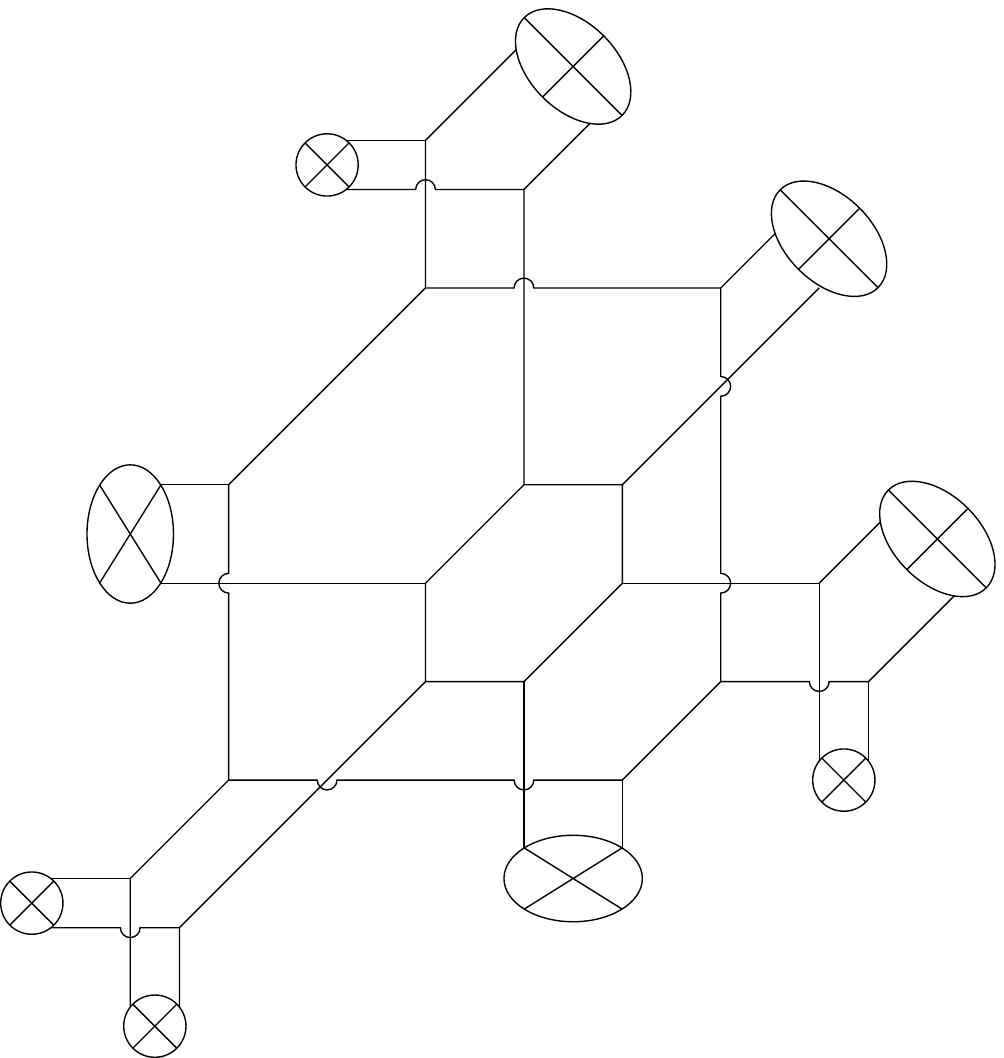}
\end{center}
\caption{The web diagram of the rank $2$ $E_6$ theory.}
\label{fig:E6}
\end{figure}

The rank $2$ $E_6$ theory can be realised as an IR theory in a Higgs branch of the $T_6$ theory. The web diagram is depicted in Figure \ref{fig:E6}. 
The full puncture $[1, 1, 1, 1, 1, 1]$ of the $T_6$ theory reduces to $[2, 2, 2]$, which gives an $SU(3)$ flavour symmetry. Hence, we have in total $SU(3) \times SU(3) \times SU(3)$ flavour symmetries which are nicely embedded in $E_6$. In general, the rank $N$ $E_6$ theory is obtained by Higgsing each of the full punctures of the $T_{3N}$ theory to $[N, N, N]$. The mass deformation of the rank $N$ $E_6$ theory gives an $Sp(N)$ gauge theory with five flavours and one anti-symmetric hypermultiplet.

\begin{figure}[t]
\begin{center}
\includegraphics[width=60mm]{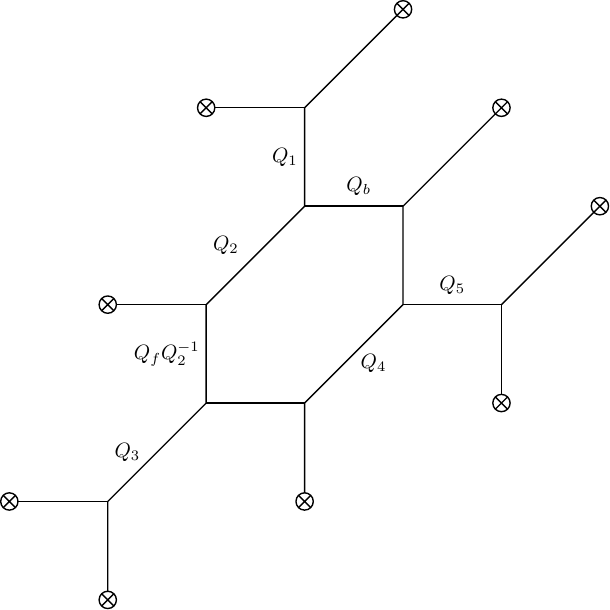}
\end{center}
\caption{The $[T_3]_1$ web diagram for the rank $1$ $E_6$ theory.}
\label{fig:rk1E6}
\end{figure}

From the web diagram in Figure \ref{fig:E6}, we can clearly see that the web diagram of the rank $2$ $E_6$ theory consists of two copies of the web diagram of the rank $1$ $E_6$ theory, which is nothing but the $T_3$ theory. Let the $T_3$ diagram with the larger and smaller closed face be $[T_3]_1$, $[T_3]_2$ diagram respectively. The $[T_3]_1$ web diagram is depicted in Figure \ref{fig:rk1E6}.
Since the topological vertex computation for Higgsed web diagrams can be carried out just by the usual rule of the topological vertex for toric web diagrams, the topological string partition function from the rank $2$ $E_6$ web diagram should be given by the product of two topological string partition functions computed from each $T_3$ web diagram, namely
\begin{equation}
Z_{top}^{\text{rk2 }E_6} = Z_{top}^{[T_3]_1}\left[P_1\right] \cdot Z_{top}^{[T_3]_2}\left[P_2\right],\label{top.E6}
\end{equation}
where $Z_{top}^{[T_3]_i}\left[P_i\right ]$ represents the topological string partition function from the $[T_3]_i$ diagram with a set of  parameters and moduli $P_i$, for $i=1, 2$. $P_{1,2}$ are related to the lengths of finite length five-branes in the $[T_3]_{1,2}$ web diagram respectively. The relation between $P_1$ and $P_2$ may be read off from the web diagram of the rank $2$ $E_6$ theory in Figure \ref{fig:E6}. Furthermore, the decoupled factor for Higgsed web diagrams can be also directly understood as the decoupled factor associated with strings between parallel external legs of the Higgsed web diagrams. Therefore, the decoupled factor for the rank $2$ $E_6$ diagram should be also the product of the two decoupled factors from each $T_3$ diagram, namely
\begin{equation}
Z_{dec}^{\text{rk2 }E_6} = Z_{dec}^{[T_3]_1}\left[P_1\right] \cdot Z_{dec}^{[T_3]_2}\left[P_2\right],
\end{equation}
where $Z_{dec}^{[T_3]_i}\left[P_i\right]$ represents the decoupled factor of the $[T_3]_i$ diagram with a set of  parameters $P_i$, for $i=1, 2$. Therefore, the partition function of the rank $2$ $E_6$ theory realised by the web diagram should be given by\footnote{In this paper, we do not take into account the perturbative contribution from the Cartan part of the vector multiplet. That contribution is not included in the topological vertex computation. The contribution is easily recovered by introducing the factor 
\begin{equation}
\prod_{i,j=1}^{\infty}\left(1 - q^{i+j-1}\right)^{-[\text{rank G}]},
\end{equation}
where $\text{rank G}$ is the rank of the gauge group $G$.}
\begin{equation}
Z^{\text{rk2 }E_6} = \frac{Z_{top}^{\text{rk2 }E_6}}{Z_{dec}^{\text{rk2 }E_6}} = \frac{ Z_{top}^{[T_3]_1}\left[P_1\right] }{ Z_{dec}^{[T_3]_1}\left[P_1\right]}\cdot\frac{ Z_{top}^{[T_3]_2}\left[P_2\right]}{Z_{dec}^{[T_3]_2}\left[P_2\right]}\,, \label{part.E6}
\end{equation}
and therefore it is the product of two partition functions of the $T_3$ theory.

Let us then see how the parameters of the $Sp(2)$ gauge theory arise in the web diagram of Figure \ref{fig:E6}. For a theory realised by a web diagram a local deformation which does not move semi-infinite 5-branes corresponds to a modulus of the theory while a global deformation which moves semi-infinite 5-branes corresponds to a parameter of the theory. In the case of the rank $2$ $E_6$ diagrams, each size of the two closed faces gives a local deformation. Hence each $[T_3]_{1, 2}$ diagram has one modulus, which is in fact a Coulomb branch modulus and 
we shall call these two moduli $\alpha_i, (i=1, 2)$ respectively. The rank $2$ $E_6$ web diagram has in general $6$ global deformations. 
Five of them are mass parameters $m_a, (a=1, \cdots, 5)$ of the five fundamental hypermultiplets, and one of them is the instanton fugacity $u$ or the gauge coupling of the $Sp(2)$ gauge group. 
One might be tempted to think that there is another parameter which corresponds to the relative distance between the centres of mass of the two closed faces. However the semi-infinite 5-brane ending on the same 7-brane should be grouped together. Hence, such a degree of freedom is frozen. One can also argue that we should have one gauge coupling after tuning off the Coulomb branch moduli. This also implies that the centres of the two closed faces should be at the same position. Hence, the parameter corresponding to the relative distance between the centres of mass of the two closed faces is absent.

When the centres of mass of the two closed faces coincide with each other, all the global deformations are given by the masses of the five flavours and the instanton fugacity of the $Sp(2)$ gauge group. Hence, the theory realised by such a web diagram should correspond to an $Sp(2)$ gauge theory with $5$ massive fundamental hypermultiplets and $1$ massless anti-symmetric hypermultiplet. Therefore, the web diagram Figure \ref{fig:E6} realises a special rank $2$ $E_6$ theory. 

For the computation of the topological string partition function from the rank $2$ $E_6$ diagram we need explicit relations between the parameters and the moduli of the theory and the lengths of finite size 5-branes in the web diagram. 
It turns out that we can use the same parameterisation as that of the $T_3$ diagram, which is determined in \cite{Hayashi:2013qwa}, but the only difference is that we use $\alpha_{1, 2}$ for the Coulomb branch moduli for the parameterisation of the $[T_3]_{1, 2}$ diagrams respectively. Namely, we choose in Figure \ref{fig:rk1E6}
\begin{eqnarray}
&&Q_1 = e^{-(m_1-\alpha_1)}, \quad Q_2 = e^{-(\alpha_1-m_2)}, \quad Q_3 = e^{-(-\alpha_1 - m_3)}, \quad Q_4 = e^{-(m_4+\alpha_1)},\nonumber \\
&& Q_5 = e^{-(m_5 - \alpha_1)}, \quad Q_f = e^{-2\alpha_1}, \quad Q_b = ue^{-\alpha_1-\frac{1}{2}\left(-m_1+m_2+m_3+m_4-m_5\right)}, \label{para.T3-1}
\end{eqnarray}
for the $[T_3]_1$ diagram. The parameterisation of the $[T_3]_2$ diagram is the same as \eqref{para.T3-1} with $\alpha_1$ exchanged with $\alpha_2$. Here $Q :=e^{-L}$ where $L$ is the length of the 5-brane or the size of the corresponding two-cycle in the dual M-theory picture. Hence, the product relation \eqref{part.E6} can be written more precisely by
\begin{equation}
Z^{\text{rk2 }E_6} = \frac{ Z_{top}^{[T_3]_1}\left[\alpha_1, m_1, \cdots, m_5, u\right] }{ Z_{dec}^{[T_3]_1}\left[ m_1, \cdots, m_5, u\right]}\cdot\frac{ Z_{top}^{[T_3]_2}\left[\alpha_2, m_1, \cdots, m_5, u\right]}{Z_{dec}^{[T_3]_2}\left[m_1, \cdots, m_5, u\right]}\,.
\end{equation}
Note that the decoupled factors do not depend on the Coulomb branch moduli.

The physical meaning of the parametrisation is also clear. Strings in the $[T_3]_1$ diagram yield particles with mass given by $\pm\alpha_1\pm m_a, a=1, \cdots, 5$ with appropriate signs. On the other hand, strings in the $[T_3]_2$ diagram yield particles with mass given by $\pm\alpha_2 \pm m_a, a=1, \cdots, 5$ with appropriate signs. Note here that the weights of the fundamental representation of the $Sp(2)$ Lie algebra are given by $\pm e_1, \pm e_2$ where $\{e_1, e_2\}$ are orthonormal basis of $\mathbb{R}^2$. Therefore, the fundamental hypermultiplets related to weights $\pm e_1$ come from strings in the $[T_3]_1$ diagram, and the fundamental hypermultiplets related to weights $\pm e_2$ come from strings in the $[T_3]_2$ diagram. By including both of them, we can form the complete components of the fundamental hypermultiplets of $Sp(2)$.

With the parameterisation \eqref{para.T3-1} and similarly that from the $[T_3]_2$ diagram, we can explicitly compute the topological string partition function as well as the decoupled factor for the rank $2$ $E_6$ theory. Due to the relation \eqref{top.E6}, it is enough to compute the topological string partition function from the $[T_3]_{1,2}$ diagrams, which is essentially the topological string partition function from the $T_3$ diagram. The computation of the topological string partition function for the $T_3$ diagram was done in \cite{Hayashi:2013qwa, Bao:2013pwa} and we make use of the result by changing the parameterisation into that of the $[T_3]_{1,2}$ diagrams. Then the topological string partition function from the $[T_3]_1$ diagram is 
\begin{eqnarray}
Z_{top}^{[T_3]_1}&=&Z_0^{[T_3]_1}\cdot Z_1^{[T_3]_1}\cdot Z_{=}^{[T_3]_1},\\
Z_0^{[T_3]_1} &=& \frac{H(e^{\pm\alpha_1-m_1}) H(e^{\pm\alpha_1+m_3})\left(\prod_{a=2,4} H(e^{-\alpha_1\pm m_a})\right)}{H(e^{-2\alpha_1})^{2}},\nonumber\\
Z_1^{[T_3]_1} &=& \sum_{\nu_1, \nu_2, \nu_3} u^{|\nu_1|+|\nu_2|}e^{-\left(|\nu_3| - \frac{1}{2}(|\nu_1|+|\nu_2|)\right)m_5} \nonumber\\
&& \prod_{i=1}^2\prod_{s \in \nu_i}\frac{\left(\prod_{a=1}^32\sinh\frac{E_{i0}-m_a}{2}\right) 2\sinh\frac{E_{i3}-m_4}{2}}{\prod_{j=1}^2\left(2 \sinh \frac{E_{ij}}{2}\right)^2}
\prod_{s\in\nu_3} \frac{\prod_{i=1}^2 2\sinh\frac{E_{3i}+m_4}{2}}{\left(2\sinh\frac{E_{33}}{2}\right)^2},\nonumber\\
 Z_{=}^{[T_3]_1} &=&  \left(H\left(e^{-(m_1-m_2)}\right)H\left(e^{-(m_2-m_3)}\right)H\left(e^{-(m_1-m_3)}\right)\right)^{-1}.\nonumber
\end{eqnarray}
We introduced some notations for the simplicity of the expressions
\begin{eqnarray}
H(Q) &=& \prod_{i,j=1}^{\infty}(1 - Qq^{i+j-1})\\
E_{ij}(s) &=& \beta_i - \beta_j - \epsilon\left(l_{\nu_i}(s)+a_{\nu_j}(s)+1\right),
\end{eqnarray}
where $\beta_0=\beta_3=0, \beta_1 = -\beta_2 = \alpha_1, \nu_0 =\emptyset$ and $q = e^{g_s} = e^{\epsilon}$. $g_s$ is the topological string coupling. Also we defined $H(e^{\pm x +y}):=H(e^{x+y})H(e^{-x +y})$. 

Similarly, the decoupled factor for the $[T_3]_1$ diagram is 
\begin{eqnarray}\label{eqq:decT3}
 Z_{dec}^{[T_3]_1} &=&  Z_{dec, =}^{[T_3]_1} \cdot Z_{dec, ||}^{[T_3]_1} \cdot Z_{dec, //}^{[T_3]_1},\\
Z_{dec, =}^{[T_3]_1} &=&  \left(H\left(e^{-(m_1-m_2)}\right)H\left(e^{-(m_2-m_3)}\right)H\left(e^{-(m_1-m_3)}\right)\right)^{-1}, \nonumber\\
Z_{dec, ||}^{[T_3]_1}  &=& \left(H\left(u\;e^{-\frac{1}{2}(m_1+m_2+m_3+m_4-m_5)}\right)H\left(e^{-(m_5-m_4)}\right)H\left(u\;e^{-\frac{1}{2}(m_1+m_2+m_3-m_4+m_5)}\right)\right)^{-1}, \nonumber\\
Z_{dec, /\!/}^{[T_3]_1}  &=& \left(H\left(u\;e^{\frac{1}{2}(m_1+m_2+m_3+m_4+m_5)}\right)H\left(e^{-(m_5+m_4)}\right)H\left(u\;e^{\frac{1}{2}(m_1+m_2+m_3-m_4-m_5)}\right)\right)^{-1}. \nonumber
\end{eqnarray}
In \eqref{eqq:decT3} we used the symbols $=, ||$ and $ /\!/$ to denote the contributions to the decoupled factor associated with parallel horizontal, vertical and diagonal legs respectively.

The partition function associated to the $[T_3]_1$ diagram is obtained by dividing $Z_{top}^{[T_3]_1}$ by $Z_{dec}^{[T_3]_1} $, which is 
\begin{eqnarray}
Z^{[T_3]_1} &=& \frac{Z_{top}^{[T_3]_1}}{ Z_{dec}^{[T_3]_1} } = Z^{[T_3]_1}_{pert}\cdot Z^{[T_3]_1}_{inst} \label{T3-1}\\
Z^{[T_3]_1}_{pert} &=& \frac{\left(\prod_{a=1,5}H(e^{\pm \alpha_1-m_a})\right) H(e^{\pm\alpha_1+m_3})\left(\prod_{a=2,4}H(e^{-\alpha_1\pm m_a})\right)}{H(e^{-2\alpha_1})^{2}},\nonumber\\
Z^{[T_3]_1}_{inst} &=& H\left(u\;e^{-\frac{1}{2}(m_1+m_2+m_3+m_4-m_5)}\right)H\left(u\;e^{-\frac{1}{2}(m_1+m_2+m_3-m_4+m_5)}\right)\nonumber\\
&&H\left(u\;e^{\frac{1}{2}(m_1+m_2+m_3+m_4+m_5)}\right)H\left(u\;e^{\frac{1}{2}(m_1+m_2+m_3-m_4-m_5)}\right)\nonumber\\
&&\frac{H\left(e^{-(m_5\pm m_4)}\right)}{H(e^{\pm \alpha_1-m_5})}\sum_{\nu_1, \nu_2, \nu_3} u^{|\nu_1|+|\nu_2|}e^{-\left(|\nu_3| - \frac{1}{2}(|\nu_1|+|\nu_2|)\right)m_5} \nonumber\\
&& \prod_{i=1}^2\prod_{s \in \nu_i}\frac{\left(\prod_{a=1}^32\sinh\frac{E_{i0}-m_a}{2}\right) 2\sinh\frac{E_{i3}-m_4}{2}}{\prod_{j=1}^2\left(2 \sinh \frac{E_{ij}}{2}\right)^2}
\prod_{s\in\nu_3} \frac{\prod_{i=1}^2 2\sinh\frac{E_{3i}+m_4}{2}}{\left(2\sinh\frac{E_{33}}{2}\right)^2}.\nonumber
\end{eqnarray}
$Z^{[T_3]_1}_{pert}$ is the perturbative part 
 and $Z^{[T_3]_1}_{inst} $ is the instanton part of the partition function. Note that $Z^{[T_3]_1}_{inst} $ does not include a non-trivial term at order $\mathcal{O}(u^0)$ due to the identity \cite{Kozcaz:2010af, Bao:2013pwa}
\begin{equation}
\sum_{\nu_3}e^{-|\nu_3|m_5}\prod_{s\in\nu_3} \frac{\prod_{i=1}^2 2\sinh\frac{E_{3i}+m_4}{2}}{\left(2\sinh\frac{E_{33}}{2}\right)^2} = \frac{H(e^{\alpha_1-m_5})H(e^{-\alpha_1-m_5})}{H\left(e^{-(m_5-m_4)}\right))H\left(e^{-(m_5+m_4)}\right)}.\label{pert.identity}
\end{equation}
The partition function $Z^{[T_3]_1}$ exactly agrees with the partition function of the $Sp(1)$ gauge theory with five fundamental hypermultiplets and this was checked up to $3$-instanton in \cite{Hayashi:2013qwa} when we regard $u$ as the instanton fugacity of the $Sp(1)$ gauge theory.

Then, the partition function from the $[T_3]_2$ diagram is essentially the same as that from the $[T_3]_1$ diagram except for the exchange of the Coulomb branch moduli, namely
\begin{equation}
Z^{[T_3]_2}\left[\alpha_2, \{m_{1,2,3,4,5}\}, u\right] = Z^{[T_3]_1}\left[\alpha_2, \{m_{1,2,3,4,5}\}, u\right], \label{T3-2}
\end{equation}
where the right-hand side of \eqref{T3-2} is \eqref{T3-1} with $\alpha_2$ used instead of $\alpha_1$. Here, $\{m_{1, 2, 3, 4, 5}\}$ means a set of mass parameters $m_1, \cdots, m_5$.

Due to the relation \eqref{part.E6}, we have obtained the partition function of the rank $2$ $E_6$ theory realised by the web diagram in Figure \ref{fig:E6},
\begin{equation}
Z^{\text{rk2 }E_6} =Z^{[T_3]_1}\left[\alpha_1, \{m_{1,2,3,4,5}\}, u\right]\cdot Z^{[T_3]_1}\left[\alpha_2, \{m_{1,2,3,4,5}\}, u\right], \label{rank2.E6}
\end{equation}
where $u$ is now identified with the instanton fugacity of the $Sp(2)$ gauge theory.

Let us see whether the result \eqref{rank2.E6} agrees with the field theory result. The perturbative part is given by 
\begin{eqnarray}
Z^{\text{rk2 }E_6}_{pert} &=& \frac{\left(\prod_{a=1,5}H(e^{\pm\alpha_1-m_a})\right) H(e^{\pm \alpha_1+m_3})\left(\prod_{a=2,4}H(e^{-\alpha_1\pm m_a})\right)}{H(e^{-2\alpha_1})^{2}}\nonumber\\
&&\frac{\left(\prod_{a=1,5}H(e^{\pm \alpha_2-m_a})\right) H(e^{\pm \alpha_2+m_3})\left(\prod_{a=2,4}H(e^{-\alpha_2\pm m_a})\right)}{H(e^{-2\alpha_2})^{2}}\nonumber\\\label{pert.rk2.E6}
\end{eqnarray}
By comparing \eqref{pert.rk2.E6} with \eqref{pert.gauge} with $N=2, N_f = 5$ inserted. \eqref{pert.rk2.E6} is precisely equal to the perturbative contribution of the $Sp(2)$ gauge theory with five massive fundamental hypermultiplets and one massless anti-symmetric hypermultiplet up to the contribution from the Cartan part of the vector multiplet and also some subtle divergent factors. Note that the factors from the massless anti-symmetric hypermultiplet  do not appear in $Z_{pert}^{\text{rk2 } E_6}$ because they are cancelled by the factors from a part of the vector multiplet.

Let us then turn to the comparison of the $1$-instanton part. Since the partition function of the rank $2$ $E_6$ theory is the product of the partition functions from the $[T_3]_{1,2}$ diagrams \eqref{rank2.E6}, the $1$-instanton part is simply given by 
\begin{equation}
Z^{\text{rk2 }E_6}_{1\text{-}inst} = Z^{[T_3]_1}\left[\alpha_1, \{m_{1,2,3,4,5}\}, u\right]|_{\mathcal{O}(u^1)} + Z^{[T_3]_1}\left[\alpha_2, \{m_{1,2,3,4,5}\}, u\right]|_{\mathcal{O}(u^1)}, \label{inst.rk2.E6}
\end{equation}
where $|_{\mathcal{O}(u^1)}$ implies taking the term at order $\mathcal{O}(u^1)$. Since the partition function of the $T_3$ theory agrees with the Nekrasov partition function of the $Sp(1)$ gauge theory with five fundamental hypermultiplet \cite{Hayashi:2013qwa}, the $1$-instanton part of $Z^{[T_3]_1}\left[\alpha_1, \{m_{1,2,3,4,5}\}, u\right]$ is the $1$-instanton part of the Nekrasov partition function of the $Sp(1)$ gauge theory with five flavours, which is \eqref{Sp1.1instanton} with $N_f=5$ and $\epsilon_+ = 0$. $\epsilon_+=0$ is due to the fact that we use the unrefined topological vertex. Therefore, \eqref{inst.rk2.E6} becomes 
\begin{equation}\label{eq:1instE6}
Z^{\text{rk2 }E_6}_{1\text{-}inst} =\frac{1}{2}\sum_{i=1}^2 \left\{\frac{\prod_{a=1}^{5}2\sinh\frac{m_a}{2}}{2\sinh\frac{\pm \epsilon}{2}2\sinh\frac{\pm\alpha_i}{2}}+\frac{\prod_{a=1}^{N_f}2\cosh\frac{m_a}{2}}{2\sinh\frac{\pm \epsilon}{2}2\cosh\frac{\pm\alpha_i}{2}}\right\}.
\end{equation}
This completely agrees with the field theory result of \eqref{SpN.sp.1instanton} with $N=2$ and $\epsilon$ identified with $\epsilon_-$. In \eqref{eq:1instE6} we introduced the notation
$\sinh(\pm x):=\sinh(x)\sinh(-x)$. 

Moreover we have checked agreement of the partition function \eqref{rank2.E6} of the rank 2 $E_6$ theory with the partition function of the $Sp(2)$ theory with 5 flavours and one massless anti-symmetric hypermultiplet at 2-instanton in the special limit where one of the masses of the fundamental hypermultiplets is set to zero\footnote{The reason why we set one mass to zero is not a fundamental issue but a technical issue. Namely our computer program checking the equality did not end in a reasonable time when we chose all the masses are general.}. The method of how to compute the $Sp(2)$ instanton partition function at the 2-instanton level is summarised at the end of appendix \ref{sec:SpN}.

\subsection{Rank $2$ $E_7$ theory}

\begin{figure}[t]
\begin{center}
\includegraphics[width=80mm]{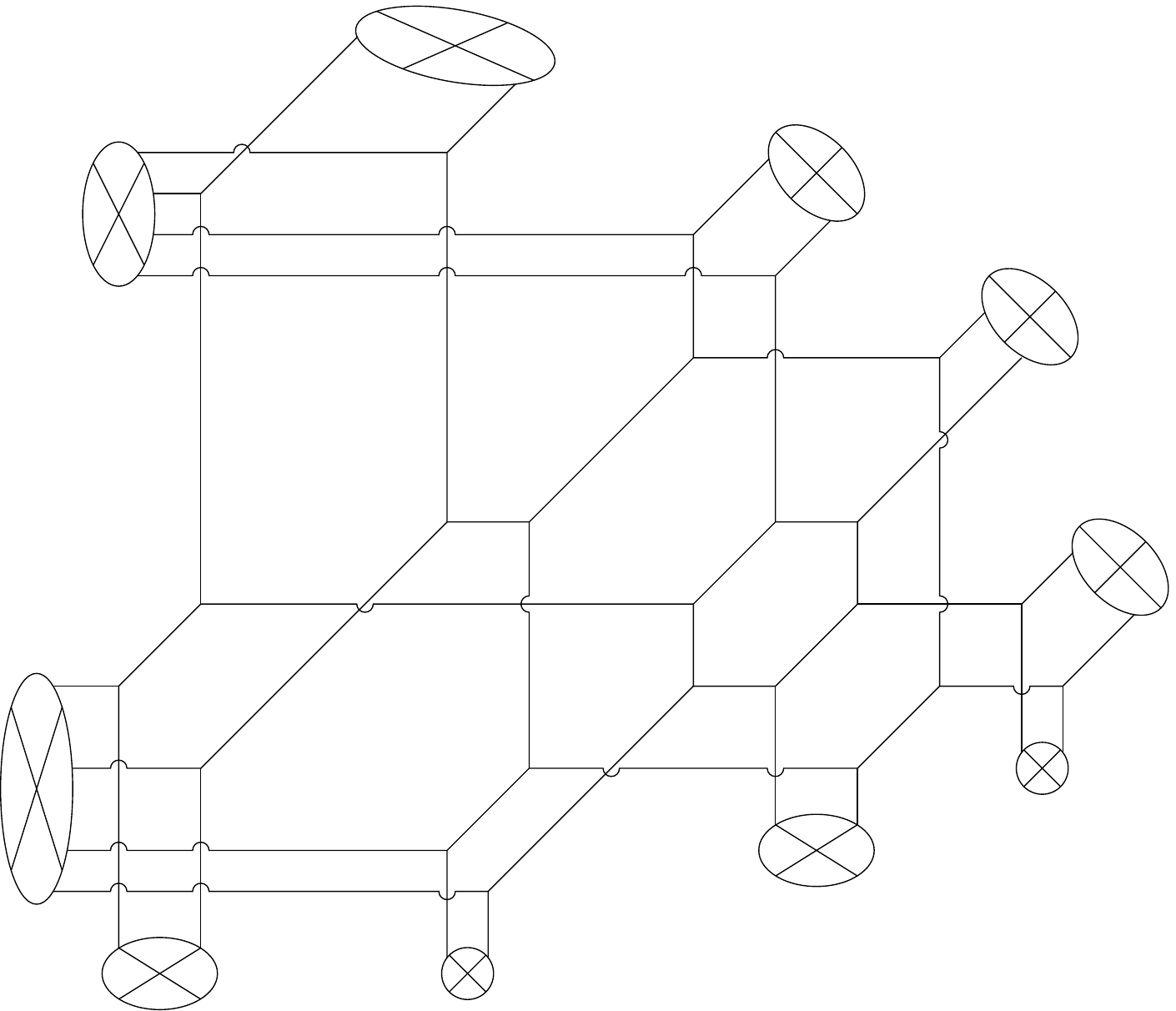}
\end{center}
\caption{The web diagram for the rank $2$ $E_7$ theory. }
\label{fig:rk2E7}
\end{figure}

Next example is the rank $2$ $E_7$ theory which is realised by the web diagram in Figure \ref{fig:rk2E7}. 
The UV theory is the $T_8$ theory with three full punctures. Then, we go to a Higgs branch where one full puncture is reduced to $[4, 4]$ and the other full punctures are reduced to $[2,2,2,2]$. The vev of the hypermultiplets induces an RG flow and we obtain the rank $2$ $E_7$ theory at low energies. The puncture with $[4, 4]$ gives an $SU(2)$ flavour symmetry and the two punctures with $[2,2,2,2]$ yields $SU(4) \times SU(4)$ flavour symmetries. The total flavour symmetries $SU(2) \times SU(4) \times SU(4)$ can be embedded in $E_7$. The rank $N$ $E_7$ theory is realised by Higgsing the three full punctures of the $T_{4N}$ theory down to one $[2N, 2N]$ and two $[N, N, N, N]$. The mass deformation of the rank $N$ $E_7$ theory is the $Sp(N)$ gauge theory with six fundamental hypermultiplets and one anti-symmetric hypermultiplet.

\begin{figure}[t]
\begin{center}
\includegraphics[width=100mm]{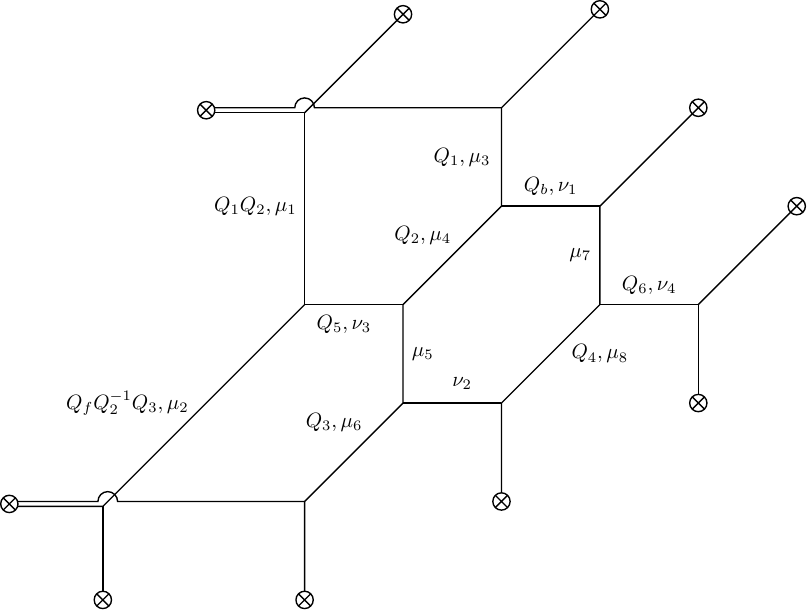}
\end{center}
\caption{The $[E_7]_1$ web diagram for the rank $1$ $E_7$ theory. }
\label{fig:rk1E7}
\end{figure}

Again, we can see that the web diagram of the rank $2$ $E_7$ theory is composed by two copies of the web diagram of the rank $1$ $E_7$ theory. The web diagram of the rank $1$ $E_7$ theory is shown in Figure \ref{fig:rk1E7}. 
We will call one copy of the rank $1$ $E_7$ web diagram with the larger closed face as $[E_7]_1$ web diagram, and another copy with the smaller closed face as $[E_7]_2$ web diagram. The topological vertex formalism of Higgsed web diagrams again implies that the topological string partition function from the rank $2$ $E_7$ diagram is given by
\begin{equation}
Z_{top}^{\text{rk2 }E_7} = Z_{top}^{[E_7]_1}\left[P_1\right] \cdot Z_{top}^{[E_7]_2}\left[P_2\right],\label{top.E7}
\end{equation}
where $Z_{top}^{[E_7]_i}\left[P_i\right ]$ represents the topological string partition function from the $[E_7]_i$ diagram with a set of  parameters and moduli $P_i$, for $i=1, 2$. The decoupled factor of the rank $2$ $E_7$ diagram is also written by a product
\begin{equation}
Z_{dec}^{\text{rk2 }E_7} = Z_{dec}^{[E_7]_1}\left[P_1\right] \cdot Z_{dec}^{[E_7]_2}\left[P_2\right],
\end{equation}
where $Z_{dec}^{[E_7]_i}\left[P_i\right]$ represents the decoupled factor of the $[E_7]_i$ diagram with a set of  parameters $P_i$, for $i=1, 2$. Therefore, the partition function of the rank $2$ $E_7$ theory realised by the web diagram Figure \ref{fig:rk2E7} can be obtained by
\begin{equation}
Z^{\text{rk2 }E_7} = \frac{Z_{top}^{\text{rk2 }E_7}}{Z_{dec}^{\text{rk2 }E_7}} = \frac{ Z_{top}^{[E_7]_1}\left[P_1\right] }{ Z_{dec}^{[E_7]_1}\left[P_1\right]}\cdot\frac{ Z_{top}^{[E_7]_2}\left[P_2\right]}{Z_{dec}^{[E_7]_2}\left[P_2\right]}. \label{part.E7}
\end{equation}
Hence, in order to compute the partition function of the rank $2$ $E_7$ theory, it is enough to compute the partition function from the $[E_7]_1$ web diagram and then multiply it by the same function with different arguments.

As in the case of the parameterisation of the rank $2$ $E_6$ theory, the parameterisation of the rank $2$ $E_7$ theory is determined by making use of the parameterisation of the rank $1$ $E_7$ theory. The relation between the gauge theory parameters and the lengths of five-branes for the rank $1$ $E_7$ theory was determined in \cite{Hayashi:2013qwa}. The only difference between the parameterisation of the $[E_7]_1$ diagram and that of the $[E_7]_2$ diagram is whether we use the Coulomb branch modulus $\alpha_1$ or $\alpha_2$. Here, $\alpha_1$ is related to the size of the larger closed face and $\alpha_2$ is related to the size of the smaller closed face. More precisely, we use the parameterisation
\begin{eqnarray}
&&Q_1=e^{-(m_1-\alpha_1)}, \quad Q_2=e^{-(\alpha_1-m_2)}, \quad Q_3=e^{-(-\alpha_1-m_3)}, \quad Q_4=e^{-(m_4+\alpha_1)},\nonumber\\
&&Q_5=e^{-(m_5-\alpha_1)}, \quad Q_6 = e^{-(m_6-\alpha_1)}, \quad Q_f = e^{-2\alpha_1}, \quad Q_b =u e^{-\alpha_1-\frac{1}{2}(-m_1+m_2+m_3+m_4-m_5-m_6)}.\nonumber\\ \label{para.E7-1}
\end{eqnarray}
for the $[E_7]_1$ web diagram. The correspondence between $Q$ and 5-branes is depicted in Figure \ref{fig:rk1E7}. $u$ is the instanton fugacity of the $Sp(2)$ gauge theory. The parameterisation of the $[E_7]_2$ diagram is the same as \eqref{para.E7-1} except that $\alpha_1$ is exchanged with $\alpha_2$. We also choose the parameterisation such that the center of mass of the larger closed face coincides with the center of mass of the smaller closed face. The theory realised by the web diagram should be the $Sp(2)$ gauge theory with six massive fundamental hypermultiplets and one massless anti-symmetric hypermultiplet.

With the parameterisation of \eqref{para.E7-1} and similarly that from the $[E_7]_2$ web diagram, we perform the explicit computation of the topological string partition function from the rank $2$ $E_7$ web diagram by making use of the technique developed in section \ref{sec:vertex}. Due to the product structure \eqref{part.E7}, we first compute the topological string partition function from the $[E_7]_1$ diagram. The refined version of the computation was essentially done in section 6.3 of \cite{Hayashi:2013qwa}, but we repeat the computation here since the discussion of the decoupled factor from the rank $1$ $E_7$ diagram was unclear in \cite{Hayashi:2013qwa}. The application of the topological vertex to the web diagram Figure \ref{fig:rk1E7} gives the topological string partition function
\begin{eqnarray}
Z_{top}^{[E_7]_1}&=&\sum_{\nu_1, \cdots, \nu_4, \mu_1, \cdots, \mu_8}(-Q_b)^{|\nu_1|}(-Q_bQ_2Q_5^{-1})^{|\nu_2|}(-Q_5)^{|\nu_3|}(-Q_6)^{|\nu_4|}(-Q_1Q_2)^{|\mu_1|}(-Q_fQ_2^{-1}Q_3)^{|\mu_2|}\nonumber\\
&&(-Q_1)^{|\mu_3|}(-Q_2)^{|\mu_4|}(-Q_fQ_2^{-1})^{|\mu_5|}(-Q_3)^{|\mu_6|}(-Q_fQ_4^{-1})^{|\mu_7|}(-Q_4)^{|\mu_8|}\nonumber\\
&&C_{\emptyset\mu_1\emptyset}(q) C_{\mu_2\mu_1^t\nu_3^t}(q) C_{\mu_2^t\emptyset\emptyset}(q) C_{\emptyset\mu_3^t\emptyset}(q)C_{\mu_4^t\mu_3\nu_1^t}(q)C_{\mu_4\mu_5\nu_3}(q)C_{\mu_6\mu_5^t\nu_2^t}(q)C_{\mu_6^t\emptyset\emptyset}(q)\nonumber\\
&&C_{\emptyset\mu_7\nu_1}(q)C_{\mu_8^t\mu_7^t\nu_4^t}(q)C_{\mu_8\emptyset\nu_2}(q)C_{\emptyset\emptyset\nu_4}(q).
\end{eqnarray}
The straightforward computation by using the formulae in appendix \ref{sec:top.vertex} gives
\begin{eqnarray}
Z_{top}^{[E_7]_1}&=&Z_0^{[E_7]_1}\cdot Z_1^{[E_7]_1}\cdot Z_{=}^{[E_7]_1},\label{E7-1-top}
\end{eqnarray}
\begin{eqnarray}
Z_0^{[E_7]_1} &=& \frac{H(e^{\pm \alpha_1-m_1}) H(e^{\pm \alpha_1+m_3})\left(\prod_{a=2,4}H(e^{-\alpha_1\pm m_a})\right)}{H\left(e^{-2\alpha_1}\right)^2},\nonumber
\end{eqnarray}
\begin{eqnarray}
Z_1^{[E_7]_1} &=& \sum_{\nu_1, \nu_2, \nu_3, \nu_4}u^{|\nu_1|+|\nu_2|}e^{-\left(|\nu_3|-\frac{1}{2}(|\nu_1|+|\nu_2|)\right)m_5}e^{-\left(|\nu_4|-\frac{1}{2}(|\nu_1|+|\nu_2|)\right)m_6}\nonumber\\
&& \prod_{i=1}^2\prod_{s \in \nu_i}\frac{\left(\prod_{a=1,3}2\sinh\frac{E_{i0}-m_a}{2}\right) 2\sinh\frac{E_{i3}-m_2}{2}2\sinh\frac{E_{i4}-m_4}{2}}{\prod_{j=1}^2\left(2 \sinh \frac{E_{ij}}{2}\right)^2}\nonumber\\
&&\prod_{s\in\nu_3} \frac{\prod_{i=1}^2 2\sinh\frac{E_{3i}+m_2}{2}}{\left(2\sinh\frac{E_{33}}{2}\right)^2}\prod_{s\in\nu_4} \frac{\prod_{i=1}^2 2\sinh\frac{E_{4i}+m_4}{2}}{\left(2\sinh\frac{E_{44}}{2}\right)^2},\nonumber
\end{eqnarray}
\begin{eqnarray}
 Z_{=}^{[E_7]_1}&=& H\left(e^{-(m_1-m_3)}\right)^{-2}.
\end{eqnarray}
We defined $\beta_0=\beta_3 = \beta_4=0, \beta_1 = -\beta_2 = \alpha_1, \nu_0 =\emptyset$.

The decoupled factor can be also directly read off from the Higgsed diagram. Namely it is associated to the contribution from strings between the parallel external legs of the Higgsed diagram, which is, in this case, the $[E_7]_1$ web diagram in Figure \ref{fig:rk1E7}. The decoupled factor from the $[E_7]_1$ diagram is given by 
\begin{eqnarray}
Z_{dec}^{[E_7]_1} &=&  Z_{dec, =}^{[E_7]_1} \cdot Z_{dec, ||}^{[E_7]_1} \cdot Z_{dec, //}^{[E_7]_1}, \label{E7-1-dec}
\end{eqnarray}
where
\begin{eqnarray}
 Z_{dec, =}^{[E_7]_1} &=& H\left(Q_1Q_3Q_f^{-1}\right)^{-2}\nonumber\\
&=& H\left(e^{-(m_1-m_3)}\right)^{-2}, \nonumber
\end{eqnarray}
\begin{eqnarray}
Z_{dec, ||}^{[E_7]_1}&=& H\left(Q_2Q_5\right)^{-1}H\left(Q_1Q_b\right)^{-1}H\left(Q_fQ_4^{-1}Q_6\right)^{-1}H\left(Q_1Q_2Q_5Q_b\right)^{-1}\nonumber\\
&&H\left(Q_1Q_4^{-1}Q_6Q_fQ_b\right)^{-1}H\left(Q_1Q_2Q_4^{-1}Q_5Q_6Q_fQ_b\right)^{-1}\nonumber\\
&=&H\left(e^{-(m_5-m_2)}\right)^{-1}H\left(ue^{-\frac{1}{2}(m_1+m_2+m_3+m_4-m_5-m_6)}\right)^{-1}H\left(e^{-(m_6-m_4)}\right)^{-1}\nonumber\\
&&H\left(ue^{-\frac{1}{2}(m_1-m_2+m_3+m_4+m_5-m_6)}\right)^{-1}H\left(ue^{-\frac{1}{2}(m_1+m_2+m_3-m_4-m_5+m_6)}\right)^{-1}\nonumber\\&&H\left(ue^{-\frac{1}{2}(m_1-m_2+m_3-m_4-m_5-m_6)}\right)^{-1}\nonumber
\end{eqnarray}
\begin{eqnarray}
 Z_{dec, /\!/}^{[E_7]_1} &=& H\left(Q_fQ_2^{-1}Q_5\right)^{-1}H\left(Q_2Q_3Q_4^{-1}Q_b\right)^{-1}H\left(Q_4Q_6\right)^{-1}H\left(Q_3Q_4^{-1}Q_5Q_fQ_b\right)^{-1}\nonumber\\
&&H\left(Q_2Q_3Q_6Q_b\right)^{-1}H\left(Q_3Q_5Q_6Q_fQ_b\right)^{-1}\nonumber\\
&=&H\left(e^{-(m_5+m_2)}\right)^{-1}H\left(ue^{\frac{1}{2}(m_1+m_2+m_3+m_4+m_5+m_6)}\right)^{-1}H\left(e^{-(m_6+m_4)}\right)^{-1}\nonumber\\
&&H\left(ue^{\frac{1}{2}(m_1-m_2+m_3+m_4-m_5+m_6)}\right)^{-1}H\left(ue^{\frac{1}{2}(m_1+m_2+m_3-m_4+m_5-m_6)}\right)^{-1}\nonumber\\&&H\left(ue^{\frac{1}{2}(m_1-m_2+m_3-m_4-m_5-m_6)}\right)^{-1}\nonumber
\end{eqnarray}
Again, we used the symbols $=, ||$ and $ /\!/$ to denote the contributions that come from strings between the parallel horizontal, vertical and diagonal legs respectively.

By combining the topological string partition function \eqref{E7-1-top} with the decoupled factor \eqref{E7-1-dec}, we obtain the partition function associated to the $[E_7]_1$ web diagram
\begin{eqnarray}
Z^{[E_7]_1} &=& \frac{Z_{top}^{[E_7]_1}}{ Z_{dec}^{[E_7]_1} } = Z^{[E_7]_1}_{pert}\cdot Z^{[E_7]_1}_{inst} \label{E7-1}
\end{eqnarray}
\begin{eqnarray}
 Z^{[E_7]_1}_{pert} &=& \frac{\left(\prod_{a=1,5,6}H(e^{\pm \alpha_1-m_a}) \right) H(e^{\pm \alpha_1+m_3})\left(\prod_{a=2,4}H(e^{-\alpha_1 \pm m_a})\right)}{H\left(e^{-2\alpha_1}\right)^2},\nonumber 
 \end{eqnarray}
 \begin{eqnarray}
Z^{[E_7]_1}_{inst} &=& H\left(ue^{-\frac{1}{2}(m_1+m_2+m_3+m_4-m_5-m_6)}\right) H\left(ue^{-\frac{1}{2}(m_1-m_2+m_3+m_4+m_5-m_6)}\right)\nonumber\\
&&H\left(ue^{-\frac{1}{2}(m_1+m_2+m_3-m_4-m_5+m_6)}\right) H\left(ue^{-\frac{1}{2}(m_1-m_2+m_3-m_4-m_5-m_6)}\right)\nonumber\\
&&H\left(ue^{\frac{1}{2}(m_1+m_2+m_3+m_4+m_5+m_6)}\right) H\left(ue^{\frac{1}{2}(m_1-m_2+m_3+m_4-m_5+m_6)}\right)\nonumber\\
&&H\left(ue^{\frac{1}{2}(m_1+m_2+m_3-m_4+m_5-m_6)}\right) H\left(ue^{\frac{1}{2}(m_1-m_2+m_3-m_4-m_5-m_6)}\right) \nonumber \\
&&\frac{H\left(e^{-(m_5\pm m_2)}\right)H\left(e^{-(m_6\pm m_4)}\right)}{H(e^{\pm \alpha_1-m_5})H(e^{\pm \alpha_1-m_6}) }\sum_{\nu_1, \nu_2, \nu_3, \nu_4}u^{|\nu_1|+|\nu_2|}e^{-\left(|\nu_3|-\frac{1}{2}(|\nu_1|+|\nu_2|)\right)m_5}e^{-\left(|\nu_4|-\frac{1}{2}(|\nu_1|+|\nu_2|)\right)m_6}\nonumber\\
&& \prod_{i=1}^2\prod_{s \in \nu_i}\frac{\left(\prod_{a=1,3}2\sinh\frac{E_{i0}-m_a}{2}\right) 2\sinh\frac{E_{i3}-m_2}{2}2\sinh\frac{E_{i4}-m_4}{2}}{\prod_{j=1}^2\left(2 \sinh \frac{E_{ij}}{2}\right)^2}\nonumber\\
&&\prod_{s\in\nu_3} \frac{\prod_{i=1}^2 2\sinh\frac{E_{3i}+m_2}{2}}{\left(2\sinh\frac{E_{33}}{2}\right)^2}\prod_{s\in\nu_4} \frac{\prod_{i=1}^2 2\sinh\frac{E_{4i}+m_4}{2}}{\left(2\sinh\frac{E_{44}}{2}\right)^2},\nonumber
\end{eqnarray}  
The instanton part $Z^{[E_7]_1}_{inst}$ of the partition function starts from $1$ at order $\mathcal{O}(u^0)$ due to the identity \eqref{pert.identity}. The expression \eqref{E7-1} exactly agrees with the partition function of the rank $1$ $E_7$ theory obtained in \cite{Hayashi:2013qwa}. The computation presented here is simplified compared to the computation in \cite{Hayashi:2013qwa} in two respects. First, the partition function has four Young diagram summations from the first whereas the partition function of the UV $T_4$ theory has six Young diagram summations which reduces to the four Young diagram summations after the Higgsing. Second, we do not need to eliminate the decoupled factor in two steps where we first eliminate the decoupled factor of the UV diagram and then eliminate the singlet hypermultiplet in the Higgsed vacuum. In this formalism, we just remove the decoupled factor associated to the IR diagram of Figure \ref{fig:rk1E7}.

The partition function associated to the $[E_7]_2$ web diagram is simply given by the same function as \eqref{E7-1} with the Coulomb branch moduli exchanged
\begin{equation}
Z^{[E_7]_2}[\alpha_2, \{m_{1, 2, 3, 4, 5, 6}\}, u] = Z^{[E_7]_1}[\alpha_2, \{m_{1, 2, 3, 4, 5, 6}\}, u].
\end{equation}

After obtaining the partition function associated to the $[E_7]_1$ diagram, it is easy to obtain the partition function of the rank $2$ $E_7$ theory realised by the web diagram in Figure \ref{fig:rk2E7} due to the product relation \eqref{part.E7},
\begin{equation}
Z^{\text{rk2 }E_7} = Z^{[E_7]_1}[\alpha_1, \{m_{1, 2, 3, 4, 5, 6}\}, u] \cdot Z^{[E_7]_1}[\alpha_2, \{m_{1, 2, 3, 4, 5, 6}\}, u].\label{rk2.E7}
\end{equation}

The perturbative part of the partition function \eqref{rk2.E7} is 
\begin{eqnarray}
Z_{pert}^{\text{rk2 }E_7} &=&\frac{\left(\prod_{a=1,5,6}H(e^{\pm \alpha_1-m_a}) \right) H(e^{\pm \alpha_1+m_3})\left(\prod_{a=2,4}H(e^{-\alpha_1 \pm m_a})\right)}{H\left(e^{-2\alpha_1}\right)^2} \nonumber\\
&&\frac{\left(\prod_{a=1,5,6}H(e^{\pm \alpha_2-m_a}) \right) H(e^{\pm \alpha_2+m_3})\left(\prod_{a=2,4}H(e^{-\alpha_2 \pm m_a})\right)}{H\left(e^{-2\alpha_2}\right)^2}.\nonumber
\end{eqnarray}
By the comparison with \eqref{pert.gauge}. this is precisely the perturbative contribution from six fundamental hypermultiplets with mass $m_{1, 2, 3, 4, 5, 6}$, one massless anti-symmetric hypermultiplet, and $Sp(2)$ vector multiplet. 

Also the $1$-instanton part of the partition function from the $[E_7]_1$ web diagram should be the $1$-instanton contribution of the $Sp(1)$ gauge theory with six flavours since \eqref{E7-1} agrees with the partition function of the rank $1$ $E_7$ theory. Therefore, its contribution is written by \eqref{Sp1.1instanton} with $\epsilon_+=0, \epsilon_-=\epsilon$ and $m=0$. Hence the $1$-instanton part of $Z_{inst}^{\text{rk2 }E_7}$ is given by the sum of the $1$-instanton part of the partition function from the $[E_7]_1$ diagram and that from the $[E_7]_2$ diagram,
\begin{equation}
Z^{\text{rk2 }E_7}_{1\text{-}inst} =\frac{1}{2}\sum_{i=1}^2 \left\{\frac{\prod_{a=1}^{6}2\sinh\frac{m_a}{2}}{2\sinh\frac{\pm \epsilon}{2}2\sinh\frac{\pm\alpha_i}{2}}+\frac{\prod_{a=1}^{6}2\cosh\frac{m_a}{2}}{2\sinh\frac{\pm \epsilon}{2}2\cosh\frac{\pm\alpha_i}{2}}\right\}.
\end{equation}
This result completely agrees with the field theory computation of the $1$-instanton part \eqref{SpN.sp.1instanton} of the partition function of the $Sp(2)$ gauge theory with six fundamental hypermultiplets and one massless anti-symmetric hypermultiplet by setting $N=2, \epsilon_- = \epsilon$. 

Moreover we have checked agreement of the partition function of the rank 2 $E_7$ theory with the partition function of a $Sp(2)$ theory with 6 flavours and one massless anti-symmetric hypermultiplet at 2-instanton in the special limit where one of the masses of the fundamental hypermultiplets is set to zero\footnote{Again, the reason why we set one mass to zero is the technical issue.}.

\subsection{Rank $2$ $E_8$ theory}
\begin{figure}[t]
\begin{center}
\includegraphics[width=60mm]{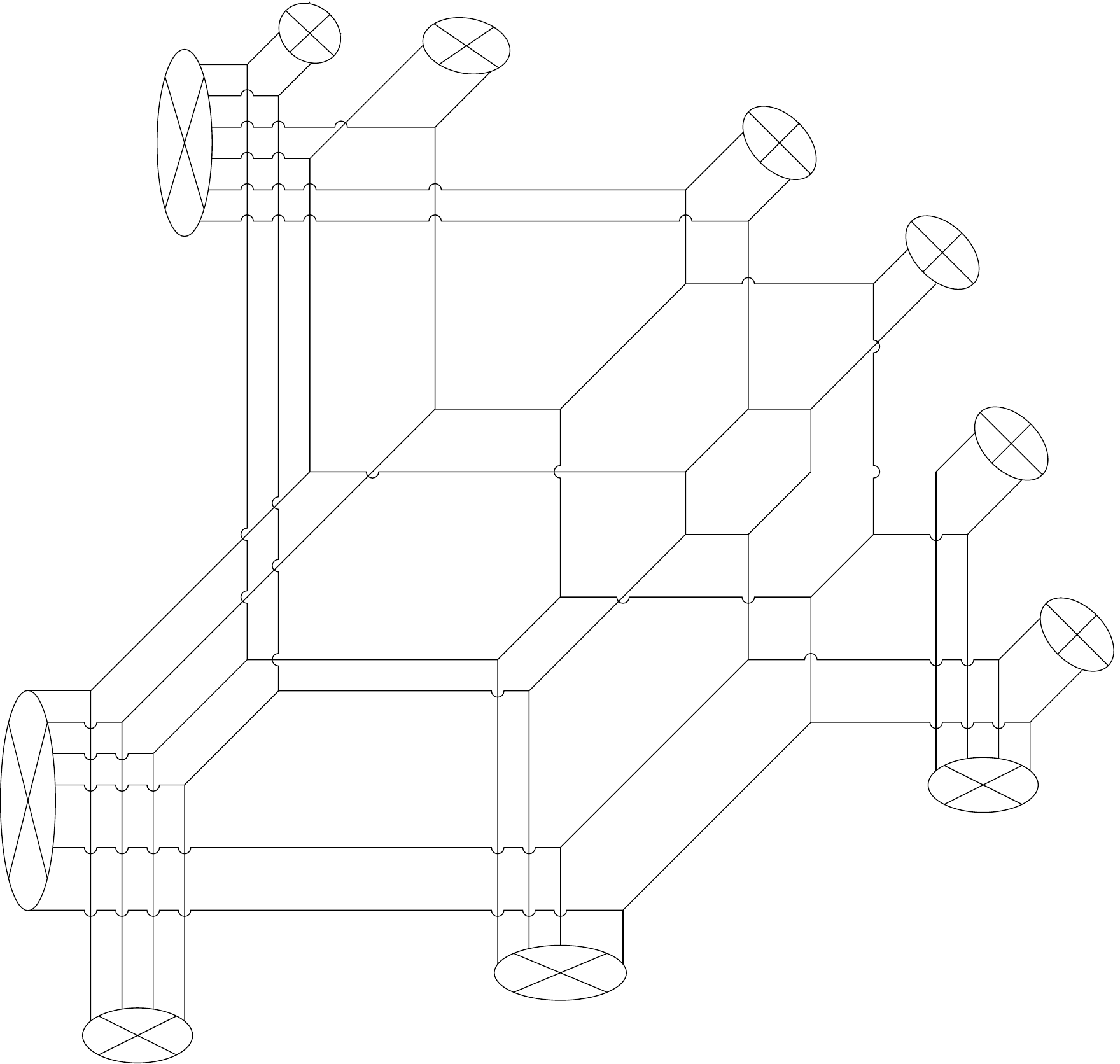}
\end{center}
\caption{The web diagram for the rank $2$ $E_8$ theory. }
\label{fig:E8}
\end{figure}

The last example we would like to discuss is the rank 2 $E_8$ which can be realised by the web diagram in Figure \ref{fig:E8}. 
In this case the UV theory is the $T_{12}$ theory and in order to realise rank 2 $E_8$ theory it is necessary
to go into the Higgs branch to have the original three full punctures reduced to $[6, 6]$, $[4, 4, 4]$ and  $[2,2,2,2,2,2]$
In this situation the flavour symmetry of the theory is $SU(2) \times SU(3) \times SU(6)$ which can be embedded in $E_8$. It is 
possible to generalise this construction to realise rank $N$ $E_8$ theory by going into the Higgs branch of $T_{6N}$ theory and reducing the original three full punctures to $[3N, 3N]$, $[2N, 2N, 2N]$
and $[N,N,N,N,N,N]$.  We will show by computing explicitly its topological string
partition function that rank 2 $E_8$ theory for a specific choice of parameters (to be discussed later) is an $Sp(2)$ gauge theory with 7 fundamental hypermultiplets and one massless antisymmetric hypermultiplet. This leads us to 
conjecture that the same happens for rank $N$ $E_8$ theory, namely that for some specific choice of parameters rank $N$ $E_8$ theory is a $Sp(N)$ gauge theory with 7 fundamental hypermultiplets and one
massless antisymmetric hypermultiplet.

\begin{figure}[t]
\begin{center}
\includegraphics[width=100mm]{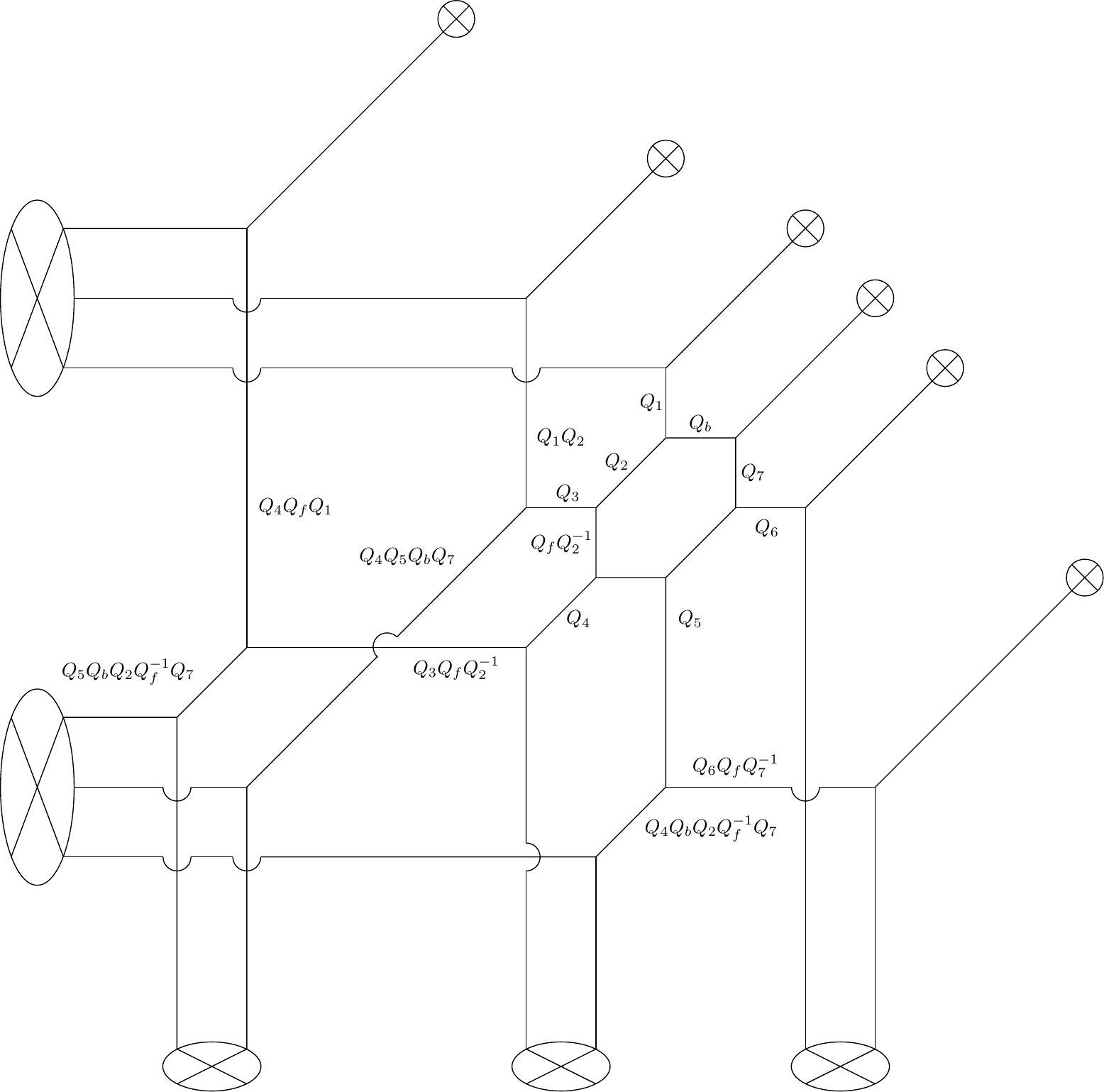}
\end{center}
\caption{The web diagram for the rank $1$ $E_8$ theory. }
\label{fig:rk1E8}
\end{figure}

Like in the previous examples discussed in this section it is possible to see that the diagram of rank 2 $E_8$ theory is made of two copies of a $E_8$ theory whose diagram is displayed in Figure \ref{fig:rk1E8}. 
We shall call $[E_8]_1$
the external copy of the rank 1 $E_8$ diagram and $[E_8]_2$ the internal copy. The fact that the diagram of the rank 2 $E_8$ theory is made of two copies of a rank 1 $E_8$ diagram implies that the topological string partition function
has the following structure
\be
Z_{top}^{\text{rk2} \, E_8} = Z_{top}^{[E_8]_1}[P_1]\cdot Z_{top}^{[E_8]_2}[P_2]
\ee
where $Z_{top}^{[E_8]_i}[P_i]$ is the topological string partition function of the rank 1 $E_8$ theory with the set of parameters and moduli $[P_i]$. Note however that, like in the previous examples, the two set of parameters
$[P_1]$ and $[P_2]$ are not independent but there are some simple relations between the two (which we will discuss later) implied by the structure of the web diagram. Moreover the particular structure of the web diagram 
of the rank 2 $E_8$ theory implies that also the decoupled factor of the theory has a product structure
\begin{equation}
Z_{dec}^{\text{rk2 }E_8} = Z_{dec}^{[E_8]_1}\left[P_1\right] \cdot Z_{dec}^{[E_8]_2}\left[P_2\right],
\end{equation}
where $Z_{dec}^{[E_8]_i}\left[P_i\right] $ is the decoupled factor of a single rank 1 $E_8$ theory with parameters $[P_i]$. Summing up we find that the partition function of the rank 2 $E_8$ theory is 
\be
Z^{\text{rk2 }E_8} = \frac{Z_{top}^{\text{rk2 }E_8}}{Z_{dec}^{\text{rk2 }E_8}} = \frac{ Z_{top}^{[E_8]_1}\left[P_1\right] }{ Z_{dec}^{[E_8]_1}\left[P_1\right]}\cdot\frac{ Z_{top}^{[E_8]_2}\left[P_2\right]}{Z_{dec}^{[E_8]_2}\left[P_2\right]}\,.
\ee
Because of this particular structure we will focus on the computation of the partition function of a single rank 1 $E_8$ theory and then reconstruct the full rank 2 $E_8$ theory partition function by multiplying two 
copies of such partition function with a different set of parameters.

In the following we will choose the parameters of the web diagram of the rank 2 $E_8$ theory so that the centres of mass of the two closed faces present in the diagram coincide: as we will show later with this particular
choice the partition function of the rank 2 $E_8$ theory for a specific choice of parameters coincides with the partition function of a $Sp(2)$ gauge theory with 7 fundamental hypermultiplets and a massless
antisymmetric hypermultiplet. This also implies that the parameters defining the $[E_8]_2$ diagram can be obtained by the parameters of the $[E_8]_1$ diagram by simply replacing the Coulomb branch modulus $\alpha_1$
with the Coulomb branch modulus $\alpha_2$. Because of this we will simply give the parametrisation for the $[E_8]_1$ diagram which was determined in \cite{Hayashi:2014wfa}
\be\begin{split}\label{para.E8-1}
&Q_1 = e^{\alpha_1 -m_1}\,, \, \, Q_2 = e^{-\alpha_1 + m_2}\,, \, \, Q_3 = e^{\alpha_1 +m_3}\,, \, \, Q_4 = e^{\alpha_1 + m_4}\,,\\
&Q_5 e^{\alpha_1 +m_5}\,, \, \, Q_6 = e^{\alpha_1 - m_7}\,, \, \, Q_7 = e^{-\alpha_1 + m_6}\,, \, \, Q_b = u\, e^{-\alpha_1 +f(m)}\,,
\end{split}\ee
where for sake of simplicity we defined $f(m) = \frac{1}{2} (m_1-m_2-m_3-m_4-m_5-m_6+m_7)$. In this parametrisation $u$ will be identified with the instanton fugacity of the $Sp(2)$ gauge theory.

We now will discuss the computation of the partition function of the rank 1 $E_8$ theory which will be the first step for the computation of the partition function of the rank 2 $E_8$ theory. The topological string
partition function for the $[E_8]_1$ diagram can be computed using the rules described in section \ref{sec:vertex} it is given by the following expression
\be\begin{split}
Z_{top}^{[E_8]_1} &= \sum_{\lambda_i,\nu_i,\mu_i} (-Q_1)^{|\mu_1|}(- Q_f Q_2^{-1})^{|\mu_2|}(-Q_4 Q_f Q_1)^{|\mu_3|}(-Q_1 Q_2)^{|\mu_4|}(-Q_7)^{|\mu_5|} (-Q_5)^{|\mu_6|}\\
&(-Q_2)^{|\lam_1|}(- Q_4)^{|\lam_2|}(-\tilde Q_b Q_5)^{|\lam_3|}(-Q_f Q_4 Q_2^{-1} \tilde Q_b Q_5)^{|\lam_4|}(-Q_fQ_7^{-1})^{|\lam_5|} (-\tilde Q_b Q_4)^{|\lam_6|}\\
& (-Q_b)^{|\nu_1|}(- \tilde Q_b)^{|\nu_2|}(-Q_3)^{|\nu_3|}(-Q_3 Q_fQ_2^{-1})^{|\nu_4|}(-Q_6)^{|\nu_5|} (-Q_6 Q_f Q_7^{-1})^{|\nu_6|}\\
& C_{\emptyset \mu_3 \emptyset}(q) C_{\lam_3 \mu_3^t \nu_4}(q) C_{\lam_3^t \emptyset\emptyset}(q)C_{\emptyset \mu_4 \emptyset}(q) C_{\lam_4 \mu_4^t \nu_3}(q) C_{\lam_4^t \emptyset\emptyset}(q)
C_{\emptyset \mu_1 \emptyset}(q) C_{\lam_1 \mu_1^t \nu_1} (q)C_{\lam_1^t \mu_2 \nu_3^t}(q) \\
&C_{\lam_2 \mu_2^t \nu_2}(q) C_{\lam_2^t \emptyset \nu_4^t}(q) C_{\emptyset \mu_5 \nu_1^t}(q) C_{\lam_5 \mu_5^t \nu_5}(q) C_{\lam_5^t \mu_6 \nu_2^5}(q) C_{\lam_6 \mu_6^t 
\nu_6}(q) C_{\lam_6^t \emptyset\emptyset}(q) C_{\emptyset\emptyset \nu_5^t}(q) C_{\emptyset\emptyset\nu_6^t}(q)\,.
\end{split}\ee
This expression can be greatly simplified using the rules described in appendix \ref{sec:top.vertex} and the result is the following
\be\label{E8-1-top}
Z_{top}^{[E_8]_1} = Z_{0}^{[E_8]_1} \cdot Z_{1}^{[E_8]_1} \cdot Z_{=}^{[E_8]_1} \,,
\ee
\be\begin{split}
Z_{0}^{[E_8]_1}& = \frac{H(e^{\pm \alpha_1 -m_1})H(e^{ -\alpha_1 \pm m_2})H(e^{\pm \alpha_1 +m_4})H(e^{\pm \alpha_1 +m_5})H(e^{ -\alpha_1 \pm m_6})}{H(e^{-2\alpha_1})^2 H(e^{m_4-m_2}) H(e^{m_5-m_6})}\\
& \frac{H(u\, e^{m_2+m_5+m_6+f(m)})H(u\, e^{m_2+m_4+m_6+f(m)})\prod_{k=2,6}H(u\, e^{m_k+m_4+m_5+f(m)})}{H(u\, e^{\pm \alpha_1 +m_2 +m_4+m_5+m_6+f(m)})}
\end{split}\ee
\be\begin{split}
Z_{1}^{[E_8]_1}& = \sum_{\nu_i} \left(u\, e^{2m_3+f(m)+m_2+m_5+m_6}\right)^{\frac{|\nu_3|+|\nu_4|}{2}} \left(u\, e^{-2m_7+f(m)+m_2+m_4+m_6}\right)^{\frac{|\nu_5|+|\nu_6|}{2}} \left(u \,e^{f(m)-m_1}\right)^{\frac{|\nu_1|+|\nu_2|}{2}}\\
&  \hat Z (\nu_3,\nu_4) \hat Z(\nu_5,\nu_6)\prod_{i=1,2}\, \prod_{s \in \nu_i} \frac{\prod_{k = 1,2,4,5,6}2 \sinh \frac{E_{i \emptyset}-m_k}{2}}{2 \sinh 
\frac{E_{i \emptyset}+f(m)-m_2-m_4-m_5-m_6+\log u}{2}\prod_{j=1,2}\left( 2 \sinh \frac{E_{ij}}{2}\right)^2}
\end{split}\ee
\be
 Z_{=}^{[E_8]_1}  = H (u\, e^{m_2+m_4+m_5+m_6-m_1+f(m)})^{-2}\,,
\ee
where $\beta_1 = - \beta_2 = \alpha_1$ and moreover we defined
\be\begin{split}
\hat Z (\nu_i ,\nu_j)& = \prod_{s \in \nu_i} \frac{2 \sinh \frac{E_{i \emptyset}-\tilde m^{ij}_{1}}{2}2 \sinh \frac{E_{i 2}-\tilde m^{ij}_{2}}{2}2 \sinh \frac{E_{i 3}-\tilde m^{ij}_{3}}{2}}{\left(2 \sinh \frac{E_{i i}}{2}\right)^2  2 \sinh \frac{E_{i j}}{2}}\\
&\prod_{s \in \nu_j} e^{\frac{\beta_i-\beta_j}{2}}\,\frac{2 \sinh \frac{E_{j \emptyset}-\tilde m^{ij}_{1}}{2}2 \sinh \frac{E_{j 2}-\tilde m^{ij}_{2}}{2}2 \sinh \frac{E_{j 3}-\tilde m^{ij}_{3}}{2}}{\left(2 \sinh \frac{E_{j j}}{2}\right)^2  2 \sinh \frac{E_{j i}}{2}}\,,
\end{split}\ee
with
\be\begin{split}
&\beta_3 = - \beta_4 = \frac{1}{2}(m_2-m_4)\,, \quad \beta_5 = - \beta_6 = \frac{1}{2}(m_7-m_5)\,,\\
&\tilde m_1^{34} = \log u +m_5+m_6+\frac{1}{2}(m_2+m_4)+f(m)\,, \quad \tilde m_2^{34} = \tilde m_3^{34} = -\frac{1}{2}(m_2+m_4)\,,\\
&\tilde m_1^{56} = \log u +m_2+m_4+\frac{1}{2}(m_5+m_6)+f(m)\,, \quad \tilde m_2^{34} = \tilde m_3^{34} = -\frac{1}{2}(m_5+m_6)\,.
\end{split}\ee

It is also important to correctly take into account the decoupled factors which can be computed from the web diagram using the rule described in section \ref{sec:decoupled}
\be\label{E8-1-dec}
Z_{dec}^{[E_8]_1} =Z_{dec, =}^{[E_8]_1}\cdot  Z_{dec, ||}^{[E_8]_1} \cdot  Z_{dec, /\!/}^{[E_8]_1}
\ee
\be
 Z_{dec,=}^{[E_8]_1}  = H (u\, e^{m_2+m_4+m_5+m_6-m_1+f(m)})^{-2}H(u\,e^{m_2+m_4+m_5+m_6+f(m)-m_1})^{-1}\,,
\ee
\be\begin{split}
Z_{dec, /\!/}^{[E_8]_1} =& H(e^{m_2+m_3})^{-1}H( u \, e^{f(m)-m_1})^{-1}H(e^{m_6-m_7})^{-1} H(u \, e^{m_2+m_3+f(m)-m_1})^{-1}\\
&H(u \, e^{m_2+m_3+m_6+f(m)-m_1-m_7})^{-1}H( u \, e^{m_6+f(m)-m_1-m_7})^{-1} H(e^{m_5-m_6})^{-1} \\
&H(e^{m_5-m_7})^{-1} H(u\,e^{m_5+f(m)-m_1-m_7}) ^{-1}H(u\, e^{m_2+m_3+m_5+f(m)-m_1-m_7})^{-1}\\
& H(u\, e^{m_3+m_4+m_5+f(m)-m_1-m_7})^{-1} H(u\,e^{m_3+m_4+f(m)-m_1})^{-1}\\
&H(u\,e^{m_3+m_4+m_6+f(m)-m_1-m_7})^{-1}H (e^{m_3+m_4})^{-1}H(e^{m_4-m_2})^{-1}\,,
\end{split}\ee
\be
Z_{dec, ||}^{[E_8]_1} = H(u\, e^{m_3+m_5+m_6+f(m)})^{-2} H(u \, e^{m_2+m_4+f(m)-m_7})^{-2} H(u^2 \, e^{m_2+m_3+m_4+m_5+m_6+f(m)-m_7})^{-2}
\ee
Note that, in the formalism in section \ref{sec:vertex}, we can simply take into account the decoupled factors from the Higgsed diagram without considering the singlet hypermultiplet contribution in the Higgsed vacuum. This is one of the advantage of the prescription of using the topological vertex rule for Higgsed diagrams instead of UV diagrams.

By combining the topological string partition function \eqref{E8-1-top} with the decoupled factor \eqref{E8-1-dec}, we obtain the partition function associated to the $[E_8]_1$ web diagram
\be\label{eq:E8part}
Z^{[E_8]_1} = \frac{Z_{top}^{[E_8]_1} }{Z_{dec}^{[E_8]_1} } = Z_{pert}^{[E_8]_1} Z_{inst}^{[E_8]_1} 
\ee
\be
Z_{pert}^{[E_8]_1} = \frac{\left(\prod_{a=2,6} H(e^{-\alpha_1\pm m_a})\right)\left(\prod_{a=3,4,5} H(e^{\pm\alpha_1+ m_a})\right)\left(\prod_{a=1,7} H(e^{\pm\alpha_1- m_a})\right)}{H(e^{-2\alpha_1})^2}
\ee
\be\begin{split}
Z_{inst}^{[E_8]_1} =& \frac{H(u\, e^{m_2+m_5+m_6+f(m)})H(u\, e^{m_2+m_4+m_6+f(m)})\prod_{k=2,6}H(u\, e^{m_k+m_4+m_5+f(m)})}{ H(e^{\pm\alpha_1+ m_3})H(e^{\pm\alpha_1- m_7})H(u\, e^{\pm \alpha_1 +m_2 +m_4+m_5+m_6+f(m)})}\\
&H(u\,e^{m_2+m_4+m_5+m_6+f(m)-m_1})H(e^{m_6-m_7}) H(u \, e^{m_2+m_3+f(m)-m_1})\\
&H(u \, e^{m_2+m_3+m_6+f(m)-m_1-m_7})H( u \, e^{m_6+f(m)-m_1-m_7}) H(e^{m_5-m_7})\\
&H(u\,e^{m_5+f(m)-m_1-m_7}) H(u\, e^{m_2+m_3+m_5+f(m)-m_1-m_7})H( u \, e^{f(m)-m_1})\\
& H(u\, e^{m_3+m_4+m_5+f(m)-m_1-m_7}) H(u\,e^{m_3+m_4+f(m)-m_1}) H(e^{m_2+m_3})\\
&H(u\,e^{m_3+m_4+m_6+f(m)-m_1-m_7})H (e^{m_3+m_4})H(u\, e^{m_3+m_5+m_6+f(m)})^{2} \\
&H(u \, e^{m_2+m_4+f(m)-m_7})^{2} H(u^2 \, e^{m_2+m_3+m_4+m_5+m_6+f(m)-m_7})^{2}\\
&\sum_{\nu_i} \left(u\, e^{2m_3+f(m)+m_2+m_5+m_6}\right)^{\frac{|\nu_3|+|\nu_4|}{2}} \left(u\, e^{-2m_7+f(m)+m_2+m_4+m_6}\right)^{\frac{|\nu_5|+|\nu_6|}{2}} \left(u \,e^{f(m)-m_1}\right)^{\frac{|\nu_1|+|\nu_2|}{2}}\\
&  \hat Z (\nu_3,\nu_4) \hat Z(\nu_5,\nu_6)\prod_{i=1,2}\, \prod_{s \in \nu_i} \frac{\prod_{k = 1,2,4,5,6}2 \sinh \frac{E_{i \emptyset}+m_k}{2}}{2 \sinh 
\frac{E_{i \emptyset}+f(m)-m_2-m_4-m_5-m_6+\log u}{2}\prod_{j=1,2}\left( 2 \sinh \frac{E_{ij}}{2}\right)^2}\,.
\end{split}\ee
The instanton part $Z_{inst}^{[E_8]_1}$ is appropriately chosen to be 1 at order $\mathcal{O}(u^0)$. Note that \eqref{eq:E8part} exactly agrees with the partition function
of the rank 1 $E_8$ theory which was computed in \cite{Hayashi:2014wfa}.

After obtaining the partition function of the rank 1 $E_8$ theory it is straightforward to obtain the partition function of the rank 2 $E_8$ theory for this is simply given
by the product
\be
Z^{\text{rk2 }E_8} = Z^{[E_8]_1}[\alpha_1, \{m_{1, 2, 3, 4, 5, 6,7}\}, u] \cdot Z^{[E_8]_1}[\alpha_2, \{m_{1, 2, 3, 4, 5, 6,7}\}, u]\,.
\ee
It follows that the perturbative part is
\be\label{eq:rank2E8pert}\begin{split}
Z_{pert}^{\text{rk}2 \, E_8} =& \frac{\left(\prod_{a=2,6} H(e^{-\alpha_1\pm m_a})\right)\left(\prod_{a=3,4,5} H(e^{\pm\alpha_1+ m_a})\right)\left(\prod_{a=1,7} H(e^{\pm\alpha_1- m_a})\right)}{H(e^{-2\alpha_1})^2}\\
&\frac{\left(\prod_{a=2,6} H(e^{-\alpha_2\pm m_a})\right)\left(\prod_{a=3,4,5} H(e^{\pm\alpha_2+ m_a})\right)\left(\prod_{a=1,7} H(e^{\pm\alpha_2- m_a})\right)}{H(e^{-2\alpha_2})^2}\,.
\end{split}\ee
By comparing this result with \eqref{pert.gauge}, one can see that this is exactly the perturbative contribution from seven fundamental hypermultiplets with masses given by $m_{1, 2, 3, 4, 5, 6,7}$, one massless antisymmetric hypermultiplet
and the $Sp(2)$ vector multiplet. Like in the previous examples discussed in this section we see that the contribution of the anti-symmetric hypermultiplet does not appear
in \eqref{eq:rank2E8pert} because it is cancelled by some factors in the $Sp(2)$ vector multiplet contribution.

It is also possible to compute the instanton contributions to the partition function of the rank 2 $E_8$ theory. At 1-instanton level the result is particularly simple because it 
is simply the sum of the 1-instanton part of the $[E_8]_1$ diagram and the 1-instanton part of the $[E_8]_2$ diagram. The 1-instanton part of each diagram agrees with
the expression \eqref{Sp1.1instanton} of the partition function of an $Sp(1)$ gauge theory with $N_f =7$ fundamental flavours
\be
Z^{\text{rk2 }E_8}_{1\text{-}inst} =\frac{1}{2}\sum_{i=1}^2 \left\{\frac{\prod_{a=1}^{7}2\sinh\frac{m_a}{2}}{2\sinh\frac{\pm \epsilon}{2}2\sinh\frac{\pm\alpha_i}{2}}+\frac{\prod_{a=1}^{7}2\cosh\frac{m_a}{2}}{2\sinh\frac{\pm \epsilon}{2}2\cosh\frac{\pm\alpha_i}{2}}\right\}\,,
\ee
which agrees with the expression \eqref{SpN.sp.1instanton} of the
1-instanton part of $Sp(2)$ with seven fundamental hypermultiplets and one massless antisymmetric hypermultiplets computed on the field
theory side. 

Moreover we have checked agreement of the partition function of the rank 2 $E_8$ theory with the partition function of a $Sp(2)$ theory with 7 flavours and one massless anti-symmetric hypermultiplet at 2-instanton in the special limit where one of the masses of the fundamental hypermultiplets is set to zero\footnote{Again, the reason why we set one mass to zero is the technical issue.}.

\subsection{Rank $N$ $E_{6, 7, 8}$ theories}

The generalisation to the partition functions of the rank $N$ $E_6, E_7, E_8$ theories is straightforward. The web diagrams of the rank $N$ $E_{6, 7, 8}$ theories are simply the superposition of $N$ copies of the web diagram of the rank $1$ $E_{6, 7, 8}$ theories respectively \cite{Benini:2009gi}. Therefore, the topological vertex formalism on the rank $N$ $E_{6, 7, 8}$ diagrams shows that it can be written by the product
\begin{equation}
Z^{\text{rkN} E_{N_f+1}} = \prod_{i=1}^N \frac{Z_{top}^{[E_{N_f+1}]_i}[P_i]}{Z_{dec}^{[E_{N_f+1}]_i}[P_i]}, \label{rkN.ENf-1}
\end{equation} 
where $Z_{top}^{[E_{N_f+1}]_i}[P_i]$ and $Z_{dec}^{[E_{N_f+1}]_i}[P_i]$ represent the topological string partition function and the decoupled factor computed from the $i$-th copy of the web diagram of the rank $N$ $E_{N_f+1}$ theory. In this subsection, $N_f$ always mean either $N_f = 5, 6, $ or $7$. We also consider special web diagrams such that the centres of mass of all the closed faces in the diagram coincide with each other. In this particular situation the web diagram realises an $Sp(N)$ gauge theory with $N_f$ fundamental hypermultiplets and with a massless anti-symmetric
hypermultiplet. Moreover, we essentially choose the same parameterisation as \eqref{para.T3-1}, \eqref{para.E7-1} and \eqref{para.E8-1} for the first copy of the web diagram of the rank $N$ $E_6, E_7, E_8$ theories respectively, The parameterisation of the $i$-th copy is simply obtained by exchanging $\alpha_1$ with $\alpha_i$. The product \eqref{rkN.ENf-1} is more precisely given by
\begin{equation}
Z^{\text{rkN }E_{N_f+1}} = \prod_{i=1}^N \frac{Z_{top}^{[E_{N_f+1}]_i}[\alpha_i, \{m_{1, \cdots, N_f}\}, u]}{Z_{dec}^{[E_{N_f+1}]_i}[\{m_{1, \cdots, N_f}\}, u]},\label{rkN.ENf-2}
\end{equation}
where each factor $Z_{top}^{[E_{N_f+1}]_i}[\alpha_i, \{m_{1, \cdots, N_f}\}, u]/Z_{dec}^{[E_{N_f+1}]_i}[\{m_{1, \cdots, N_f}\}, u]$ for $N_f = 5, 6, 7$ is essentially given by \eqref{T3-1}, \eqref{E7-1} and \eqref{eq:E8part} respectively with $\alpha_1$ exchanged with $\alpha_i$.

One can easily see that \eqref{rkN.ENf-2} correctly realises the perturbative part and the $1$-instanton part of the partition function of the $Sp(N)$ gauge theory with $N_f$ fundamental hypermultiplets and one massless anti-symmetric hypermultiplet. The perturbative part of \eqref{rkN.ENf-2} is given by
\begin{equation}
Z^{\text{rkN }E_{N_f+1}}_{pert} = \prod_{i=1}^N Z_{pert}^{[E_{N_f+1}]_i}[\alpha_i, \{m_{1, \cdots, N_f + 1}\}].
\end{equation}
This indeed agrees with the perturbative partition function \eqref{pert.gauge} of the $Sp(N)$ gauge theory with $N_f$ flavours and a massless anti-symmetric hypermultiplet.

Also, the $1$-instanton part of \eqref{rkN.ENf-2} is given by
\begin{equation}
Z^{\text{rkN }E_{N_f+1}}_{1\text{-}inst}  = \sum_{i=1}^{N}Z^{[E_{N_f+1}]_i}_{inst}[\alpha_i, \{m_{1, \cdots, N_f}\}, u]|_{\mathcal{O}(u)}\label{rkN.ENf.1instanton}
\end{equation}
Since the $1$-instanton part of the partition function $Z^{[E_{N_f+1}]_i}_{inst}$ reproduces the $1$-instanton part of the partition function of the $Sp(1)$ gauge theory with $N_f$ flavours and a massless anti-symmetric hypermultiplet with the Coulomb branch modulus $\alpha_i$, \eqref{rkN.ENf.1instanton} precisely agrees with \eqref{SpN.sp.1instanton}. 

We have seen that the partition function of the $Sp(N)$ gauge theory with $N_f$ fundamental hypermultiplets and a massless anti-symmetric hypermultiplet always shows the product structure. A physical reason why the factorisation happens may be that we are studying an IR theory in a Higgs branch of the $Sp(N)$ gauge theory with a non-vanishing vev for a ``$0$'' weight of the anti-symmetric hypermultiplet\footnote{The anti-symmetric hypermultiplet can acquire a non-vanishing vev because it is massless.}. When it acquires a vev along the ``$0$" weight of the anti-symmetric representation of $Sp(N)$, then the Cartan parts remain massless but some (but not all) of the root components  of the adjoint representation of the $Sp(N)$ vector multiplet become massive. In fact the $Sp(N)$ gauge group is broken to $Sp(1)^N$ by the vev.

The factorisation of the partition function of the $Sp(N)$ gauge theories with massless anti--symmetric hypermultiplet is consistent with the fact that the Seiberg--Witten curves of the theories also factorise \cite{Douglas:1996js, Kim:2014nqa}. Although the product structure of the Seiberg--Witten curves imply that the prepotential is written by the sum of $N$ copies of $Sp(1)$ prepotential, our analysis shows that the full Nekrasov partition function itself also factorises.

\section{Towards refined topological vertex for Higgsed 5d $T_N$ theories}
\label{sec:refined}
\begin{figure}[t]
\begin{center}
\includegraphics[width=50mm]{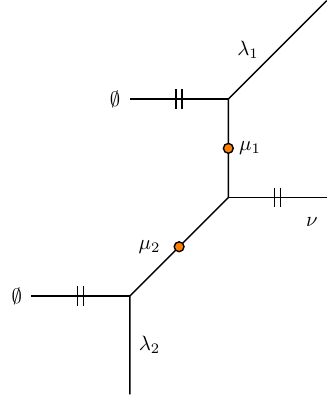}
\end{center}
\caption{Higgsing of parallel horizontal legs in a $T_N$ diagram. The orange dots indicate the curves that are shrunk to zero length and the double lines the preferred directions.}
\label{fig:horref}
\end{figure}
In this section, we extend the analysis in section \ref{sec:vertex} to the refined topological vertex for Higgsed 5d $T_N$ theories. 
In some special cases, we can derive new refined topological vertex that can be directly applied to Higgsed web diagrams. In all the computation of the refined topological string partition function, we will choose the horizontal directions as the preferred directions. 

\subsection{Refined topological vertex, external horizontal legs}

We will start by considering the case in which the external legs that we put on top of each other are horizontal. We show in Figure \ref{fig:horref} the diagram we consider. The local part of the topological string partition function can be easily computed using the refined topological vertex and the result is
\be
Z = \sum_{\mu_1,\mu_2} \left(-\left(\frac{q}{t}\right)^{\frac{1}{2}}\right)^{|\mu_1|+|\mu_2|} C_{\lam_1 \mu_1 \emptyset}(q,t) C_{\mu_2 \mu_1^t \nu}(t,q)
C_{\mu_2^t \lam_2 \emptyset}(q,t)\,.
\ee
Here we used the tuning condition \eqref{refined.higgs1}. It is possible to perform also in this case the Young diagrams summations over $\mu_1$ and $\mu_2$ arriving at the following expression
\be\begin{split}
Z =& \sum_{\eta_i, \xi_i} \left(-1\right)^{|\xi_1|+|\xi_2|} \left(\frac{q}{t}\right)^{\frac{|\lam_1|-|\lam_2|-|\xi_1|-|\xi_2|}{2}} q^{-\frac{||\lam_2^t||^2+||\nu||^2}{2}}t^{\frac{||\lam_2||^2}{2}}\tilde Z_{\nu}(t,q) \\
&s_{\lam_1^t/\eta_1}(q^{-\rho}) s_{\lam_2/\eta_3}(t^{-\rho}) s_{\eta_1^t /\xi_1}(q^{-\rho}t^{-\nu^t}) s_{\eta_2^t /\xi_1^t}(t^{-\rho})s_{\eta_2^t /\xi_2}(q^{-\rho})s_{\eta_3^t
/\xi_2^t}(t^{-\rho}q^{-\nu})
\\&\prod_{i,j=1}^\infty \frac{(1-q^{i}t^{j-1-\nu_i^t})(1-q^{i-\nu_j}t^{j-1})}{(1-q^{i+1}t^{j-2})}\,.
\end{split}\ee
In this case looking at the infinite product factor we see that the result will be zero unless $\nu= \emptyset$, and so in the following we will focus in this case.
In this situation using \eqref{eq:schur2} it is possible to perform some additional simplifications and arrive at the following simple result
\be\label{eq:refhor}
Z  = q^{-\frac{||\lam_2^t||^2}{2}}t^{\frac{||\lam_2||^2}{2}}\left(\frac{q}{t}\right)^{|\lam_1|+|\lam_2|} \sum_\kappa \left(\frac{t}{q}\right)^{\frac{3|\kappa|+|\lam_1|-|\lam_2|}{2}}s_{\lam_1^t /\kappa}(q^{-\rho}) s_{\lam_2 /\kappa}(t^{-\rho})\prod_{i,j=1}^\infty\frac{(1-q^i t^{j-1})^2}{(1-q^{i+1}t^{j-2})}\,.
\ee
Motivated by the form of the result we will define the new vertex
\be
\tilde C_{\lam \mu \nu} (q,t)= q^{-\frac{||\mu^t||^2}{2}}t^{\frac{||\mu||^2+||\nu||^2}{2}}\tilde Z_\nu(q,t) \sum_\eta 
\left(\frac{t}{q}\right)^{\frac{3|\eta|+|\lam|-|\mu|}{2}} s_{\lam^t/\eta}(q^{-\rho}t^{-\nu}) s_{\mu/\eta} (t^{-\rho}q^{-\nu^t})
\ee
which allows us to rewrite \eqref{eq:refhor} as
\be
\tilde{Z} = \left(\frac{q}{t}\right)^{|\lam_1|+|\lam_2|} \tilde C_{\lam_1 \lam_2 \emptyset}(q,t)\,, \label{newvertex}
\ee
where we also dropped the infinite product terms which only contribute to the decoupled factor. Therefore we see that in the case of putting a pair of parallel horizontal external legs on top of each other we 
can use the new refined topological vertex \eqref{newvertex} to compute the partition function for a Higgsed diagram.

\subsection{Refined topological vertex, external vertical and diagonal legs}
\begin{figure}[t]
\begin{center}
\includegraphics[width=70mm]{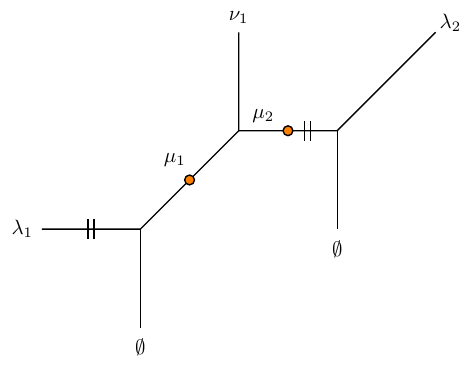}
\end{center}
\caption{Higgsing of parallel vertical legs in a $T_N$ diagram. The orange dots indicate the curves that are shrunk to zero length and the double lines the preferred directions.}
\label{fig:vertref}
\end{figure}
We then consider a Higgs branch realised by a tuning that places parallel external NS5-branes on top of each other. Since the refined topological vertex is not symmetric under the exchange between the three Young diagrams, one needs to work on this case separately. 

The corresponding local diagram is depicted in Figure \ref{fig:vertref}. The  refined topological string partition function for the local diagram is
\begin{equation}
Z(\lambda_1, \nu_1, \lambda_2) = \sum_{\mu_1, \mu_2} \left(-\left(\frac{t}{q}\right)^{\frac{1}{2}}\right)^{|\mu_1|+|\mu_2|}C_{\mu_1\emptyset\lambda_1}(q, t)C_{\mu_1^t\nu_1\mu_2}(t, q)C_{\lambda_2\emptyset\mu_2^t}(q,t). \label{refined.vertical1}
\end{equation}
Here we used the tuning condition \eqref{refined.higgs2}. It is possible to sum over the Young diagram of $\mu_1$ in \eqref{refined.vertical1},
\begin{eqnarray}
Z(\lambda_1, \nu_1, \lambda_2) &=& \sum_{\mu_2, \eta} \left(-\left(\frac{t}{q}\right)^{\frac{1}{2}}\right)^{|\mu_2|}q^{\frac{1}{2}(||\nu_1||^2+||\mu_2||^2)}t^{\frac{1}{2}(||\lambda_1|| - ||\nu_1^t||^2 + ||\mu_2^t||^2)}\tilde{Z}_{\lambda_1}(q, t)\tilde{Z}_{\mu_2}(t, q)\tilde{Z}_{\mu_2^t}(q, t)\nonumber\\
&&s_{\nu_1/\eta}(t^{-\mu_2^t+\frac{1}{2}}q^{-\rho-\frac{1}{2}})s_{\lambda_2^t}(t^{-\mu_2^t+\frac{1}{2}}q^{-\rho-\frac{1}{2}})s_{\eta^t}(-t^{-\lambda_1+\frac{1}{2}}q^{-\rho-\frac{1}{2}})\nonumber\\
&&\prod_{i,j=1}^{\infty}\left(1- q^{i-\mu_{2,j}-1}t^{j-\lambda_{1,i}}\right). \label{refined.vertex2}
\end{eqnarray}
The last term of \eqref{refined.vertex2} indicates that the $\mu_2$ summation of \eqref{refined.vertex2} is bounded by the Young diagram $\lambda_1$. More precisely, \eqref{refined.vertex2} is zero unless $\lambda_{1, i} \leq \mu^t_{2, i}$ for each $i$. In the $\mu_2$ summation, we can proceed further for a special case where $\mu_2 = \lambda_1^t$. When $\mu_2 = \lambda_1^t$, \eqref{refined.vertex2} further reduces to 
\begin{equation}
Z(\lambda_1, \emptyset, \lambda_2)|_{\mu_2 = \lambda_1^t} = \left(\frac{t}{q}\right)^{|\lambda|}C_{\lambda_2\emptyset\lambda_1}(q, t)\; \prod_{i,j=1}^{\infty}\left(1 - q^{i-1}t^j\right),
\end{equation}
where $|_{\mu_2 = \lambda_1^t}$ indicates extracting the term of $\mu_2 = \lambda_1^t$ in the $\mu_2$ summation. In this case, $\nu_1$ is restricted to $\emptyset$.

For general Young diagrams of $\nu_1, \lambda_2$, it is difficult to get an analytic expression after performing the summations of $\mu_2$ and $\eta$. However, we can still perform the summation in a special case where $\nu_1=\emptyset, \lambda_2 = \emptyset$. In fact in this particular situation we find that
\begin{equation}
\frac{\sum_{\mu_2}(-Q)^{|\mu_2|}q^{\frac{1}{2}||\mu_2||^2}t^{\frac{1}{2}||\mu_2^t||^2}\tilde{Z}_{\mu_2}(t, q)\tilde{Z}_{\mu_2^t}(q, t)\prod_{i,j=1}^{\infty}\left(1- q^{i-\mu_{2,j}-1}t^{j-\lambda_{1,i}}\right)}{\sum_{\mu_2}(-Q)^{|\mu_2|}q^{\frac{1}{2}||\mu_2||^2}t^{\frac{1}{2}||\mu_2^t||^2}\tilde{Z}_{\mu_2}(t, q)\tilde{Z}_{\mu_2^t}(q, t)\prod_{i,j=1}^{\infty}\left(1- q^{i-\mu_{2,j}-1}t^{j}\right)} = \left(\frac{t}{q}\right)^{\frac{|\lambda_1|}{2}}Q^{|\lambda_1|}. \label{terminate1}
\end{equation}
This identity has been checked order by order in $Q$ up to $|\mu_2| = 4$. The denominator of the left-hand side of \eqref{terminate1} can be evaluated exactly,
\begin{eqnarray}
&&\sum_{\mu_2}(-Q)^{|\mu_2|}q^{\frac{1}{2}||\mu_2||^2}t^{\frac{1}{2}||\mu_2^t||^2}\tilde{Z}_{\mu_2}(t, q)\tilde{Z}_{\mu_2^t}(q, t)\prod_{i,j=1}^{\infty}\left(1- q^{i-\mu_{2,j}-1}t^{j}\right) \nonumber\\
&=& \prod_{i,j=1}^{\infty}\left(1 - Q\left(\frac{t}{q}\right)^{\frac{1}{2}}q^{i-1}t^j\right)\left(1 - \left(\frac{t}{q}\right)^{\frac{1}{2}}q^{i-\frac{1}{2}}t^{j-\frac{1}{2}}\right)\left(1 - Qq^{i-\frac{1}{2}}t^{j-\frac{1}{2}}\right). \label{U1inst}
\end{eqnarray}
By using \eqref{refined.vertex2}, \eqref{terminate1} and \eqref{U1inst}, we obtain
\begin{equation}
Z(\lambda_1, \emptyset, \emptyset) = \left(\frac{t}{q}\right)^{|\lambda_1|}C_{\emptyset\emptyset\lambda_1}(q, t)\prod_{i,j=1}^{\infty}\frac{(1-q^{i-1}t^j)^2}{(1-q^{i-2}t^{j+1})}. \label{refined.vertical.special}
\end{equation}
The three factors at the end of \eqref{refined.vertical.special} may be a part of the decoupled factor. Hence, we can use a different new refined topological vertex
\begin{equation}
\tilde{Z}(\lambda_1, \emptyset, \emptyset) = \left(\frac{t}{q}\right)^{|\lambda_1|}C_{\emptyset\emptyset\lambda_1}(q, t), 
\end{equation}
when $\nu_1 = \lambda_2 = \emptyset$ in Figure \ref{fig:vertref}.

Note that \eqref{refined.vertical.special} is obtained by setting $\lambda_2 = \emptyset$ in \eqref{refined.vertex2} up to decoupled factors. This is not a coincidence. Since the numerator of \eqref{terminate1} starts from $\mathcal{O}(Q^{|\lambda_1|})$, the coefficient at this order is dictated by the Young diagram $\mu_2 = \lambda_1^t$. Therefore, \eqref{refined.vertical.special} agrees with \eqref{refined.vertex2} when $\lambda_2= \emptyset$ up to the decoupled factor.
\begin{figure}[t]
\begin{center}
\includegraphics[width=60mm]{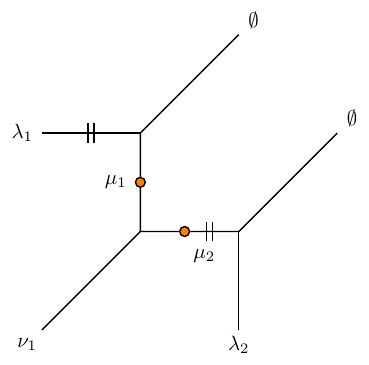}
\end{center}
\caption{Higgsing of parallel diagonal legs in a $T_N$ diagram. The orange dots indicate the curves that are shrunk to zero length and the double lines the preferred directions.}
\label{fig:diagref}
\end{figure}

 It is also possible to consider a Higgs branch realised by placing external (1,1)-branes on top of each other. The diagram is shown in Figure \ref{fig:diagref}. In this case the local part of the partition function
 is
 \be\label{eq:higvert}
 Z(\lambda_1,\nu_1,\lambda_2) = \sum_{\mu_1,\mu_2} \left(-\left(\frac{q}{t}\right)^{\frac{1}{2}}\right)^{|\mu_1|+|\mu_2|} C_{\emptyset\mu_1\lambda_1}(q,t) C_{\nu_1 \mu_1^t \mu_2}(t,q) C_{\emptyset \lambda_2 \mu_2^t}(q,t)\,,
 \ee
where we have used the tuning condition  \eqref{refined.higgs1}. It is possible to perform the $\mu_1$ summation in \eqref{eq:higvert} and the result is
\be\begin{split}
 Z(\lambda_1,\nu_1,\lambda_2) &= \sum_{\mu_2,\eta} \left(-\left(\frac{q}{t}\right)^{\frac{1}{2}}\right)^{|\mu_2|} t^{\frac{1}{2}(||\lambda_2||^2+||\mu^t_2||^2+||\lambda_1||^2)}q^{\frac{1}{2}(||\mu_2||^2-||\lambda^t_2||^2)} \tilde Z_{\lambda_1}(q,t) \tilde Z_{\mu_2}(t,q) \tilde Z_{\mu_2^t}(q,t)\\
 & s_{\nu_1/\eta}(t^{-\rho-\frac{1}{2}}q^{-\mu_2+\frac{1}{2}}) s_{\lambda_2}(t^{-\rho+\frac{1}{2}}q^{-\mu_2^t-\frac{1}{2}}) s_{\eta^t}(q^{-\lam_1^t+\frac{1}{2}} t^{-\rho-\frac{1}{2}})\\
 & \prod_{i,j=1}^\infty (1- t^{i-\mu_{2,j}^t-1}q^{j-\lambda_{1,i}^t})\,.
\end{split}\ee
Note that like in the previous case there is a bound on the $\mu_2$ summation which in this case is $\lambda_{1,i}\leq\mu_{2,i} $. Again it is difficult to obtain an analytic expression after performing the $\mu_2$ and $\eta$ summations for general $\lambda_2$ and $\nu_1$, however in the special case $\lambda_2 = \nu_1 = \emptyset$ there are some great simplifications. In particular
we find that
\be
\frac{\sum_{\mu_2} (-Q)^{|\mu_2|} t^{\frac{1}{2}||\mu^t_2||^2}q^{\frac{1}{2}||\mu_2||^2}\tilde Z_{\mu_2}(t,q) \tilde Z_{\mu_2^t} (q,t)\prod_{i,j=1}^\infty(1-t^{i-\mu_{2,j}^t-1}q^{j-\lambda_{1,i}^t})}{\sum_{\mu_2} (-Q)^{|\mu_2|} t^{\frac{1}{2}||\mu^t_2||^2}q^{\frac{1}{2}||\mu_2||^2}\tilde Z_{\mu_2}(t,q) \tilde Z_{\mu_2^t} (q,t)\prod_{i,j=1}^\infty(1-t^{i-\mu_{2,j}^t-1}q^{j})} = \left(\frac{q}{t}\right)^{\frac{|\lambda_1|}{2}} Q^{|\lambda_1|}\,.
\ee
Therefore like in the case of placing parallel vertical legs on top of each other we find that, up to some decoupled factors, it is possible to write the local contribution to the partition function using
a new vertex
\be
\tilde Z(\lambda_1,\emptyset,\emptyset) = \left(\frac{q}{t}\right)^{|\lambda_1|} C_{\emptyset \emptyset \lambda_1}(q,t)\,.
\ee
\section{Conclusion and discussion}

In this paper we have developed a computational method that allows to apply the topological vertex directly to a general Higgsed $T_N$ web diagram which is not dual to a toric geometry. By carefully dealing with the Higgsing prescription in a local part of a strip diagram, the topological string partition function of this strip diagram with non-trivial representations on all the external leg can be replaced by the standard topological vertex with a specific relation between the representations, a fact that nicely fits with the intuition from the Higgsed web diagram. We also argued that we can simply remove the decoupled factors directly read off from the Higgsed web diagram to reproduce the partition function of the IR theory in the Higgs vacuum without removing singlet hypermultiplets in the vacuum. Summing up, we can simply apply the method of the standard topological vertex as well as the decoupled factor to a Higgsed $T_N$ diagram although it does not corresponds to a toric geometry in the dual picture. 

Let us highlight the advantages of our method compared to the known prescription in \cite{Hayashi:2013qwa, Hayashi:2014wfa}. Since we directly apply the topological vertex to a Higgsed diagram the computation of the topological string partition function is greatly simplified compared to the computation of the topological string partition function for its UV theory. In particular the number of Young diagram summations is reduced compared to the UV computation. From the viewpoint of the UV partition function it is necessary to see the trivialisation of the Young diagram summations by carefully looking at the tuning conditions in the partition function. However this process is automatic when we make use of our method. Moreover, in the previous prescription it was necessary to remove  the contributions of singlet hypermultiplets in the Higgs vacuum which are completely absent using our method.
The only contribution which we need to remove is the decoupled factor which can be easily read off from the Higgsed diagram. Therefore with our method it is possible to skip two of the steps of the 
previous method giving therefore a great simplification.

In order to exemplify our method we have explicitly computed the partition function of the rank $2$ $E_{6,7,8}$ theories directly from their Higgsed diagrams. Although the diagrams are quite complicated, our method greatly simplifies the calculation and moreover, by directly applying the topological vertex to the Higgsed web diagram, it is clear that the partition function of the special higher rank $E_ {N_f+1}$ theories with a massless hypermultiplet has the product structure. We have also found the complete agreement with their field theory results. 

Finally, we extended our method to some special cases of the refined topological vertex. When we consider putting parallel external D5-branes together, it is possible to reduce the refined topological string partition function on the corresponding strip diagram to a function associated to a trivalent vertex. The function is similar to the refined topological vertex but it is slightly different. Therefore, we defined the new refined topological vertex, and it can be used to compute the partition function associated to the Higgsing which arises from coincident D5-branes.

We can proceed in the refined topological vertex for the case of coincident NS5-branes. In this case, we can successfully sum up the Young diagrams for a very special case where all but one Young diagrams are trivial on the external legs.  Even this special case leads to another new refined topological vertex. It would be very interesting to generalise the computation by allowing more general Young diagrams on the external legs. In the special case, it was important to note that the expression \eqref{terminate1} gives only one term. When one generalises it with a non-trivial $\nu_1$ or $\lambda_2$ (but either of them is trivial), it gives more terms but still we observed that the series expansion $Q$ terminate at finite order. More specifically, it starts from $\mathcal{O}(Q^{|\lambda_1|})$ and ends at  $\mathcal{O}(Q^{|\lambda_1| + |\nu_1|-|\eta|})$ or $\mathcal{O}(Q^{|\lambda_1| + |\lambda_2|})$ respectively. It is interesting to understand the essential reason why this termination occurs, and also find any analytic expression for their coefficients. So far, the termination occurs with general $Q$ in \eqref{terminate1}. It would be interesting to see if setting $Q = \left(\frac{t}{q}\right)^{\frac{1}{2}}$ could simplify the computation.

In this paper, we have focused on the standard Higgsing with a vev which is constant over the spacetime. In fact, one can consider different Higgsing with a position--dependent vev. The position--dependent vev triggers an RG flow and the low energy theory may contain a surface defect \cite{Gaiotto:2012xa}. In \cite{Gaiotto:2014ina}, the tuning condition corresponding to the different Higgsing was used to compute the five-dimensional superconformal index of theories with surface defects. It would be interesting to extend our analysis so that it can be applied to the cases with surface defects. 

Finally, there has recently been a lot of progress in computing the elliptic genera of strings in six-dimensional superconformal field theories \cite{Haghighat:2013gba, Haghighat:2013tka, Haghighat:2014pva, Hosomichi:2014rqa, Kim:2014dza, Haghighat:2014vxa, Kim:2015jba, Gadde:2015tra} and also the elliptic genera of strings in five-dimensional superconformal field theories \cite{Haghighat:2015coa, Hohenegger:2015cba}. It would be interesting to see if our prescription can be applied to such computation.

\section*{Acknowledgments}

We thank Hee-Cheol Kim and  Futoshi Yagi for useful discussions. H.H. would like to thank Ewha woman's university and Max Planck Institute for Physics during a part of the project, for kind hospitality and financial support. 
G.Z. would like to thank Max-Plack-Institut f\"ur Physik, Munich, for kind hospitality during the course of this work. The work of H.H. and G.Z. is supported by the grant FPA2012-32828 from the MINECO, the REA grant agreement PCIG10-GA-2011-304023 from the People Programme of FP7 (Marie Curie Action), the ERC Advanced Grant SPLE under contract ERC-2012-ADG-20120216-320421 and the grant SEV-2012-0249 of the ``Centro de Excelencia Severo Ochoa" Programme. The work of G.Z. is also supported through a grant from ``Campus Excelencia Internacional UAM+CSIC" and partially by the COST Action MP1210.

\appendix

\section{Partition function from the refined topological vertex}
\label{sec:app.top}

We summarise the rules of the refined topological vertex and also the decoupled factor to set the convention used in this paper. 

\subsection{Refined topological vertex}
\label{sec:top.vertex}

The topological string partition function can be defined as an exponential of a generating function of the Gromov--Witten invariants of a Calabi--Yau threefold $X$,
\begin{equation}
Z_{top} = \exp\left(\sum_{g=0}^{\infty}\mathcal{F}_g g_s^{2g-2}\right),
\end{equation}
where $g_s$ is the topological string coupling and $\mathcal{F}_g$ is the genus $g$ topological string amplitude 
\begin{equation}
\mathcal{F}_g = \sum_{C \in H_2(X, \mathbb{Z})} N_C^g \; Q_C.
\end{equation}
$Q_C$ is defined by $Q_C := e^{-\int_C J}$ where $J$ is the K\"ahler form of $X$ , and $N_C^g$ is the genus $g$ Gromov--Witten invariant. Although it is difficult to compute the higher genus topological string amplitudes directly, the M-theory interpretation of the topological string amplitude enables us to write down the all genus answer \cite{Gopakumar:1998ii, Gopakumar:1998jq}, Moreover, if the Calabi--Yau threefold is toric, the refined topological vertex \cite{Awata:2005fa, Iqbal:2007ii} can compute the refined version of the all genus topological string partition function up to constant map contributions in a diagrammatic way. We call such a partition function the refined topological string partition function.

In order to apply the refined topological vertex to a toric Calabi--Yau threefold, we first choose the preferred direction in the toric diagram. In this paper, we choose the horizontal direction as the preferred direction unless stated. For other lines, we assign two parameters $t$ and $q$. At each trivalent vertex, one leg is in the preferred direction, $t$ is assigned to another leg, and $q$ is assigned to the other leg. For each internal line we assign a Young diagram with an orientation and for each external leg we assign a trivial diagram. 

\begin{figure}[t]
\begin{center}
\includegraphics[width=60mm]{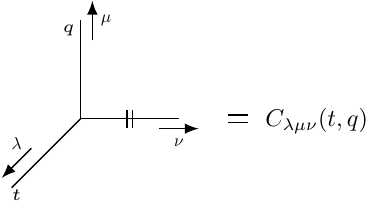}
\end{center}
\caption{A trivalent vertex with a particular assignment of Young diagrams, the preferred direction and $t, q$. $||$ represents an line in the preferred direction.}
\label{fig:vertex}
\end{figure}

We then assign the refined topological vertex for each trivalent vertex shown in Figure \ref{fig:vertex}
\begin{equation}
C_{\lambda\mu\nu}(t, q) := t^{-\frac{||\mu^t||^2}{2}}q^{\frac{||\mu||^2+||\nu||^2}{2}}\tilde{Z}_{\nu}(t, q)\sum_{\eta}\left(\frac{q}{t}\right)^{\frac{|\eta| + |\lambda| -|\mu|}{2}}s_{\lambda^t/\eta}\left(t^{-\rho}q^{-\nu}\right)s_{\mu/\eta}\left(t^{-\nu^t}q^{-\rho}\right), \label{topological.vertex}
\end{equation}
where $|\lambda| := \sum_i \lambda_i$ and $||\lambda||^2 := \sum_i \lambda_i^2$. $\lambda_i$ stands for the height of $i$-th column\footnote{Our convention of writing a Young diagram is that the height of a column is equal to or higher than the height of its right column.}. $s_{\lambda/\eta}(x)$ is the skew Schur function defined as \eqref{skew.schur}, and $\rho := - i + \frac{1}{2}, (i = 1, 2, \cdots)$. $\tilde{Z}_{\nu}(t, q)$ is defined as
\begin{equation}
\tilde{Z}_{\nu}(t, q) := \prod_{(i, j) \in \nu} \left(1 - q^{l_{\nu}(i, j)}t^{a_{\nu}(i, j) + 1}\right)^{-1},
\end{equation}
where $l_{\nu}(i,j) := \nu_i - j$ is the leg length and $a_{\nu}(i, j) := \nu^t_j - i$ is the arm length of the Young diagram $\nu$. 

\begin{figure}[t]
\begin{center}
\includegraphics[width=100mm]{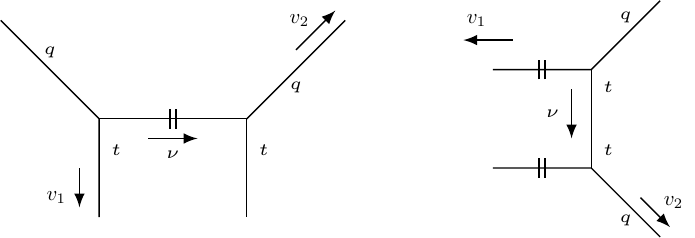}
\end{center}
\caption{Two examples of a propagator connecting two vertices.  The left figure shows a propagator in the preferred direction. The right figure shows a propagator in the non-preferred direction. }
\label{fig:propagator}
\end{figure}

For each propagator with a Young diagram $\nu$ in Figure \ref{fig:propagator}, 
which connects two vertices, we assign the K\"ahler parameter $Q = e^{-\int_C J}$ associated to the size of the two-cycle $C$ corresponding to the internal line and also the framing factor 
\begin{equation}
(-Q)^{|\nu|} f_{\nu}(t, q)^{n} \quad \text{or} \quad (-Q)^{|\nu|} \tilde{f}_{\nu}(t, q)^{n} \label{propagator}
\end{equation}
$f_{\nu}(t, q)$ is the framing factor for the preferred direction
\begin{equation}
f_{\nu}(t, q) := (-1)^{|\nu|}t^{\frac{||\nu^t||^2}{2}}q^{-\frac{||\nu||^2}{2}}
\end{equation}
$\tilde{f}_{\nu}(t, q)$ is the framing factor for the non-preferred direction
\begin{equation}
\tilde{f}_{\nu}(t, q) := (-1)^{|\nu|}q^{\frac{||\nu^t||^2}{2}}t^{-\frac{||\nu||^2}{2}}\left(\frac{q}{t}\right)^{\frac{|\nu|}{2}}.
\end{equation}
$n$ is defined as $n = \det(v_1, v_2)$ where $v_1, v_2$ are vectors depicted in Figure \ref{fig:propagator}. 

The refined topological string partition function up to the constant map contributions can be computed by combining the vertices, propagators and then sum up all the Young diagrams.

In fact, the toric diagrams in the M-theory compactification are dual to webs of $(p, q)$ 5-branes \cite{Leung:1997tw}. In the dual picture, the five-dimensional theory is effectively realised on the worldvolume theory on the 5-branes compactified on segments. 

The unrefined topological string partition function can be obtained is simply by setting $t = q$. Therefore, the rule of the topological vertex is given by setting $t = q$ in \eqref{topological.vertex} and \eqref{propagator}, which is essentially the same rule obtained in \cite{Iqbal:2002we, Aganagic:2003db}. 

\subsection{Decoupled factor}

The refined topological string partition itself does not reproduce the Nekrasov partition function of the five-dimensional theory realised by the M-theory compactification on the toric Calabi--Yau threefold $X$, but we need to eliminate some factors which come from BPS states with no gauge charge \cite{Bergman:2013ala, Hayashi:2013qwa, Bao:2013pwa, Bergman:2013aca}. In the dual brane picture, those BPS states come from strings between parallel external legs. We call such a factor as decoupled factor. The decoupled factor can be obtained by the refined topological string partition function of a strip diagram with parallel external legs.

\begin{figure}[t]
\begin{center}
\includegraphics[width=100mm]{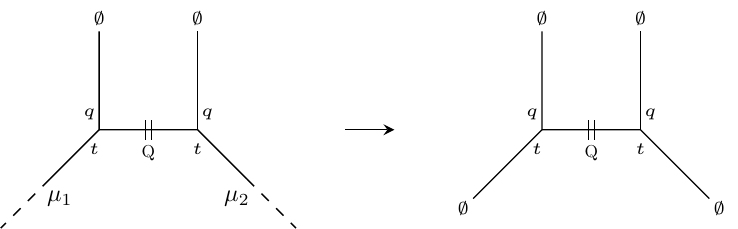}
\end{center}
\caption{A diagram with parallel external legs. The right-hand side stands for a diagram whose refined topological string partition function reproduce the decoupled factor of the theory from the diagram in the left.}
\label{fig:strip}
\end{figure}

Let us consider a simple example of Figure \ref{fig:strip}. 
Left figure in Figure \ref{fig:strip} represents a diagram with parallel external vertical legs. The refined topological string partition function from the diagram contains the decoupled factor associated to the parallel external legs. The contribution can be extracted by computing the refined topological string partition function on the right figure in Figure \ref{fig:strip}, which is the toric diagram of a local Calabi--Yau threefold $\mathcal{O}(-2) \oplus \mathcal{O}$ fibered over $\mathbb{P}^1$. The factor can be explicitly written as
\begin{equation}
Z_{dec, ||} = \prod_{i,j=1}^{\infty}\left(1 -  Q q^{i}t^{j-1}\right)^{-1}.  \label{O(-2)+O}
\end{equation}

Eq.~\eqref{O(-2)+O} may be seen as a part of the perturbative partition function of a vector multiplet. Namely, it does not contain the full spin content of a vector multiplet. From the perspective of 5d BPS states labeled by $(j_l, j_r)$, which are quantum numbers under $SU(2)_l \times SU(2)_r \subset SO(4)$, a vector multiplet may be written as $\left[2(0, 0) + \left(\frac{1}{2}, 0\right)\right]\otimes \left(0, \frac{1}{2}\right)$. Then the contribution \eqref{O(-2)+O} comes from components of the tensor product between a half--hypermultiplet, {\it i.e.} $\left[2(0, 0) + \left(\frac{1}{2}, 0\right)\right]$, and a lower component of $\left(0, \frac{1}{2}\right)$. The non--full spin content appears since the moduli space of the $\mathbb{P}^1$ inside  $\mathcal{O}(-2) \oplus \mathcal{O}$ fibered over $\mathbb{P}^1$ is non--compact. Therefore, as for removing the decoupled factor, we may also say that we remove the contribution of components which do not fill the full spin content of the vector multiplet or recover the invariance under the exchange between $q$ and $t$ \cite{Bergman:2013ala, Bao:2013pwa, Bergman:2013aca}.

After removing the decoupled factor, we can obtain the 5d Nekrasov partition function of the theory realised by a web diagram,
\begin{equation}
Z_{Nek} = \frac{Z_{top}}{Z_{dec}},
\end{equation}
with appropriate parametrisation. 

\section{Schur functions}\label{app:Schur}

In this appendix we will recall some facts about Schur functions that will be used in the main text. Recall that Schur functions $s_\mu(x)$ are certain symmetric polynomials in $n$ variables $x_i$ indexed by a Young diagram $\mu$. Among the different ways of expressing these functions we will choose to use the first Jacobi-Trudi formula to relate them to complete homogeneous symmetric polynomials
\be
s_\mu(x) = \text {det}\, [ h_{\mu_i +j -i}]\,, \quad 1\leq i,j \leq n\,,
\ee
where the complete homogeneous symmetric polynomials of degree $k$ in $n$ variables are defined as
\be
h_k (x) = \sum_{1\leq i_1 \leq i_2 \leq \dots \leq i_k \leq n} \prod_{j=1}^k x_{i_j}\,.
\ee
It is also possible to define a generalisation of Schur functions (called skew Schur functions) in terms of two partitions $\lambda$ and $\mu$ such that $\lambda$ interlaces
$\mu$ as
\be
s_{\mu/\lambda} (x) = \text {det}\, [ h_{\mu_i -\lambda_j+j -i}]\,, \quad 1\leq i,j \leq n\,. \label{skew.schur}
\ee
Since Schur functions form an orthonormal basis of symmetric polynomials it is possible to express skew Schur functions in terms of Schur functions using the 
Littlewood-Richardson coefficients $c_{\mu \nu}^\lambda$
\be
s_{\mu/\lambda} (x) = \sum_{\nu} c_{\mu \nu}^\lambda \, s_{\nu}(x)\,.
\ee
In the computations of topological string partition functions it is often necessary to use the following identities of Schur functions
\be\label{eq:schurid1}
\sum_\mu s_{\mu}(x) s_{\mu}(y) = \prod_{i,j=1}^\infty (1- x_i y_j)^{-1}\,,
\ee
\be\label{eq:schurid2}
\sum_\mu s_{\mu^t}(x) s_{\mu}(y) = \prod_{i,j=1}^\infty (1+ x_i y_j)\,.
\ee
Moreover in the main text we will also use two additional identities: the first one is the following one\footnote{See for instance \cite{Zhou:2003} for a proof of this identity.}
\be\label{eq:schur1}
\sum_{\nu} (-1)^{|\nu|}s_{\mu/\nu}(x) s_{\nu^t/\lambda^t}(x) = (-1)^{|\mu|} \delta_{\mu \lambda}\,,
\ee
and the second one which readily follows from \eqref{eq:schur1} is the following one
\be\label{eq:schur2}
\sum_{\eta,\kappa} (-1)^{|\eta|+|\kappa|}s_{\lambda/\eta}(x)s_{\eta^t/\kappa}(x)s_{\kappa^t/\xi}(x) = s_{\lambda/\xi}(x)\,.
\ee

\section{$Sp(N)$ Nekrasov partition functions}

In section \ref{sec:examples}, we computed the partition functions of rank $2$ $E_6, E_7, E_8$ theories by directly applying the new rules of the topological vertex to the web diagrams. The mass deformation of the rank $2$ $E_6, E_7, E_8$ theories triggers an RG flow to the $Sp(2)$ gauge theories with $N_f = 5, 6, 7$ fundamental and $N_A = 1$ anti-symmetric hypermultiplets respectively. In this appendix, we recall the results of the Nekrasov partition functions of the $Sp(N)$ gauge theories with $N_f$ flavours and an anti-symmetric hypermultiplet.

\subsection{$Sp(N)$ instanton partition functions}
\label{sec:SpN}

The 5d $Sp(N)$ gauge theories with $N_f$ flavours and one anti-symmetric hypermultiplet are realised on the worldvolume theory on $N$ D4-branes with $N_f$ D8-branes and one O8-plane. The Witten index of the ADHM quantum mechanics on $k$ D0-branes in the system gives the $k$ instanton partition function up to a factor which is a contribution of D0-branes moving on the worldvolume of the D8-branes and the O8-plane \cite{Hwang:2014uwa}. The index of the ADHM quantum mechanics is given by the Witten index with some fugacities associated to symmetries of the theory
\begin{equation}
Z^k_{QM}\left(\epsilon_1, \epsilon_2, \alpha_i, z\right) = \text{Tr}\left[(-1)^Fe^{-\beta\{Q, Q^{\dagger}\}}e^{-\epsilon_1(J_1 + J_2)}e^{-\epsilon_2(J_2 + J_R)}e^{-\alpha_i\Pi_i}e^{-m_aF^{\prime}_a}\right]. \label{witten.index}
\end{equation}
$Q, Q^{\dagger}$ are supercharges that commute with the fugacities. $F$ is the Fermion number operator, $J_1, J_2$ are the Cartan generators of the spacetime rotation $SO(4)$, $J_R$ is the Cartan generator of the $SU(2)$ R-symmetry, $\Pi_i$ are the Cartan generators of the gauge group, and $F^{\prime}_a$ are the Cartan generators of the flavour symmetry. $\epsilon_1, \epsilon_2, \alpha_i, m_a$ are chemical potentials associated to the symmetries. In particular, $\alpha_i$ are the Coulomb branch moduli, $m_a$ are the mass parameters. In comparison with the refined topological string partition function, we choose $q = e^{-\epsilon_2}, t=e^{\epsilon_1}$. The instanton partition function is then basically given by $Z_{QM} = \sum_{k}Z_{QM}^ku^k$ up to the factor from the D0-D8-O8 bound states. Here $u$ is the instanton fugacity of the gauge group.

We will simply quote the final result of \eqref{witten.index} for the $Sp(N)$ gauge theories with $N_f$ flavours and one anti-symmetric hypermultiplet. The dual gauge group on the $k$ D0-branes for the $Sp(N)$ gauge theory is $O(k)$, which have two components. Hence, we have two contributions $Z_{\pm}^k$ from $O(k)_{\pm}$ at each instanton order, and the end result from $k$-instanton  is written by \cite{Kim:2012gu, Hwang:2014uwa}
\begin{equation}
Z^k_{QM} = \frac{Z^k_+ + Z^k_-}{2}.
\end{equation}
$Z_{\pm}^k$ consists of the contribution from the vector multiplet $Z_{vec}^{\pm}$, the fundamental hypermultiplet $Z_{fund}^{\pm}$ and the anti-symmetric hypermultiplet $Z_{anti}^{\pm}$
\begin{equation}
Z_{\pm}^k = \frac{1}{|W|}\oint\prod_{I}\frac{d\phi_I}{2\pi i} Z^{\pm}_{vec}\; Z^{\pm}_{fund}\; Z^{\pm}_{anti}. \label{contour.integral}
\end{equation} 
The integration region is a unit circle. $I$ runs from $1$ to $n$ for $O(k)_{+}$ with $k=2n, 2n+1$. As for $O(k)_{-}$, $I$ runs from $1$ to $n$ for $k=2n+1$, and $1$ to $n-1$ for $k=2n$. $|W|$ is the order of the Weyl group of $O(k)_{\pm}$,
\begin{equation}
|W|^{\chi=0}_{+}=\frac{1}{2^{n-1}n!}, \quad |W|^{\chi=1}_{+}=\frac{1}{2^{n-1}(n-1)!}, \quad |W|^{\chi=1}_{-}=\frac{1}{2^{n-1}n!}, \quad |W|^{\chi=0}_{-}=\frac{1}{2^{n}n!},
\end{equation}
where $\chi$ is defined as $k=2n+\chi, \; (\chi=0, 1)$. The integrands from the $O(k)_+$ component is 
\begin{eqnarray}
Z_{vec}^{+} &=&\prod_{I < J}^n 2\sinh\frac{\pm\phi_I\pm\phi_J}{2}\left(\frac{\prod_I^n2\sinh\frac{\pm \phi_I}{2}}{2\sinh\frac{\pm\epsilon_-+\epsilon_+}{2}\prod_{i=1}^N2\sinh\frac{\pm\alpha_i+\epsilon_+}{2}}\prod_{I=1}^n\frac{2\sinh\frac{\pm\phi_I+2\epsilon_+}{2}}{2\sinh\frac{\pm\phi_I\pm\epsilon_-+\epsilon_+}{2}}\right)^{\chi}\nonumber\\
&&\prod_{I=1}^n\frac{2\sinh\epsilon_+}{2\sinh\frac{\pm\epsilon_-+\epsilon_+}{2}2\sinh\frac{\pm 2\phi_I\pm\epsilon_-+\epsilon_+}{2}\prod_{i=1}^N2\sinh\frac{\pm\phi_I\pm\alpha_i+\epsilon_+}{2}}\prod_{I<J}^n\frac{2\sinh\frac{\pm\phi_I\pm\phi_j+2\epsilon_+}{2}}{2\sinh\frac{\pm\phi_I\pm\phi_J\pm\epsilon_-+\epsilon_+}{2}}\nonumber\\
\end{eqnarray}
\begin{eqnarray}
Z_{anti}^+ &=&\left(\frac{\prod_{i=1}^N2\sinh\frac{m\pm\alpha_i}{2}}{2\sinh\frac{m\pm\epsilon_+}{2}}\prod_{I=1}^n\frac{2\sinh\frac{\pm\phi_I\pm m-\epsilon_-}{2}}{2\sinh\frac{\pm\phi_I\pm m -\epsilon_-}{2}}\right)^{\chi}\prod_{I=1}^n\frac{2\sinh\frac{\pm m -\epsilon_-}{2}\prod_{i=1}^N 2\sinh\frac{\pm\phi_I\pm\alpha_i -m}{2}}{2\sinh\frac{\pm m -\epsilon_+}{2}2\sinh\frac{\pm 2\phi_I\pm m -\epsilon_+}{2}}\nonumber\\
&&\prod_{I< J}^n\frac{2\sinh\frac{\pm \phi_I \pm \phi_J \pm m -\epsilon_-}{2}}{2\sinh\frac{\pm\phi_I\pm\phi_J\pm m -\epsilon_+}{2}}
\end{eqnarray}
\begin{eqnarray}
Z_{fund}^+ = \prod_{a=1}^{N_f}\left\{\left(2\sinh\frac{m_a}{2}\right)^{\chi}\prod_{I=1}^{n}2\sinh\frac{\pm\phi_I + m_a}{2}\right\}.
\end{eqnarray}
Here we defined $2\sinh(a\pm b) = 2\sinh(a) \; 2\sinh(b)$. We also defined $\epsilon_+ = \frac{\epsilon_1 + \epsilon_2}{2}, \epsilon_- = \frac{\epsilon_1 - \epsilon_2}{2}$. On the other hand, the integrands from the $O(k)_-$ component depend on whether $k$ is even or odd. When $k=2n+1$, we have
\begin{eqnarray}
Z_{vec}^{-} &=&\prod_{I < J}^n 2\sinh\frac{\pm\phi_I\pm\phi_J}{2}\left(\frac{\prod_I^n2\cosh\frac{\pm \phi_I}{2}}{2\sinh\frac{\pm\epsilon_-+\epsilon_+}{2}\prod_{i=1}^N2\cosh\frac{\pm\alpha_i+\epsilon_+}{2}}\prod_{I=1}^n\frac{2\cosh\frac{\pm\phi_I+2\epsilon_+}{2}}{2\cosh\frac{\pm\phi_I\pm\epsilon_-+\epsilon_+}{2}}\right)\nonumber\\
&&\prod_{I=1}^n\frac{2\sinh\epsilon_+}{2\sinh\frac{\pm\epsilon_-+\epsilon_+}{2}2\sinh\frac{\pm 2\phi_I\pm\epsilon_-+\epsilon_+}{2}\prod_{i=1}^N2\sinh\frac{\pm\phi_I\pm\alpha_i+\epsilon_+}{2}}\prod_{I<J}^n\frac{2\sinh\frac{\pm\phi_I\pm\phi_j+2\epsilon_+}{2}}{2\sinh\frac{\pm\phi_I\pm\phi_J\pm\epsilon_-+\epsilon_+}{2}}\nonumber\\
\end{eqnarray}
\begin{eqnarray}
Z_{anti}^- &=&\frac{\prod_{i=1}^N2\cosh\frac{m\pm\alpha_i}{2}}{2\sinh\frac{m\pm\epsilon_+}{2}}\prod_{I=1}^n\frac{2\cosh\frac{\pm\phi_I\pm m-\epsilon_-}{2}}{2\cosh\frac{\pm\phi_I\pm m -\epsilon_-}{2}}\frac{2\sinh\frac{\pm m -\epsilon_-}{2}\prod_{i=1}^N 2\sinh\frac{\pm\phi_I\pm\alpha_i -m}{2}}{2\sinh\frac{\pm m -\epsilon_+}{2}2\sinh\frac{\pm 2\phi_I\pm m -\epsilon_+}{2}}\nonumber\\
&&\prod_{I< J}^n\frac{2\sinh\frac{\pm \phi_I \pm \phi_J \pm m -\epsilon_-}{2}}{2\sinh\frac{\pm\phi_I\pm\phi_J\pm m -\epsilon_+}{2}}
\end{eqnarray}
\begin{eqnarray}
Z_{fund}^- = \prod_{a=1}^{N_f}\left(2\cosh\frac{m_a}{2}\prod_{I=1}^{n}2\sinh\frac{\pm\phi_I + m_a}{2}\right).
\end{eqnarray}
When $k=2n$, we have
\begin{eqnarray}
Z_{vec}^{-} &=&\prod_{I < J}^n 2\sinh\frac{\pm\phi_I\pm\phi_J}{2}\prod_I^{n-1}2\sinh(\pm \phi_I)\nonumber\\
&&\frac{2\cosh(\epsilon_+)}{2\sinh\frac{\pm\epsilon_-+\epsilon_+}{2}2\sinh(\pm\epsilon_-+\epsilon_+)\prod_{i=1}^N2\sinh(\pm\alpha_i+\epsilon_+)}\prod_{I=1}^{n-1}\frac{2\sinh(\pm\phi_I+2\epsilon_+)}{2\sinh(\pm\phi_I\pm\epsilon_-+\epsilon_+)}\nonumber\\
&&\prod_{I=1}^{n-1}\frac{2\sinh\epsilon_+}{2\sinh\frac{\pm\epsilon_-+\epsilon_+}{2}2\sinh\frac{\pm 2\phi_I\pm\epsilon_-+\epsilon_+}{2}\prod_{i=1}^N2\sinh\frac{\pm\phi_I\pm\alpha_i+\epsilon_+}{2}}\prod_{I<J}^{n-1}\frac{2\sinh\frac{\pm\phi_I\pm\phi_j+2\epsilon_+}{2}}{2\sinh\frac{\pm\phi_I\pm\phi_J\pm\epsilon_-+\epsilon_+}{2}}\nonumber\\
\end{eqnarray}
\begin{eqnarray}
Z_{anti}^- &=&\frac{2\cosh\frac{\pm m -\epsilon_-}{2}\prod_{i=1}^N2\sinh(m\pm\alpha_i)}{2\sinh\frac{m\pm \epsilon_+}{2}2\sinh(m\pm\epsilon_+)}\nonumber\\
&&\prod_{I=1}^{n-1}\frac{2\sinh(\pm\phi_I\pm m-\epsilon_-)}{2\sinh(\pm\phi_I\pm m -\epsilon_-)}\frac{2\sinh\frac{\pm m -\epsilon_-}{2}\prod_{i=1}^N 2\sinh\frac{\pm\phi_I\pm\alpha_i -m}{2}}{2\sinh\frac{\pm m -\epsilon_+}{2}2\sinh\frac{\pm 2\phi_I\pm m -\epsilon_+}{2}}\prod_{I< J}^{n-1}\frac{2\sinh\frac{\pm \phi_I \pm \phi_J \pm m -\epsilon_-}{2}}{2\sinh\frac{\pm\phi_I\pm\phi_J\pm m -\epsilon_+}{2}}\nonumber\\
\end{eqnarray}
\begin{eqnarray}
Z_{fund}^- = \prod_{a=1}^{N_f}\left(2\sinh m_a\prod_{I=1}^{n-1}2\sinh\frac{\pm\phi_I + m_a}{2}\right).
\end{eqnarray}
Note that we included the Haar measure into the integrands.

The summation $Z_{QM}$ is not exactly the instanton partition function of the $Sp(N)$ gauge theories but we need to factor out the contribution from the D0-D8-O8 bound states. The contribution is written by the Plethystic exponential
\begin{equation}
Z_{\text{D0-D8-O8}}=PE[f_{N_f}(x, y, v, w_i, u)] = \exp\left[\sum_{n=1}^{\infty} \frac{f_{N_f}(x^n, y^n, v^n, w_i^n, u^n)}{n}\right],
\end{equation}
where $f_{N_f}$ is \cite{Hwang:2014uwa}
\begin{eqnarray}
f_0 &=& -\frac{t^2}{(1-tu)\left(1-\frac{t}{u}\right)(1-tv)\left(1-\frac{t}{v}\right)}q,\\
f_{N_f} &=&  -\frac{t^2}{(1-tu)\left(1-\frac{t}{u}\right)(1-tv)\left(1-\frac{t}{v}\right)}q\chi(y_i)^{SO(2N_f)}_{2^{N_f-1}}\quad \text{for $1 \leq N_f \leq 5$},\\
f_{7} &=& -\frac{t^2}{(1-tu)\left(1-\frac{t}{u}\right)(1-tv)\left(1-\frac{t}{v}\right)}\left(q\chi(y_i)^{SO(12)}_{32}+q^2\right),\\
f_{8} &=&  -\frac{t^2}{(1-tu)\left(1-\frac{t}{u}\right)(1-tv)\left(1-\frac{t}{v}\right)}\left(q\chi(y_i)^{SO(14)}_{64}+q^2\chi(y_i)^{SO(14)}_{14}\right),
\end{eqnarray}
We defined $x = e^{-\epsilon_+}, y= e^{-\epsilon_-}, v= e^{-m}, w_i = e^{\frac{m_a}{2}}$ for $a = 1, \cdots, N_f$. $\chi(w_i)^{SO(2N_f)}_{2^{N_f-1}}$ is the character of the positive chirality spinor representation of $SO(2N_f)$. The explicit expression is 
\begin{equation}
\chi(w_i)^{SO(2N_f)}_{2^{N_f-1}} = \frac{1}{2}\prod_{a=1}^{N_f}2\sinh \frac{m_a}{2} +  \frac{1}{2}\prod_{a=1}^{N_f}2\cosh \frac{m_a}{2}.
\end{equation}
The instanton partition functions of the $Sp(N)$ gauge theories are then given by
\begin{equation}
Z_{inst}^{Sp(N)} = \frac{Z_{QM}}{Z_{\text{D0-D8-O8}}}. \label{SpN}
\end{equation}

Let us explicitly write down the $1$-instanton part of \eqref{SpN}. At the $1$-instanton order, we do not have the contour integral in \eqref{contour.integral} and it is straightforward to obtain
\begin{eqnarray}
Z_{1\text{-}inst}^{Sp(N)} &=& \frac{1}{2}\Big\{\frac{\prod_{a=1}^{N_f}2\sinh\frac{m_a}{2}}{2\sinh\frac{\epsilon_{+}\pm \epsilon_-}{2}\prod_{i=1}^N2\sinh\frac{\pm\alpha_i + \epsilon_+}{2}}\frac{\prod_{i=1}^N2\sinh\frac{m\pm \alpha_i}{2} - \prod_{i=1}^N2\sinh\frac{\pm\alpha_i +\epsilon_+}{2}}{2\sinh\frac{m\pm \epsilon_+}{2}}\nonumber\\
&&+\frac{\prod_{a=1}^{N_f}2\cosh\frac{m_a}{2}}{2\sinh\frac{\epsilon_{+}\pm \epsilon_-}{2}\prod_{i=1}^N2\cosh\frac{\pm\alpha_i + \epsilon_+}{2}}\frac{\prod_{i=1}^N2\cosh\frac{m\pm \alpha_i}{2} - \prod_{i=1}^N2\cosh\frac{\pm\alpha_i +\epsilon_+}{2}}{2\sinh\frac{m\pm \epsilon_+}{2}}\Big\}.\nonumber \\ \label{SpN.1instanton}
\end{eqnarray}
When, $N=1$, then \eqref{SpN.1instanton} reduces to 
\begin{equation}
Z_{1\text{-}inst}^{Sp(1)} = \frac{1}{2}\left\{\frac{\prod_{a=1}^{N_f}2\sinh\frac{m_a}{2}}{2\sinh\frac{\epsilon_{+}\pm \epsilon_-}{2}2\sinh\frac{\pm\alpha + \epsilon_+}{2}}+\frac{\prod_{a=1}^{N_f}2\cosh\frac{m_a}{2}}{2\sinh\frac{\epsilon_{+}\pm \epsilon_-}{2}2\cosh\frac{\pm\alpha + \epsilon_+}{2}}\right\}.\label{Sp1.1instanton}
\end{equation}
In the special cases with $m=0, \epsilon_+=0$, the $1$-instanton part of the $Sp(N)$ gauge theory, \eqref{SpN.1instanton}, can be written by the $N$ copies of the $1$-instanton part of the $Sp(1)$ gauge theory, \eqref{Sp1.1instanton},
\begin{equation}
Z_{1\text{-}inst}^{Sp(N)}|_{m= \epsilon_+ = 0} = \frac{1}{2}\sum_{i=1}^N \left\{\frac{\prod_{a=1}^{N_f}2\sinh\frac{m_a}{2}}{2\sinh\frac{\pm \epsilon_-}{2}2\sinh\frac{\pm\alpha_i}{2}}+\frac{\prod_{a=1}^{N_f}2\cosh\frac{m_a}{2}}{2\sinh\frac{\pm \epsilon_-}{2}2\cosh\frac{\pm\alpha_i}{2}}\right\}. \label{SpN.sp.1instanton}
\end{equation}
One can show \eqref{SpN.sp.1instanton} for arbitrary $N$ by induction.

From the $2$-instanton order, we need to carefully evaluate the contour integral in \eqref{contour.integral} \cite{Hwang:2014uwa}. In this case it is necessary to evaluate the contour integrals only 
for the $O(2)_+$ component. In particular we find that the integrand has simple poles at the zeroes of the hyperbolic sines in the denominator with the general form
\be
\frac{1}{ \sinh \frac{Q \phi+\dots}{2}}\,,
\ee
where $Q$ is an integer. The prescription of \cite{Hwang:2014uwa} to compute $Z^2_+$ is to take the sum of the residues of the integrand at the poles with $Q>0$. Equivalently we can take
the contour of integration to be the unit circle in the variable $z = e^\phi$ and replace $t=e^{-\epsilon_+}$ in $Z_{vec}^+$ and $T = e^{-\epsilon_+}$ in $Z_{anti}^+$ and taking $t \ll1$ and
$T\gg1$. These two procedures are equivalent because if $t$ is taken sufficiently small and $T$ sufficiently large only the poles with $Q>0$ will lay inside the unit circle in $z$. We find that for 
an $Sp(N)$ gauge theory with one anti-symmetric hypermultiplet and $N_f$ fundamental flavours there are $8 + 2 N$ poles, 4 coming from $Z_{anti}^+$ and $4+2N$ coming from $Z_{vec}^+$.
In particular the poles are located at
\be\begin{split}
&\phi_1 = -\frac{1}{2}(\epsilon_+ + \epsilon_-)\,,\quad \phi_2 = -\frac{1}{2}(\epsilon_+ + \epsilon_-)+i \pi\,,\\& \phi_3 = -\frac{1}{2}(\epsilon_+ - \epsilon_-)\,,\quad \phi_4 = -\frac{1}{2}(\epsilon_+ - \epsilon_-)+i \pi\,,\\
& \phi_5 =  \frac{1}{2}(\epsilon_+ - m)\,,\quad \phi_6 =  \frac{1}{2}(\epsilon_+ - m)+i \pi\,,\\
& \phi_7 =  \frac{1}{2}(\epsilon_+ + m)\,,\quad \phi_8 =  \frac{1}{2}(\epsilon_+ + m)+i \pi\,,\\
& \phi_{8+i} = -\epsilon_+- \alpha_i\,, \quad \phi_{8+N+i} = -\epsilon_++ \alpha_i\,, \quad i = 1, \dots,N\,.
\end{split}\ee
By evaluating the residues at the poles. we obtain the $Sp(N)$ partition function from the $O(2)_+$ component. The $Sp(N)$ partition function from the $O(2)_-$ component is obtained straightforwardly since it does not involve the contour integrals.

\subsection{Perturbative partition functions}

We summarise the results of the perturbative contribution to the partition function. 

The perturbative contribution from a vector multiplet of a gauge group $G$ is \cite{Kim:2012gu}
\begin{equation}
\prod_{i,j=1}^{\infty} \prod_{r}\left(2\sinh\frac{(i-1)\epsilon_1 + (j-1)\epsilon_2 + r\cdot\alpha}{2}\right)^{\frac{1}{2}}\left(2\sinh\frac{i\epsilon_1 + j\epsilon_2 + r\cdot\alpha}{2}\right)^{\frac{1}{2}}, \label{pert.vec.field}
\end{equation}
where $r$ represents all the root vectors of the Lie algebra of $\mathfrak{g}$. $\mathbb{\alpha}$  is a vector of the Coulomb branch moduli. On the other hand, the perturbative contribution from a hypermultiplet in a representation $R$ and a mass $m$ is \cite{Kim:2012gu} 
\begin{equation}
\prod_{i,j=1}^{\infty} \prod_{w \in R}\left(2 \sinh\frac{\left(i-\frac{1}{2}\right)\epsilon_1+\left(j-\frac{1}{2}\right)\epsilon_2 + w\cdot \alpha - m}{2}\right)^{-1}, \label{pert.hyper.field}
\end{equation}
where $w$ is all the weight vectors of the representation $R$. To obtain \eqref{pert.vec.field} and \eqref{pert.hyper.field} on the field theory side, constant divergent factors are factored out. Therefore, we will ignore any constant prefactor in the perturbative partition functions.

Then, we move on to the unrefined case of $\epsilon_1 = -\epsilon_2 = \epsilon$ and obtain the expression of the perturbative partition function which is appropriate for the comparison with the perturbative contribution in the topological string partition function. We also focus on the $Sp(N)$ gauge theory with five fundamental hypermultiplets with mass $m_a, a=1, \cdots N_f$ and an anti-symmetric hypermultiplet with mass $m$. The vector multiplet contribution can be written as
\begin{eqnarray}
Z_{pert}^{vec} &&= \prod_{i,j=1}^{\infty} \Big[\left(1 - q^{i+j-1}\right)^{-N}\nonumber\\ &&\prod_{1 \leq k < l \leq N}\left(1 - e^{\alpha_k + \alpha_l }q^{i+j-1}\right)^{-1}\left(1 - e^{-\alpha_k + \alpha_l }q^{i+j-1}\right)^{-1}\left(1 - e^{\alpha_k - \alpha_l }q^{i+j-1}\right)^{-1}\left(1 - e^{-\alpha_k -\alpha_l }q^{i+j-1}\right)^{-1}\nonumber \\ 
&&\prod_{1 \leq k \leq N}\left(1 - e^{2\alpha_k }q^{i+j-1}\right)^{-1}\left(1 - e^{-2\alpha_k }q^{i+j-1}\right)^{-1}\Big] \label{pert.vec}
\end{eqnarray}
where the explicit expressions of the root vectors $r = \pm e_{k} \pm e_{l}, \pm 2e_k$ with $1 \leq k < l \leq N$ are used. The first line of \eqref{pert.vec} is the contribution from the Cartan part of the $Sp(N)$ vector multiplet. To obtain the expression \eqref{pert.vec}, we used analytic continuation
\begin{eqnarray}
&&\prod_{i,j=1}^{\infty}\left(1 - Qt^{i-1}q^{-j+1}\right)^{\frac{1}{2}} =   \exp\left(\sum_{k=1}^{\infty}\frac{Q^k}{2k}\frac{1}{\left(1-t^{k}\right)}\frac{q^k}{\left(1-q^{k}\right)}\right)  = \prod_{i,j=1}^{\infty}\left(1 - Qt^{i-1}q^{j}\right)^{-\frac{1}{2}},\nonumber\\
&&\prod_{i,j=1}^{\infty}\left(1 - Qt^{i}q^{-j}\right)^{\frac{1}{2}} =   \exp\left(\sum_{k=1}^{\infty}\frac{Q^k}{2k}\frac{t^k}{\left(1-t^{k}\right)}\frac{1}{\left(1-q^{k}\right)}\right)  = \prod_{i,j=1}^{\infty}\left(1 - Qt^{i}q^{j-1}\right)^{-\frac{1}{2}},\nonumber
\end{eqnarray}
and then setting $t = q$. Similarly, the perturbative contribution from the fundamental hypermultiplets is 
\begin{equation}
Z_{pert}^{fund} =  \prod_{i,j=1}^{\infty} \prod_{a=1}^{N_f}\prod_{k=1}^N\left(e^{-m_a}\right)\left(1 - e^{\alpha_k - m_a}q^{i+j-1}\right)\left(1 - e^{-\alpha_k - m_a}q^{i+j-1}\right). \label{pert.fund}
\end{equation}
We used the fact that the weights of the fundamental representation are $\pm e_k, (k=1, \cdots, N)$, and also analytic continuation 
\begin{equation}
\prod_{i,j=1}^{\infty}\left(1 - Qt^{i-\frac{1}{2}}q^{-j+\frac{1}{2}}\right)^{-1} =   \exp\left(-\sum_{k=1}^{\infty}\frac{Q^k}{k}\frac{t^{\frac{k}{2}}}{\left(1-t^{k}\right)}\frac{q^{\frac{k}{2}}}{\left(1-q^{k}\right)}\right)  = \prod_{i,j=1}^{\infty}\left(1 - Qt^{i-\frac{1}{2}}q^{j-\frac{1}{2}}\right),\label{analytic1}
\end{equation}
and then setting $t=q$. Note that we have an overall divergent factor in \eqref{pert.fund}. In the comparison with the perturbative contribution from the topological string partition function, we ignore the factor. 
The perturbative contribution from the anti-symmetric hypermultiplet is 
\begin{eqnarray}
Z_{pert}^{anti\text{-}sym} =  \prod_{i,j=1}^{\infty}\prod_{1 \leq k < l \leq N}&&\left(-e^{-\frac{5}{2}m}\right)\left(1 - e^{\alpha_k + \alpha_l - m}q^{i+j-1}\right)\left(1 - e^{-\alpha_k + \alpha_l - m}q^{i+j-1}\right)\nonumber\\
&&\left(1 - e^{\alpha_k - \alpha_l - m}q^{i+j-1}\right)\left(1 - e^{-\alpha_k -\alpha_l - m}q^{i+j-1}\right)\nonumber\\&&\left(1-e^{-m} q^{i+j-1}\right), \label{pert.antisym}
\end{eqnarray}
where we used the fact that the weights of the anti-symmetric representation are $\pm e_k \pm e_l, \; (1 \leq k < l \leq  N)$ and $0$, and also the analytic continuation \eqref{analytic1} again. 
Hence, the perturbative partition function of the $Sp(N)$ gauge theory with $N_f$ fundamental hypermultiplets and one anti-symmetric hypermultiplet is simply given by the product of \eqref{pert.vec}, \eqref{pert.fund} and \eqref{pert.antisym}.

When we further concentrate on the case where mass of the anti-symmetric hypermultiplet is zero, $i.e. \; m=0$, there is a cancellation between the contribution of the anti-symmetric hypermultiplet and that of a part of the vector multiplet. In fact, the perturbative partition function of the anti-symmetric hypermultiplet disappears up to the divergent factor. More explicitly, the perturbative partition function becomes 
\begin{eqnarray}
Z_{pert} = \prod_{i,j=1}^{\infty}\left[\left(1 - q^{i+j-1}\right)^{-N+1}\prod_{k=1}^{N_f}\frac{ \prod_{a=1}^{N_f} \left(1 - e^{\alpha_k - m_a}q^{i+j-1}\right)\left(1 - e^{-\alpha_k - m_a}q^{i+j-1}\right)}{\left(1 - e^{2\alpha_k }q^{i+j-1}\right)\left(1 - e^{-2\alpha_k }q^{i+j-1}\right)} \right],\nonumber\\\label{pert.gauge}
\end{eqnarray}
up to the divergent factors. We compare \eqref{pert.gauge} with the perturbative part of the partition function obtained from the topological vertex computation for rank $N$ $E_6, E_7, E_8$ theories. 

There is another subtle point when one tries to compare \eqref{pert.gauge} with the topological string result. By using the identity
\begin{equation}
\prod_{i,j=1}^{\infty}\left(1 - Qq^{i+j-1}\right) = \prod_{n=1}^{\infty}\left(1 - Qq^n\right)^n,
\end{equation}
 one can obtain
\begin{eqnarray}
\prod_{n=1}^{\infty}\left(1 - Qq^n\right)^n &=& \prod_{n=1}^{\infty}\left(-Q\right)^nq^{n^2}\left(1 - Q^{-1}q^{-n}\right)^n\nonumber\\
&=& \prod_{n=1}^N \left(-Q\right)^n q^{n^2} \left(1 - Q^{-1} q^n\right)^n\nonumber\\
&=&\prod_{n=1}^N \left(-Q\right)^n \left(1 - Q^{-1} q^n\right)^n. \label{eq.transform}
\end{eqnarray}
From the first line to the second line of \eqref{eq.transform}, we used analytic continuation 
\begin{equation}
 \prod_{n=1}^{\infty}\left(1 - Q^{-1} q^{-n}\right)^n =  \exp\left(-\sum_{k=1}^{\infty}\frac{Q^k}{k}\frac{q^k}{\left(1-q^k\right)^2}\right) = \prod_{n=1}^{\infty}\left(1 - Q^{-1} q^n\right)^n.
\end{equation}
From the second line to the third line of \eqref{eq.transform} we used the zeta function regularisation $\zeta(-2) = 0$. Therefore, as for the perturbative factor, the factor with $Q$ is basically the same as the factor with $Q^{-1}$ up to the regularisation and also the divergent factor. In the comparison between \eqref{pert.gauge} and the perturbative contribution from the topological string partition function, we will neglect the subtlety regarding the regularisation, the divergent factor and also the analytic continuation.

\bibliographystyle{JHEP}
\bibliography{refs}

\end{document}